\documentclass[11pt, a4paper]{article}
\usepackage{jheppub, amsmath, amssymb, bbm, color, graphicx, hyperref, pgf, tikz}
\usepackage{braket,mathtools, subcaption, textcomp}
\usepackage[export]{adjustbox}
\usepackage{algorithm}
\usepackage{algpseudocode}
\renewcommand\[{\begin{equation}}
\renewcommand\]{\end{equation}}
\newcommand{\ba}{\begin{eqnarray}}
\newcommand{\ea}{\end{eqnarray}}

\renewcommand{\eqref}[1]{Eq.\,(\ref{#1})}

\def\i{\mathrm i}

\begin{document}

\title{A Monte Carlo algorithm for spin foam intertwiners}

\author[a]{Sebastian Steinhaus}
\emailAdd{sebastian.steinhaus@uni-jena.de}

\affiliation[a]{Theoretisch-Physikalisches Institut, Friedrich-Schiller-Universit\"{a}t Jena\\ Max-Wien-Platz 1, 07743 Jena, Germany, EU}

\begin{abstract}{
Monte Carlo algorithms are barely considered in spin foam quantum gravity. Due to the quantum nature of spin foam amplitudes one cannot readily apply them, and the present sign problem is a threat to convergence and thus efficiency. Yet, ultimately the severity of the sign problem in spin foams is not known. In this article we propose a new probability distribution for coherent (boundary) intertwiners, which we use to define a Markov Chain Monte Carlo algorithm. We apply this algorithm to the SU$(2)$ coherent vertex amplitude for various Regge-type boundary data and find convergent, accurate results at far lower costs than the explicit calculation. The resources are instead used to increase the size of boundary spins, bridging the gap to the asymptotic formulae. While the sign problem is not solved, it is under control in the vast majority of cases. We close by discussing how this algorithm can be extended to larger triangulations with boundary and 4d Lorentzian spin foam models. More speculatively, we discuss how this algorithm can be used to sample (Regge-geometric) bulk intertwiners.
}
\end{abstract}

\setcounter{tocdepth}{2}

\maketitle

\section{Introduction}
Spin foam quantum gravity~\cite{Perez:2013uz,Engle:2023qsu} is a non-perturbative, background independent path integral approach of quantum gravity, which is often referred to as a covariant formulation of loop quantum gravity~\cite{Rovelli:2004wb,Thiemann:2007wt}. It constructs quantum space-time by gluing and superimposing quantum geometric building blocks~\cite{Baez:1999tk}. These building blocks and their associated amplitudes are derived from general relativity written as a constrained topological quantum field theory~\cite{Baez:2000kp,Plebanski:1977zz}. To regularize the path integral, the theory is discretrised on a 2-complex, which is typically chosen to be dual to a triangulation, and then colored with group theoretic data encoding the quantum geometry. Eventually, the dynamics is implemented by summing over these data weighted by quantum amplitudes.

Beyond the theoretical elevator pitch, we face the challenge of turning spin foam models into a computational formalism: we need to be able to efficiently and reliably extract results from the theory, e.g. expectation values of observables describing a semi-classical geometry. Part of this question is to understand the impact of the discretisation and how to remove it in a suitable continuum limit~\cite{Dittrich:2014ui,Steinhaus:2020lgb,Asante:2022dnj}. Thus, we must perform explicit non-perturbative calculations on large triangulations with many degrees of freedom for which efficient numerical techniques are vital.

However, the before mentioned quantum, i.e. non-Wick-rotated, amplitudes pose a challenge: they are not positive (semi-)definite and can be complex, e.g. after introducing coherent states~\cite{Livine:2007bq}, and are typically highly oscillatory. Many methods in spin foams developed and used so far are adapted to this. The derivations of asymptotic formulae of spin foam vertex amplitudes~\cite{Conrady:2008mk,Barrett:2009ci,Barrett:2009mw,Kaminski:2017eew,Liu:2018gfc,Simao:2021qno}, which oscillate with the Regge action~\cite{Regge:1961ct}, exploit this oscillatory nature by employing stationary phase analysis. In recent years, this was generalized to larger triangulations using so-called complex critical points~\cite{Han:2021kll,Han:2023cen,Han:2024ydv,Dona:2022yyn}. In this context complex refers to relevant configuration in addition to the dominating real critical points, where the former include curved configurations. Thus, this method suggest a route to avoid the so-called flatness problem~\cite{Hellmann:2013gva,Engle:2020ffj,Engle:2021xfs}.

However, asymptotic methods are valid approximationg only for large representations, they are not accurate for small representations where geometry exhibits a more quantum behavior.
The gap to the small representation regime is being bridged numerically in recent years~\cite{Dona:2017dvf,Dona:2019dkf,Dona:2022dxs,Dona:2022yyn}, with explicit calculations tackling triangulations with multiple simplices~\cite{Dona:2020tvv,Dona:2022vyh,Dona:2023myv}. These works rely on an explicit numerical implementaion of the representation theoretic calculations utilizing suitable analytical identities and efficient recoupling symbol libraries (mostly) avoiding truncations. The outstanding achievement is the creation of the library \verb|sl2cfoam-next|\footnote{\url{https://github.com/qg-cpt-marseille/sl2cfoam-next}}~\cite{Gozzini:2021kbt} with which amplitudes of the Lorentzian EPRL-FK model~\cite{Engle:2008fj,Freidel:2008fv} can be computed. These have been applied to a variety of physical systems~\cite{Sarno:2018ses,Frisoni:2022urv,Frisoni:2023agk}. Nevertheless, these calculations are costly, and what is more that these costs grow rapidly with the size of (boundary) representations as well as the size of triangulation, i.e. more degrees of freedom to explicitly sum over.

In an attempt to forego these costly computations and focus on the explicit path integral evaluation for larger triangulations, effective spin foams~\cite{Asante:2020qpa,Asante:2020iwm,Asante:2021phx,Asante:2022lnp,Dittrich:2023rcr} were developed: instead of using the full amplitude, one assigns the exponentiated Regge action to each vertex restricting each vertex individually to a geometric $4$-simplex. Such amplitudes are rapid to compute and the free resources can instead be used to explicitly sum over geometries. This sum is defined as a sum over all areas using the area Regge action~\cite{Barrett:1999fa,Dittrich:2008hg,Asante:2018wqy}. To return to length Regge calculus second class simplicity constraints are modeled as Gaussian-shaped gluing constraints peaked on shape matching tetrahedra\footnote{The Gaussians are peaked on matching 3d dihedral angles, more precisely two non-opposite ones. The four triangle areas agree by definition.}, such that two glued $4$-simplices agree on the lengths assigned to the shared tetrahedron. This particular realization also suggests a reason (and how to avoid) the flatness problem~\cite{Asante:2020iwm}. One goal of this approach is then to investigate the emergence of semi-classical physics from the dynamics of the quantum theory, e.g. cosmology from Regge calculus~\cite{Liu:2015gpa,Dittrich:2021gww,Jercher:2023csk}. One promising numerical method has been recently rediscovered~\cite{Dittrich:2022yoo}, so-called acceleration operators~\cite{Schmidt41,Shanks55}, that are efficient for computing sums of highly oscillating functions as they appear in spin foams. Related to effective spin foams are also so-called restricted spin foams~\cite{Bahr:2016co,Bahr:2017bn,Assanioussi:2020fml}, which restrict the spin foam path integral to particular configurations, e.g. cuboids, and utilizes the asymptotic formula to accelerate numerical calculations. This strategy was e.g. successful to investigate coarse graining flows~\cite{Bahr:2016dl,Bahr:2017kr,Bahr:2018gwf}, observables like the spectral dimension~\cite{Steinhaus:2018aav,Jercher:2023rno} and matter coupling~\cite{Ali:2022vhn}.

However, summing / integrating over a large numbers of variables explicitly becomes inefficient as the number of configurations in general grows exponentially. Monte Carlo methods instead attempt to find an approximation by considering a finite number of samples of configurations, hence the costs do not scale with the number of variables. These samples are generated (semi-)randomly from a probability distribution, which for statistical systems is often derived from the partition function; this is then used to approximate expectation values. If this approximation is accurate for small sample sizes, the algorithm is often more efficient than explicit summation. Unfortunately, there is justified doubt that Monte Carlo methods are suitable for spin foams.
The root of this doubt is that spin foams are quantum theories, i.e. they assign complex quantum amplitudes to each configuration. There is no obvious way how to analytically continue spin foams to purely statistical weights\footnote{See~\cite{Han:2021rjo,Dona:2021ldn} for analytical continuations that relate Lorentzian and Euclidean signature models.}. This is in contrast to causal dynamical triangulations~\cite{Ambjorn:2012vc} and causal set theory~\cite{Surya:2019ndm}, where the theory can be continued to a purely statistical one, which can then be investigated by Markov chain Monte Carlo techniques. 
While this is a disadvantage for spin foams at the practical level, there are reasons not to Wick-rotate, since Euclidean quantum gravity models suffer from the conformal mode problem, where the Euclidean Einstein Hilbert action is not bounded from below\footnote{Note that causal dynamical triangulations do not suffer from this by realizing that the space of (Wick-rotated) Lorentzian configurations is different from the set of Euclidean configurations.}. Indeed, quantum amplitudes might be beneficial as they can lead to destructive interference, e.g. there are indications in causal set theory that quantum amplitudes suppress non-manifold-like causal sets~\cite{Carlip:2022nsv}.

However, if we aim to implement importance sampling Monte Carlo techniques in spin foams we face a technical and a fundamental problem. On the technical side we require a probability distribution, which is at best directly derived from the dynamics. Due to complex amplitudes, the spin foam partition function cannot be used directly to define such a distribution. Still, it is possible to ``guess'' a distribution, which can then be used to sample configurations. The simplest choice is the constant one, leading to random sampling, which is however not at all informed by the dynamics. It is the goal of this article to propose a new distribution for coherent intertwiners and apply this to the coherent spin foam vertex amplitude. Choosing a probability distribution however does not address the fundamental problem: the sign problem. If the functions we are investigating are alternating in sign / oscillating, contributions from different configurations will interfere destructively and cancel. However when sampling configurations we are likely to miss these cancellations. This might lead to slow convergence (if at all) and thus require a larger number of samples, which renders the algorithm less efficient. In our opinion, it is important to distinguish these two issues clearly: even for complex amplitudes we can define and apply Monte Carlo techniques by introducing a new probability distribution. This however does not address the sign problem, thus it does not guarantee that these methods efficiently produce reliable results.
%In our opinion, it is important to differentiate these issues. The first one is a more practical issue, which can be overcome by proposing a new probability distribution that is suited to the problem at hand. It is the goal of this article to propose such a distribution. Still, the thus defined Monte Carlo algorithm might fail if the sign problem is too severe and one obtains non-convergent results. However, we do not know how severe the sign problem is in spin foam models. We will show that at least for the coherent SU$(2)$ BF vertex amplitude we find convergent results.

Still, the sign problem does not rule out the usefulness of Monte Carlo methods for spin foams. Particular probability distributions, e.g. one derived from a Lefshetz thimble, do not suffer from the sign problem and can be used to investigate expectation values of observables~\cite{Han:2020npv}. A Lefshetz thimble is defined by deforming an integration contour such that the imaginary part of the integrand is constant on that contour, making the expression non-oscillatory. Also, in particular situations, e.g. symmetry restrictions, the spin foam partition function can be positive semi-definite and be used for importance sampling~\cite{Bahr:2016co,Bahr:2016dl,Bahr:2017kr,Ali:2022vhn,Frisoni:2022urv}. Furthermore, a Monte Carlo algorithm can still provide acceptable results even if the sign problem is present. Indeed, there exist examples in the spin foam literature: in~\cite{Dona:2023myv} a constant probability distribution, i.e. random sampling, is used to sample bulk representations and intertwiners. The results are convergent and the algorithm more efficient than simply summing over all possible values, which suggests that the sign problem may not be severe. We will confirm this impression in this article for the coherent SU$(2)$ vertex amplitude~\cite{Barrett:2009as,Dona:2017dvf}, at least in most cases. However, note that the total number of variables sampled over is still fairly low; whether the sign problem remains tame for larger triangulations is not clear, but the indications is encouraging.

For completeness, a method that can be applied to systems suffering from the sign problem is tensor network renormalization~\cite{Levin2007,Gu2009,Evenbly2015}. Its approach to studying the dynamics is opposite to Monte Carlo: instead of considering the entire system through (probable) samples, one considers local amplitudes, the tensors, which are locally manipulated and coarse grained into effective amplitudes of composite degrees of freedom. Built into the coarse graining process is a truncation method to avoid exponentially growing numbers of degrees of freedom, where the truncations are done with respect to the relevance of the degrees of freedom derived from a singular value decomposition. Since the sums over degrees of freedom are performed without approximations, the sign problem does not manifest itself. However, challenges remain, e.g. algorithms for higher dimensional systems are intricate and are typically computationally costly. Moreover, one must work with tensors that have a finite index range; systems with continuous variables must be appropriately transformed, see~\cite{Delcamp:2020hzo} as an example of 2d lattice field theory. Quantum gravitational applications exist, e.g. in Lorentzian quantum Regge calculus~\cite{Ito:2022ycc}, 2d analogue spin foam models~\cite{Dittrich:2012he,Dittrich2013,Dittrich2014,Dittrich:2016dc,Steinhaus2015} and 3d lattice gauge theories~\cite{Dittrich2016,Delcamp2017}. The most recent work for 3d lattice gauge theories~\cite{Cunningham:2020uco} utilizes the so-called fusion basis~\cite{Dittrich2017,Delcamp2017a}, which is stable under coarse graining and well suited for studying expectation values of coarse (grained) observables.

The goal of this article is to propose an importance sampling algorithm for coherent intertwiners and use it to approximately compute spin foam amplitudes, here concretely the coherent SU$(2)$ BF theory vertex amplitude. To do so, we define a probability distribution from the absolute value of the coefficents of coherent intertwiners expressed in the orthonormal spin network basis. For coherent tetrahedra peaked on classical shapes, i.e. the coherent data satisfy the closure condition, these coefficients are sharply peaked and almost Gaussian. Sampling from such a distribution is straightforward. Since each intertwiner has its own independent distribution, generalizing this algorithm to the vertex amplitude (or larger triangulations with boundary) is immediate. We apply the algorithm the coherent BF vertex amplitude for different boundary data, e.g. the equilateral $4$-simplex, and compare the results to the full calculation and the asymptotic formula. Overall, we find a good agreement of the results for a moderate number of samples. We see that the sign problem is present in determining the phase, but it is tame for most boundary data. Only when the amplitude is actually small, i.e. close to a root of the oscillations of the Regge action, we do observe convergence issues. In terms of computational times, the full calculation is superior only at small representations, where actually only few configurations need to be summed over. Yet due to growing costs, the Monte Carlo method becomes efficient above spin $j \sim 10$ and provides a good approximation. The thus freed computational resources are invested into larger boundary spins.

This article is organized as follows: section~\ref{sec:spin-foam-nutshell} provides a brief introduction of spin foam models with a particular focus on the computation of the coherent vertex amplitude. Secion~\ref{sec:Monte-Carlo} discusses the sign problem and describes the derivation of the probability distribution for coherent intertwiners and how intertwiner values are sampled. In section~\ref{sec:approx-coherent} we present the results for various boundary data, provide a brief comparison to random sampling and show measurements of computational time. Lastly, we summarize the results in section~\ref{sec:discussion} and discuss how the algorithm can be generalized to larger 2-complexes, the Lorentzian EPRL model and bulk intertwiners.

\section{Spin Foam numerics and coherent amplitudes} \label{sec:spin-foam-nutshell}

Since the main focus of this article is to introduce and discuss a new method to compute / approximate coherent spin foam vertex amplitudes, we will provide the necessary context to explain the calculation. For more detailed introductions and presentations of already existing numerical methods, we recommend several insightful reviews~\cite{Dona:2022yyn,Asante:2022dnj}.

As it is frequently the case, the definition of a path integral requires regularization; in spin foams this is done via the introduction of a discretization, a so-called 2-complex, which in most cases is chosen to be dual to a triangulation. Such a 2-complex is a collection of vertices $v$, edges $e$ and faces $f$, to which we assign algebraic data. In 4d, a vertex is dual to a $4$-simplex, an edge dual to a tetrahedron and a face dual to a triangle. Irreducible representations $\rho_f$ are attached to the faces and describe their area, while intertwiners $\iota_e$, invariant tensors in the tensor product of representations of faces sharing said edge, are assigned to the edges. The latter (partially) encode the shape of polyhedra, e.g. tetrahedra in a 4d triangulation~\cite{Baez:1999tk}. One assignment of data to the 2-complex is a spin foam state, which prescribes the quantum geometry of this configuration. The path integral is implemented via the sum over all possible assignments of these data. 

The dynamics are encoded in local amplitudes: we associate such amplitudes to vertices, edges and faces, where they only depend on the data assigned to this object. Generically, the spin foam partition function reads:
\begin{equation}
    Z = \sum_{\rho_f} \sum_{\iota_e} \prod_f \mathcal{A}_f \prod_e \mathcal{A}_e \prod_v \mathcal{A}_v \quad ,
\end{equation}
where the sum is performed over all possible assignments of representations $\rho_f$ and intertwiners $\iota_e$.
$\mathcal{A}_f$, $\mathcal{A}_e$ and $\mathcal{A}_v$ denote face, edge and vertex amplitudes respectively. While there are differences between the various spin foam models, similarities across the models exist, e.g. the face amplitude is usually given by the dimension of the representation and the edge amplitude is the inverse of the intertwiner norm. The most important amplitude is the vertex amplitude as it is associated with the (dual of) the fundamental building blocks of quantum space-time, e.g. $4$-simplices in 4d. It is at the center of attention of this article and we will discuss it in more detail now. 

For concreteness, we will from now one specify our discussion to SU$(2)$ BF theory in 4d: while it is not a theory of quantum gravity its structure and coherent states are similar and relevant to the 4d (Lorentzian and Euclidean) EPRL model~\cite{Engle:2008fj}, such that the tools developed here for BF theory should be applicable and transferable to this more relevant model.
The vertex amplitude is usually defined as follows: each edge in a spin foam encodes a projector onto the invariant subspace~\cite{Perez:2013uz}. These projectors are written in terms of an orthonormal intertwiner basis, where one intertwiner is associated to each vertex of the edge. Each intertwiner has as many indices as the polyhedron has faces, so four in case of a tetrahedron. Then, we contract all intertwiners of a vertex with each other, i.e. we identify their indices when two intertwiners share a face and sum over them. This gives a number, the so-called vertex amplitude, which depends on spins and intertwiner basis elements. For a $4$-simplex, we have five 4-valent intertwiners, where each intertwiner has one index contracted with any of the other four intertwiners. 4-valent SU$(2)$ intertwiners are not unique and can be labelled with one representation label using recoupling theory. The associated vertex amplitude is called the SU$(2)$ BF $\{15j\}$-symbol; it depends on ten SU$(2)$ representations (called spins $j_i$) and five intertwiner labels.

Contracting intertwiners to compute the vertex amplitude is intuitive as it matches how we would combinatorially $4$-simplex from tetrahedra. However, this explicit contraction is computationally costly, as we have to sum over ten magnetic indices. Moreover, this range of indices grows with the size of the spins, too, increasing the costs further. Instead, the $\{15j\}$-symbol can be concisely expressed in terms of SU$(2)$ $\{6j\}$-symbols~\cite{Dona:2020tvv}:
\begin{align}
    & \includegraphics[width = 0.4 \textwidth, valign = c]{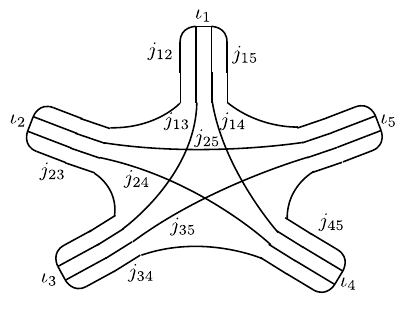} = (-1)^{\sum_i j_i + \sum_i \iota_i } \sum_x (2x + 1) \; \times \nonumber \\
    \quad & \times \begin{Bmatrix}
	\iota_1 & j_{25} & x \\
	\iota_5 & j_{14} & j_{15}
    \end{Bmatrix}
    \begin{Bmatrix}
	j_{14} & \iota_5 & x \\
	j_{35} & \iota_4 & j_{45}
    \end{Bmatrix} % \\ \nonumber
    %& \times 
    \begin{Bmatrix}
	\iota_4 & j_{35} & x \\
        \iota_3 & j_{24} & j_{34}
    \end{Bmatrix}
    \begin{Bmatrix}
	j_{24} & \iota_3 & x \\
	j_{13} & \iota_2 & j_{23}
    \end{Bmatrix}
    \begin{Bmatrix}
	\iota_2 & j_{13} & x \\
	\iota_1 & j_{25} & j_{12}
    \end{Bmatrix} \quad .
    \label{eq:15jsymbol}
\end{align}
This expression is highly efficient for numerical implementations: highly optimized libraries for computing the $\{6j\}$-symbol exist, e.g. \verb|WignerSymbols|\footnote{\url{https://github.com/Jutho/WignerSymbols.jl} .} in \verb|Julia|, and the sum over one auxiliary label $x$ is bounded by the spins.% This is far more cost effective than e.g. explicitly contracting five Wigner $4j$-symbols.

So far, we only concerned ourselves with a single $\{15j\}$-symbol. When studying the large spin behavior of spin foams, which is often referred to as the semi-classical limit (at least for a single $4$-simplex), we are rather interested in states peaked on Regge geometries, i.e. classical simplices, as these play a dominant role in this limit~\cite{Barrett:2009as}. Such Regge geometries are encoded via coherent intertwiners~\cite{Livine:2007bq}, a group averaged tensor product of SU$(2)$ Perelomov coherent states~\cite{Perelomov1986}. These coherent states form an overcomplete basis and the group integration can be written as a sum over orthonormal intertwiners. This generalizes straightforwardly to the coherent vertex amplitude, which can be written as a superposition of $\{15j\}$-symbols. We briefly review this in the next section.

\subsection{Coherent SU$(2)$ vertex amplitude}

SU$(2)$ coherent states play a crucial role in the asymptotic analysis. They are defined as maximum or minimum weight states, where the states $|j,j\rangle$ or $|j,-j\rangle$ respectively are used as reference states. As these states are eigenstates of $J_z$ for  maximum / minimum weight, they are associated with the direction $\vec{e}_z$. We will use the maximum weight eigenstate through the entirety of this paper.

We can peak this state on a different direction by acting on it with a group element, via the associated Wigner matrix.
\begin{equation}
    |j,\vec{n}\rangle = D^j(g_{\vec{n}}) |j,j\rangle = \includegraphics[width = 0.1 \textwidth, valign = c]{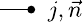} \quad .
\end{equation}
$g_{\vec{n}}$ is a group element in SU$(2)/$U$(1)$, which encodes the rotation from $\vec{e}_z$ (the reference state) to $\vec{n}$; the coherent states are only defined up to a phase. Combining several of these coherent states with a group integration defines a coherent Livine-Speziale intertwiner~\cite{Livine:2007bq}. For concreteness, we consider a 4-valent one to describe tetrahedra:
\begin{equation}
    \iota(\{j_i\},\{\vec{n}_i\}) = \int_{\text{SU}(2)} dg \bigotimes_{i=1}^4 D^{j_i}(g) |j_i,\vec{n}_i\rangle = \includegraphics[width = 0.2 \textwidth, valign = c]{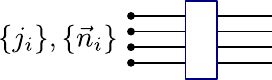} \quad ,
\end{equation}
where we represent the group integration by a box over all strands.

The coherent vertex amplitude is typically defined as the contraction of five such intertwiners:
\begin{align}
	\includegraphics[width = 0.25 \textwidth, valign = c]{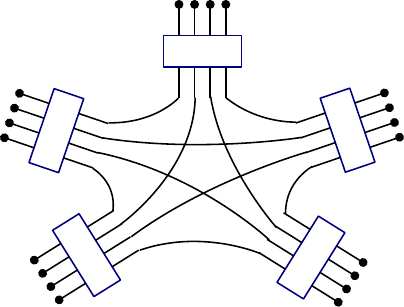} = & \int_{\text{SU}(2)} \prod_a dg_a \prod_{a < b} \left \langle j_{ab}, -\vec{n}_{ba} \left | D^{j_{ab}}(g^{-1}_b) D^{j_{ab}}(g_a) \right | j_{ab}, \vec{n}_{ab} \right \rangle \\ \nonumber 
	= & \int_{\text{SU}(2)} \prod_a dg_a \prod_{a < b} \left( \left \langle \tfrac{1}{2}, -\vec{n}_{ba} \left | g^{-1}_b g_a \right | \tfrac{1}{2}, \vec{n}_{ab} \right \rangle \right)^{2 j_{ab}} \quad ,
\end{align}
where we have used the fact that the coherent states are maximum weight states in the second line. This explicit representation in terms of group integrations is ideally suited for applying a stationary phase approximation as the inner products are highly oscillatory, from which spin foam asymptotics were derived in a plethora of contexts~\cite{Barrett:2009as,Barrett:2009ci,Barrett:2009mw,Kaminski:2017eew,Liu:2018gfc,Simao:2021qno}. On the other hand, the oscillatory nature is challenging for explicit numerical calculations.

One attempt would be to perform four SU$(2)$ integrations (one can be dropped due to gauge invariance) numerically, using e.g. the Cuba package~\cite{Hahn:2005pf}, which would be a 12-dimensional integration of a highly oscillatory function. However, high dimensional integrations suffer from poor convergence, in particular if the functions are oscillatory. This is only exacerbated for larger spins, where the functions become even more oscillatory. Instead, we can compute each intertwiner explicitly by performing one SU$(2)$ integration for each component of the intertwiner. Since the integrations are lower dimensional, convergence of the integrations is improved, yet there are far more calculations to perform. As we increase the spins uniformly the number of components grows exponentially. Additionally, the contraction of these intertwiners is costly as well. For reference, this method was used in~\cite{Allen:2022unb} for higher valent spin foam vertices.

Another way to avoid explicit integrations is to expand the projector in terms of orthonormal intertwiners and thus express the amplitude as a sum over SU$(2)$ $\{15j\}$-symbols.
\begin{equation}
    \includegraphics[width = 0.4 \textwidth, valign = c]{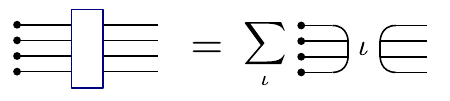} \quad .
\end{equation}
The sum runs over a basis of intertwiners, which are labelled with a single representation $\iota$ in case of 4-valent SU$(2)$ intertwiners. The diagram to the left is the coefficient $c_\iota(\{j_i\},\{\vec{n}_i\})$ of the coherent intertwiner expressed in terms of the basis:
\begin{equation} \label{eq:coefficient_c}
    \includegraphics[width = 0.1 \textwidth, valign = c]{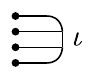} =: c_\iota(\{j_i\},\{\vec{n_i}\}) = \sum_{\{m_i\}} 
    \begin{pmatrix}
        j_1 & j_2 & j_3 & j_4 \\
        m_1 & m_2 & m_3 & m_4
    \end{pmatrix}^{(\iota)} \prod_i \langle j_i, m_i | j_i, \vec{n}_i \rangle \quad ,
\end{equation}
where the object in brackets denotes Wigner's $4jm$-symbol.
This coefficient, more precisely its absolute value, plays a central role in the Monte Carlo algorithm we will introduce below.

Applying this expansion to each coherent intertwiner in the vertex amplitude yields:
\begin{equation} \label{eq:coherent_vertex_int}
	\includegraphics[width = 0.225 \textwidth, valign = c]{drawings/vertex} \; = \; \sum_{\{\iota_i\}}	\includegraphics[width = 0.325 \textwidth, valign = c]{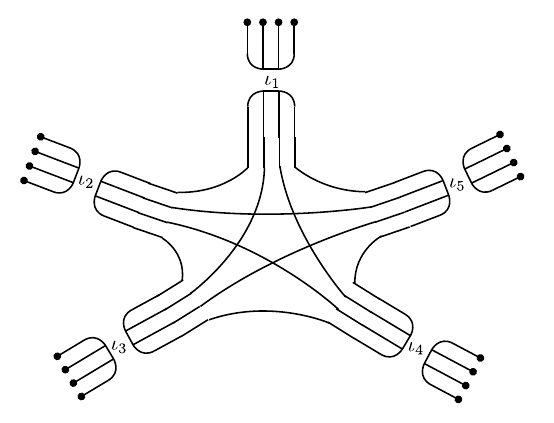} \quad .
\end{equation}
Now the coherent amplitude is rewritten as a sum over (many) $\{15j\}$-symbols. This method is far more efficient than any of those sketched above and it is a major reason for many numerical advancements in recent years~\cite{Dona:2017dvf,Dona:2019dkf,Gozzini:2021kbt}. The calculation can now also be written as a contraction of a tensor network, where the 5-valent tensor of all $\{15j\}$-symbols is contracted with five vectors of coefficients $c_{\iota_i}$ of coherent intertwiners~\cite{Gozzini:2021kbt,Asante:2022lnp}. However, as the allowed intertwiner range grows when uniformly scaling the boundary spins, also this method suffers from exponentially growing numerical costs. Let us briefly explain this.

For simplicity we consider the case where all spins $j$ are the same. For a 4-valent intertwiner, there are $2j+1$ basis vectors. To compute the coherent amplitude numerically, we need to compute the value of the $\{15j\}$-symbol for all possible orthonormal intertwiner labels, i.e. $(2j+1)^5$ possibilities. To compute each $\{15j\}$-symbol we have to sum over an auxiliary label with range proportional to $j$. Thus, overall computing and saving the array of all variants of the $\{15j\}$-symbol scales proportional to $j^6$. The following contractions of this array with five vectors of coherent intertwiner components is subleading in comparison: the first contraction scales with $(2j+1)^5$ since we are summing over one label while keeping four ones fixed. The following contractions scale more favourably. In summary, the full calculation scales exponentially when homogeneously scaling up all spins for general boundary data. To illustrate the scope, for all spins $j=50$, the array storing the $\{15j\}$-symbols for all intertwiner labels contains $\sim 3 \cdot 10^8$ entries, which requires a significant computational time to compute and a large amount of memory to store\footnote{Writing the contraction as for loops reduces the memory costs, but is typically far slower. Tensor network contractions make use of highly optimized linear algebra operations. Further speed up can be achieved by utilizing GPUs~\cite{Gozzini:2021kbt}.}. This poses a challenge to computing the amplitude for larger 2-complexes.

Hence the question arises, whether we can optimize these calculations further or find suitable approximations. In this article we pursue the question whether we can truncate the sum over intertwiners, therefore reducing the numerical costs. Indeed, for sufficiently large spins and coherent states peaked on geometric tetrahedra, the absolute value of the coherent intertwiner coefficients $c_\iota$ are typically sharply peaked, such that labels far away from the peak should be safe to ignore. This idea was first formulated in \cite{Dona:2019dkf}, where labels were truncated if they are less relevant than a chosen cut-off relative to the peak. Choosing this cut-off then controls the quality of the approximation versus the costs of the calculation. Yet, explicit testing of different cut-offs is necessary to see whether the approximation can be trusted.

In this article, we build upon this idea is and use the absolute value of $c_\iota$ to define a probability distribution for the intertwiner labels for a Markov Chain Monte Carlo algorithm. From this distribution, we sample intertwiner labels according to their relevance inferred from the coherent states and approximate the coherent vertex amplitude without introducing a cut-off by hand. The approximation should then improve by increasing the number of samples and eventually converge to the actual result. If the algorithm is converging quickly, it should provide a good approximation utilizing a fraction of configurations compared to the full calculation. Therefore, a convergent Monte Carlo algorithm promises to free computational resources, which can instead be used to investigate spin foams defined on larger 2-complexes. However, the main obstacle for convergence is the so-called sign problem. In the next section, we briefly review Monte Carlo methods and the sign problem, explain why it is present in spin foams and briefly discuss the sampling procedure for coherent intertwiner labels.

\section{Monte Carlo and the sign problem} \label{sec:Monte-Carlo}

In physics Monte Carlo methods are an efficient and useful tool to study statistical systems with a large number of degrees of freedom. These systems are described by high dimensional integrals or sums over all possible configurations of its variables, which typically cannot be evaluated analytically or calculated numerically by brute force. Instead, Markov chain Monte Carlo methods sample typical configurations from the probability distribution defining the statistical theory. With sufficiently many such samples one may well approximate expectation values of observables of the system and extract physical information. The key advantage is that this algorithm scales with the number of samples rather than exponentially with the number of variables.

These samples are generated in a random process, take e.g. the Metropolis-Hastings algorithm~\cite{Metropolis1949,Hastings1970}. Starting from a random configuration, a new configuration is proposed by randomly deviating from the previous one by a given set of moves. Then the probabilities of these two configurations are compared; if the new configuration is more probable it is always accepted. If it is not, then it is accepted only with a certain probability. This method must be implemented with care to obtain correct and reliable results, e.g. one must implement detailed balance to ensure that one is indeed sampling with respect to the probability distribution of the system. Moreover, the proposal of new configurations must be ergodic, i.e. it must be possible to reach any configuration of the system (given enough time). Finally, one must ensure that the algorithm explores a sufficiently large part of the configuration space. This is often gauged by the acceptance rate of new configurations.

\subsection{Complex amplitudes and the sign problem}

Not all physical theories are statistical theories. While many theories, e.g. lattice field theories (at zero temperature and without fermions) and causal dynamical triangulations~\cite{Ambjorn:2012vc}, can be analytically continued from a quantum to a statistical theory, this is not the case for many quantum theories including spin foam models (unless some particular symmetry restrictions are considered~\cite{Bahr:2015gxa}). The consequences are two-fold: first, the partition function does not define a probability distribution and thus cannot be used to sample configurations. The second and more severe consequence is the so-called sign problem: alternating / oscillating amplitudes can lead to contributions cancelling each other. If the sign problem is severe, we must take a lot of samples to capture this effect and hopefully obtain a convergent result. Of course, more samples come with larger numerical costs, such that the algorithm is less efficient.

It is important to distinguish these two aspects: for complex amplitudes we have no probability distribution readily available and the sign problem is present. Still, Monte Carlo methods might be useful for exploring the dynamics of the model if the sign problem is tame. To do so, one must propose a new probability distribution. A simple possibility is the constant distribution, which assigns the same weight to all configurations, but usually it is beneficial to use a distribution fitting the dynamics of the system, which again can be challenging for oscillating functions. Note however, that defining a new distribution does not solve the sign problem.

To illustrate this consider the so-called reweighting procedure~\cite{Chandrasekharan:1999cm,Kieu:1993gw}. Consider a physical system with complex amplitude $\mathcal{A}$ for variables $\phi_i$. For concreteness, we consider the $\phi_i$ to have a discrete spectrum; then the partition function and expectation values of observables are (formally) defined  by
\begin{equation}
    Z = \sum_{\{\phi_i\}} \mathcal{A}(\{\phi_i\}) \quad , \quad \langle \mathcal{O} \rangle_\mathcal{A} = \frac{1}{Z} \sum_{\{\phi_i\}} \mathcal{O}(\{ \phi_i \}) \; \mathcal{A}(\{\phi_i\}) \quad .
\end{equation}
From $\mathcal{A}$, we can define a probability distribution by taking its absolute value with normalization $\tfrac{1}{Z'}$. With this probability distribution, we compute $\langle \mathcal{O} \rangle_\mathcal{A}$ as follows:
\begin{equation}
    \langle \mathcal{O} \rangle_\mathcal{A} = \frac{1}{Z} \frac{Z'}{Z'} \sum_{\{\phi_i\}} \mathcal{O}(\{ \phi_i \})\; e^{i \varphi_\mathcal{A}} \; |\mathcal{A}(\{\phi_i\})| = \frac{\langle e^{i \varphi_\mathcal{A}} \; \mathcal{O} \rangle_{|\mathcal{A}|}}{\langle e^{i \varphi_{\mathcal{A}}} \rangle_{|\mathcal{A}|}} \quad ,
\end{equation}
where $\varphi_\mathcal{A}$ denotes the phase of $\mathcal{A}$ (for a specific configuration $\{ \phi_i \}$), and we denote by $\langle \mathcal{O} \rangle_{|\mathcal{A}|}$ the expectation value of $\mathcal{O}$ computed in the distribution defined by $|\mathcal{A}|$. Essentially, we absorb the phase $\varphi_\mathcal{A}$ into the observable and compute it in the distribution given by $|\mathcal{A}|$. Since this is not the original expression, we must correct this by dividing by the expectation value of the phase. Both expectation values are now in principle computable using importance sampling Monte Carlo techniques, but the sign problem enters here: if the sign problem is severe, the expectation value of the phase is small or zero. Then, the expectation value computed in this way will not converge and the result cannot be trusted.

Eventually, we do not know how severe the sign problem in spin foams is. In the following we propose a probability distribution for coherent (boundary) intertwiners, with which we approximate the coherent vertex amplitude. The results suggest that the sign problem for this amplitude is not severe (in most cases).

\subsection{Probability distribution from coherent intertwiners}

In eq. \eqref{eq:coherent_vertex_int}, the coherent vertex amplitude is written as the contraction of the $\{15j\}$-symbol with the components of coherent intertwiners $c_\iota$, see eq. \eqref{eq:coefficient_c}. The goal is to approximate the sum over intertwiners using Monte Carlo methods by a number of samples, which hopefully converges for significantly fewer samples than the total number of configurations. Towards defining a probability distribution suitable for this amplitude, we consider the coherent intertwiner expanded in othonormal intertwiners more closely.
%The goal is to define a probability distribution from the coherent intertwiners and use it to approximate the amplitude with Monte Carlo methods. 

\begin{figure}
    \centering
    \includegraphics[width = 0.45 \textwidth]{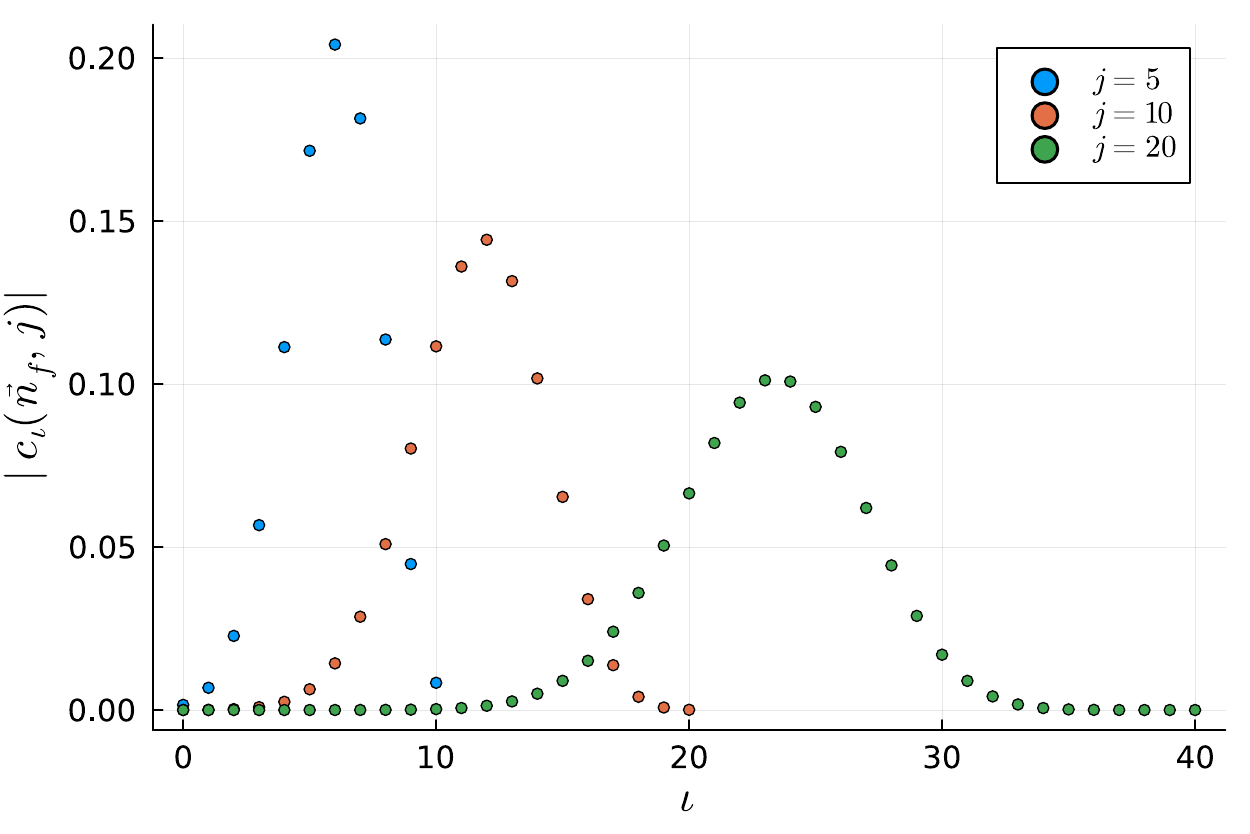}
    \includegraphics[width=0.45 \textwidth]{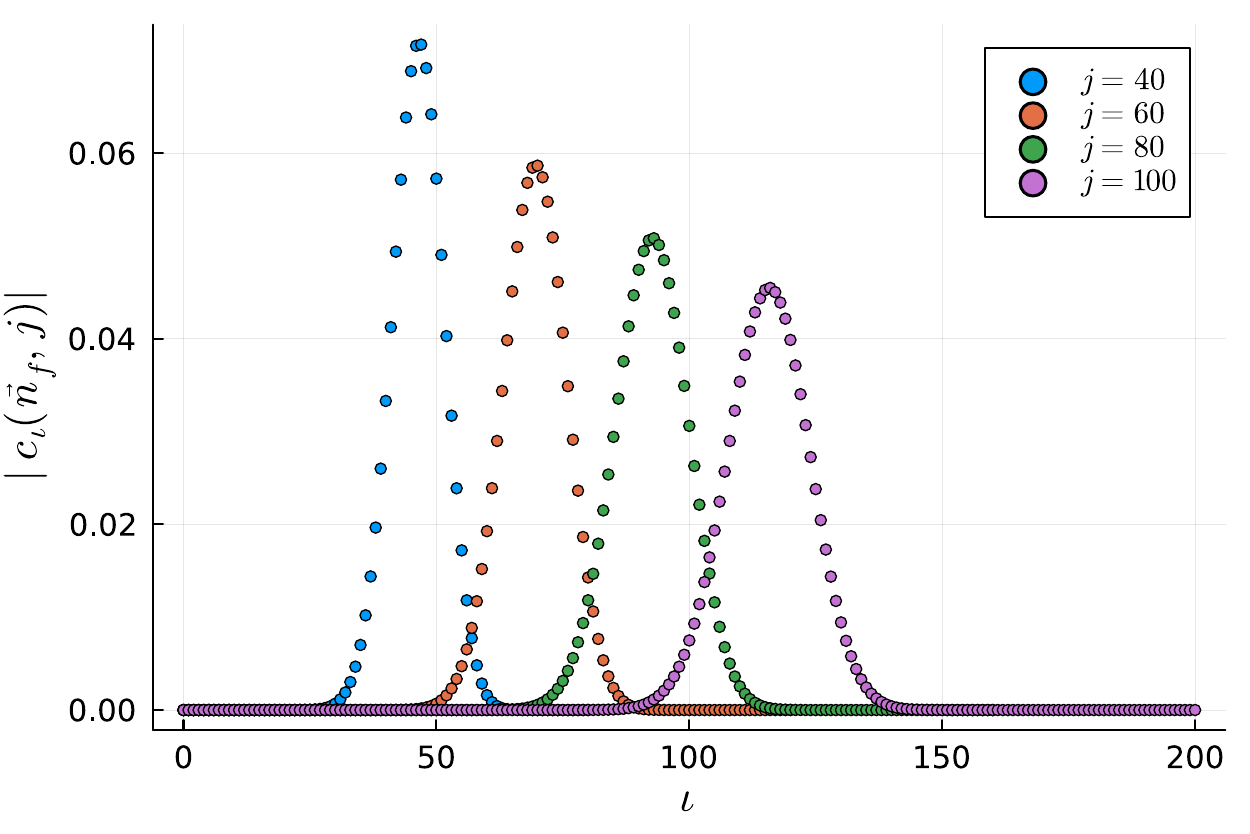} 
    \caption{Coherent intertwiner coefficients for equilateral tetrahedra. \textit{Left:} spins $j = 5, 10, 20$. \textit{Right:} spins $j = 40, 60, 80, 100$}
    \label{fig:coherent_coefficient}
\end{figure}

The intertwiner components are in general complex, as can be expected from an overcomplete basis. Their absolute value on the other hand, see fig. \ref{fig:coherent_coefficient}, has a pronounced peak, which becomes sharper (relative to the total range of intertwiner labels) as the spins are uniformly increased. Thus, for large spins, this is an excellent choice to define a probability distribution, yet it also reveals a weakness. For small spins, the peak is so broad that it covers the entire range of all intertwiner labels. Hence, all intertwiners labels are similarly relevant (roughly same order of magnitude) and the potential to save computational time by sampling is low. Fortunately, the range of intertwiner labels is then still low such that explicit summation is viable. Therefore, we should rather consider the Monte Carlo methods to complement the full calculations at large spins. The final step to obtain the probability distribution is to fix the normalization, which is simply given by the inverse of the sum of all $|c_\iota|$. In the next section we briefly discuss the sampling procedure.

\subsubsection{Brief introduction to importance sampling}

Let us give here a brief introduction to importance sampling and adapt it to sampling coherent intertwiners. The general idea is to perform a random walk through configuration space guided by the probability distribution, such that we more frequently accept probable configurations and reject improbable ones. Typically we start with a random configuration, then propose a new one and compare its probability to the old one; if it is more probable it is always accepted. If it is less probable, a random number $r \in [0,1]$ is drawn; if $r$ is smaller than the relative probability, the new configuration is accepted and rejected otherwise. Thus, if the relative probability is low, i.e. the new configuration is far less probable than the previous one, it is highly unlikely that this move will be accepted.

A priori, the proposal method for new configurations only has to ensure that it is ergodic, i.e. any configuration of the system can be reached in principle. So, e.g. one could always propose a new random configuration, yet in particular in high dimensional systems, this might be an inefficient way of identifying the relevant configurations. Instead one frequently opts for small deviations away from the previous configuration, such that one randomly ``walks'' through configuration space. Several subtleties must be considered here: first, before collecting samples, we must ensure that we are close to the probable configurations. Thus, starting from a random configuration, one lets the system ``thermalize'' and checks whether it thermalizes to the same region of configurations in multiple runs from different random starting locations. Second, we must ensure that the samples taken are independent of one another, i.e. they are not correlated. By only making small deviations, a newly proposed configuration is obviously correlated with the one it was generated from. Thus, it is best to do several Monte Carlo steps before taking another, uncorrelated sample. Third, we must ensure that a sufficiently large part of configuration space is explored and improbable configurations are proposed. This helps to avoid getting stuck in local maxima of the probability; this could e.g. be noted during thermalization runs. In high dimensional systems, this is gauged by considering the acceptance rate, which is supposed to be well above 10\% but below 50\%; indeed, if it is too low, we propose highly improbable configurations and do not explore configuration space. Vice versa, if it is too high, e.g. 90\%, we may just explore configurations around a local maximum.

These considerations are crucial when investigating systems with a large number of degrees of freedom, where one has to rely on coarse grained observables and consistency checks to ensure a proper exploration of the system. Here we are considering the comparatively simple case where we sample one variable for an almost Gaussian probability distribution, thus it is very easy to check and verify whether the sampling is satisfactory. We give an example of the possible issues one might encounter below.

\subsubsection{Sampling coherent intertwiners and minor issues}

The almost Gaussian distribution of the probability measure is ideal for sampling. Since we are implementing a fairly standard Metropolis-Hastings algorithm~\cite{Metropolis1949,Hastings1970} for sampling, detailed in the mock algorithms \ref{alg:Sampling_int} and \ref{alg:MC_step}, we keep the discussion brief and focus on the proposal method for new configurations and which issues we encountered during sampling.

\begin{algorithm}[htb!]
\small
\caption{{Sampling coherent tetrahedral intertwiners}}\label{alg:Sampling_int}
\algnotext{EndIf}
\algnotext{EndFor}
\algnotext{EndWhile}
\begin{flushleft}
\textbf{Input: }\\ \quad { \texttt{Spins}: a 4-tuple of spin assignments $j_{1},j_{2}, j_{3}, j_4$ }  \\
\quad { \texttt{Normals}: set of normal vectors $\vec{n}_1, \vec{n}_2, \vec{n}_3, \vec{n}_4$ associated to the triangles}  \\
\quad { \texttt{Number of thermalization steps}: number of MC iterations $n_t$ before sampling starts}  \\
\quad { \texttt{Number of steps between samples}: number of MC iterations $m_s$ between taking samples}  \\
\quad { \texttt{Number of samples}: number of samples $N$}  \\
\textbf{Output: }{List of sampled intertwiner values $\{\iota^{(1)}, \iota^{(2)}, \dots, \iota^{(N})\}$} \\
\end{flushleft}
\begin{algorithmic}[1]
\State Compute $r_\iota(\{j_i\})$ (range of $\iota$), coefficients $c_{\iota}(\{j_i\},\{\vec{n}_i\})$, their absolute values and the probability distribution $P_\iota$.
\State Randomly pick an $\iota \in r_\iota$.
\For {$n$ in $1:n_t$}  Monte Carlo Steps \EndFor \Comment{Thermalization}
\For {$i$ in $1:N$} \For {$m$ in $1:m_s$} Monte Carlo steps \EndFor Store sample $\iota^{(i)}$ \EndFor \Comment{Collecting samples}
\State \Return{ Vector of samples $\{\iota^{(1)}, \iota^{(2)}, \dots, \iota^{(N})\}$}
\end{algorithmic}
\end{algorithm}

\begin{algorithm}[htb!]
\small
\caption{{Monte Carlo step intertwiner}}\label{alg:MC_step}
\algnotext{EndIf}
\algnotext{EndFor}
\algnotext{EndWhile}
\begin{flushleft}
\textbf{Input: }\\ \quad { \texttt{Range of $\iota$}: $r_{\iota}$}  \\
\quad { \texttt{Probability distribution}: $P_\iota$ derived from $c_\iota$}  \\
\quad { \texttt{Current intertwiner label}: $\iota \in r_\iota$}  \\
\textbf{Output: }{New label $\tilde{\iota}$} \\
\end{flushleft}
\begin{algorithmic}[1]
\State Propose $\tilde{\iota} = \iota + a$, where $a$ random integer with $0 < |a| < z$. \Comment{E.g. choose $|a| = \frac{r_\iota}{n}, \, n \in \mathbb{N}$}
\State Check whether $\tilde{\iota} \in r_\iota$: \Comment{Alternatively implement periodic boundary conditions}
\If {$\tilde{\iota} > \text{max}(r_\iota)$}  $\tilde{\iota} = \text{max}(r_{\iota})$ \EndIf
\If {$\tilde{\iota} < \text{min}(r_\iota)$}  $\tilde{\iota} = \text{min}(r_{\iota})$ \EndIf
\State Draw random number $x \in [0,1]$
\If {$x < \text{min}\left(1, \frac{P_\iota(\tilde{\iota})}{P_\iota(\iota)}\right)$} $\iota = \tilde{\iota}$ \EndIf
\State \Return{ $\iota$}
\end{algorithmic}
\end{algorithm}

When proposing a new intertwiner configuration, we essentially shift the previous intertwiner label by a random integer. The simplest variant is to shift it by $\pm 1$, but larger (random) shifts are possible. Either variant clearly leads to an ergodic algorithm, yet they will differ in the acceptance rate of new proposals. If the range of possible shifts is larger, there will be more proposals that shift the intertwiner labels to a value far away from the peak. These are improbable, therefore they will be rejected in most cases. In particular in systems with a large number of degrees of freedom, the acceptance rate is a good gauge to see whether one is exploring a sufficiently large part of the configuration space and choose the proposal of new configurations appropriately. This is important, as one does not want to get stuck in a local maximum. %Here we are sampling a one-dimensional system with a simple distribution, where this issue {\color{red} is absent}.

\begin{figure}
    \centering
    \includegraphics[width=0.45 \textwidth]{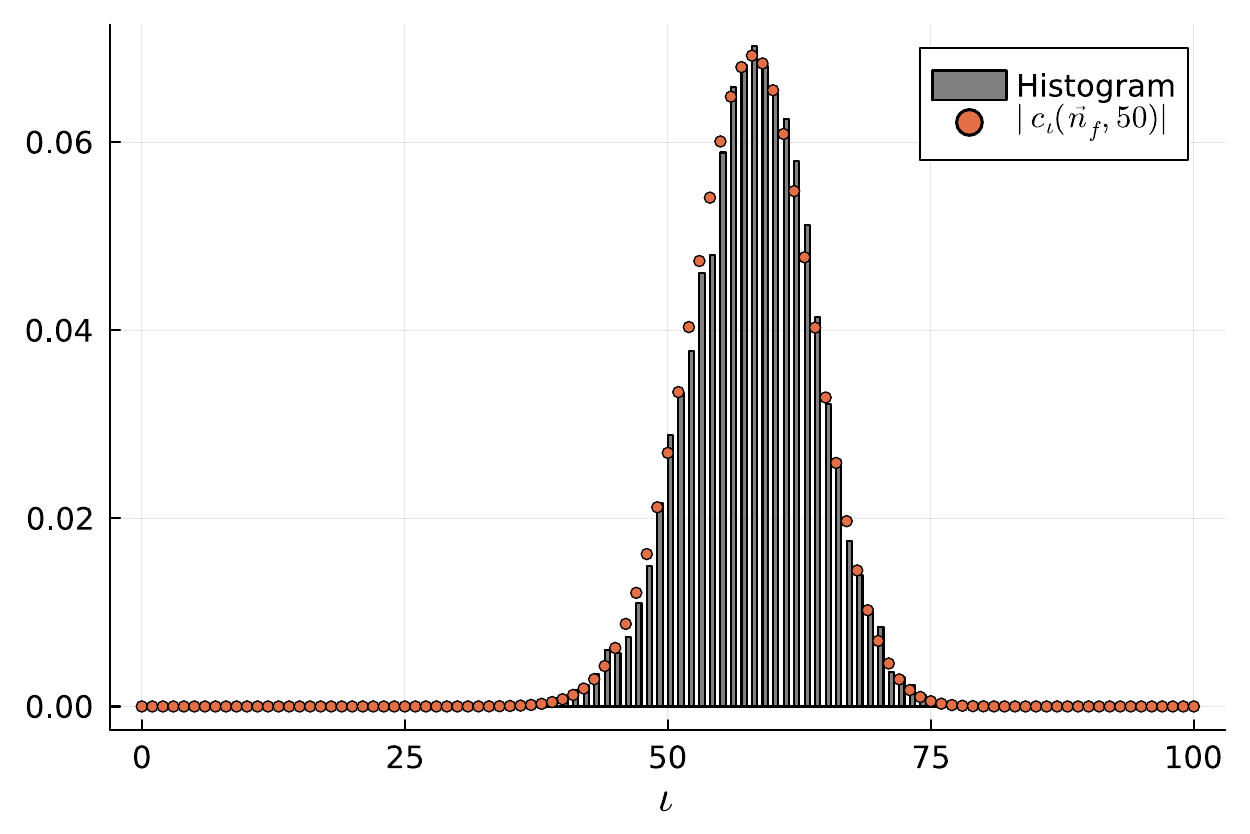}
    \includegraphics[width=0.45 \textwidth]{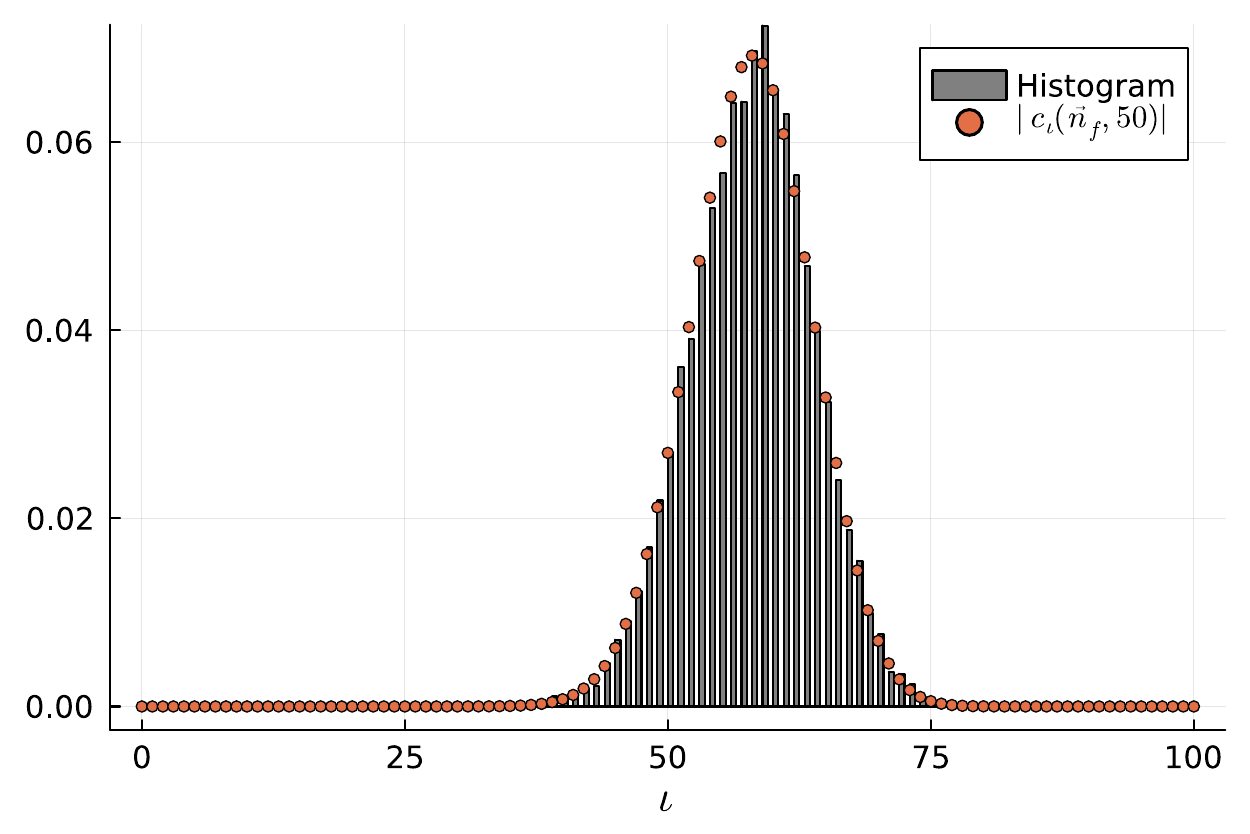}
    \caption{Normalized histogram sampling equilateral tetrahedra for $j=50$. $10^4$ thermalization steps, $500$ steps between taking samples. \textit{Left:} Step size $1$ when proposing new configuration. \textit{Right:} Maximal step size $\sim 50$ when propsing new configuration.}
    \label{fig:histogram_large_spin}
\end{figure}

For the one-dimensional almost Gaussian probability distribution proposed here, these considerations appear to be far less relevant. In fig. \ref{fig:histogram_large_spin} we plot histograms of sampled intertwiner labels for the distribution given by an equilateral tetrahedron for all spins $j=50$. The two sampling methods differ in the proposed shift size, for one it is $\pm 1$, for the other it is a random number between $1$ and $\frac{1}{3}$ of the total intertwiner range. Qualitatively the histograms look similar and are able to sample labels of the whole peak and the beginning of the tails. A larger step size, with a lower acceptance rate, should be more suitable at sampling the tails of the distribution. However, due to the low probability of these values, we expect this to become noticeable only for a larger number of samples.

\begin{figure}
    \centering
    \includegraphics[width=0.45 \textwidth]{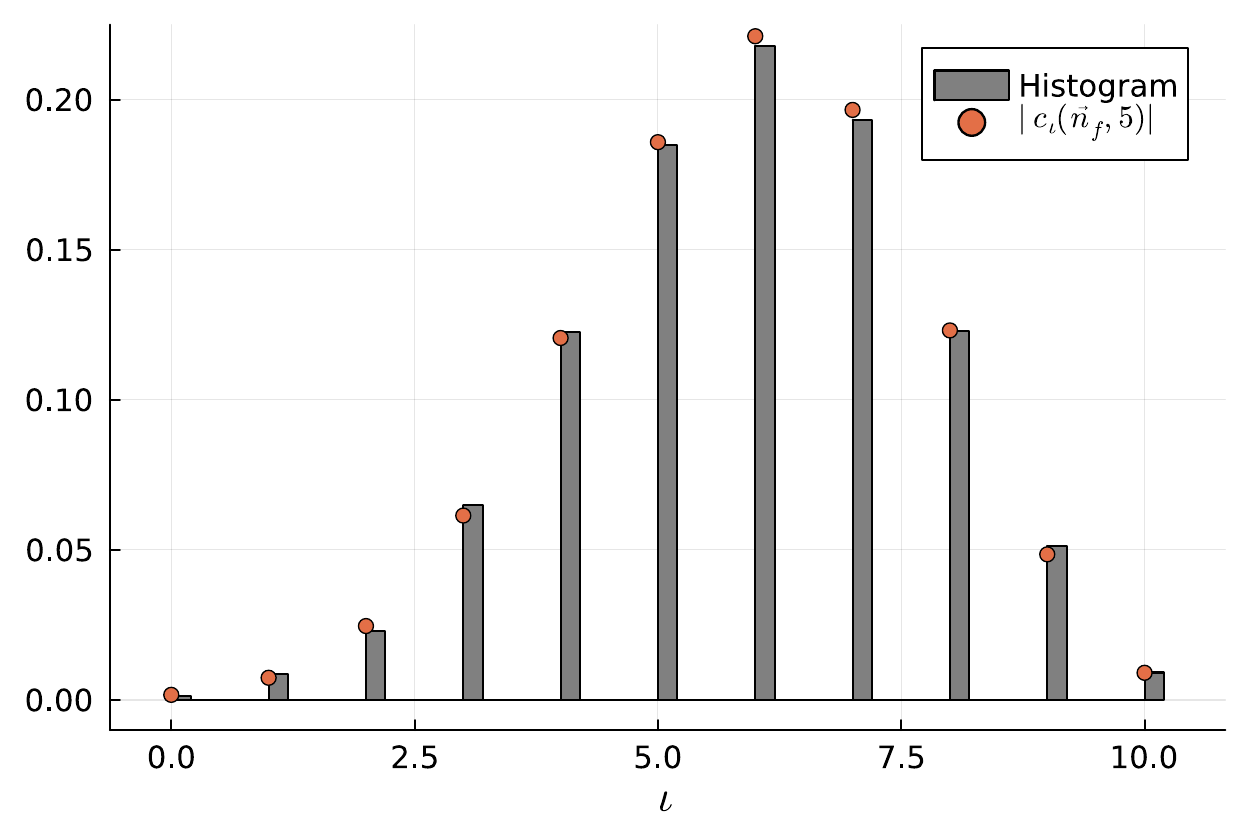}
    \includegraphics[width=0.45 \textwidth]{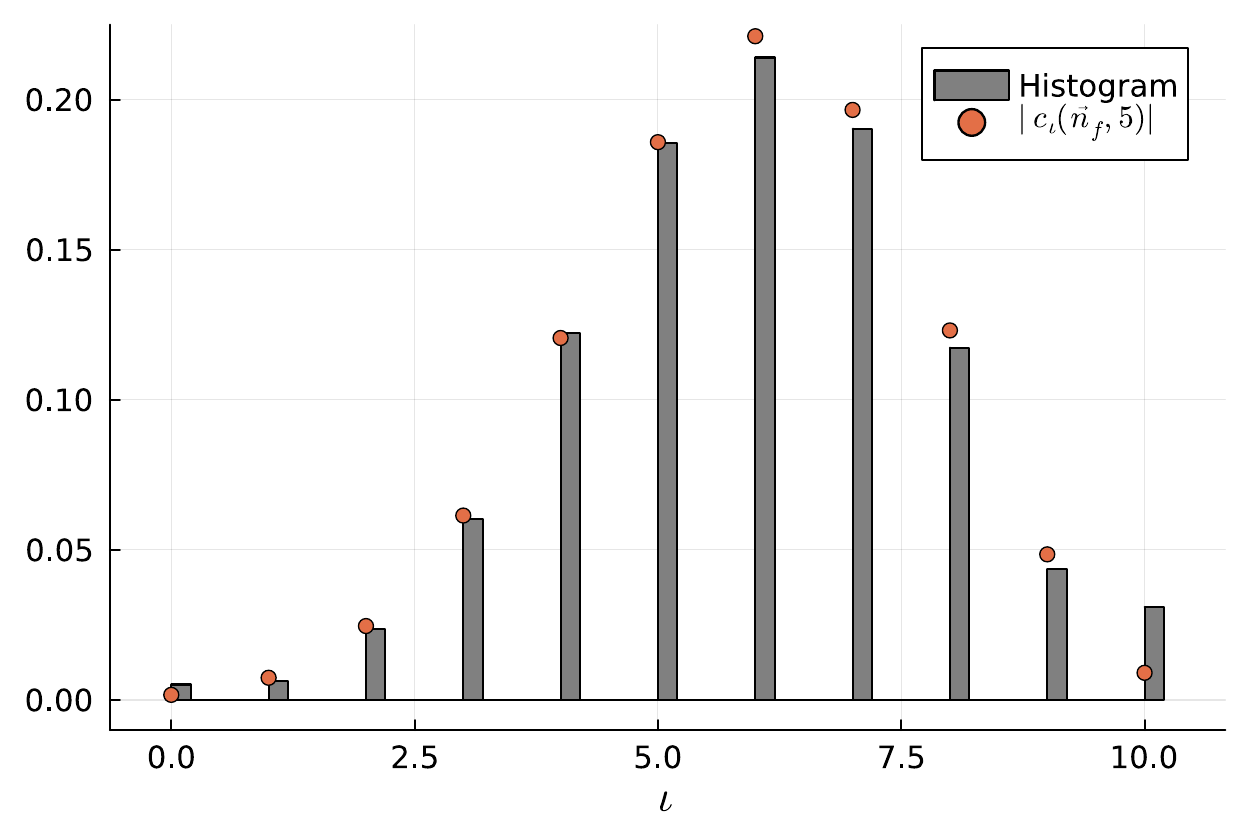}
    \caption{Normalized histogram sampling equilateral tetrahedra for $j=5$. $10^4$ thermalization steps, $500$ steps between taking samples. \textit{Left:} Step size $1$ when proposing new configuration. \textit{Right:} Maximal step size $\sim 50$ when propsing new configuration. The overemphasis of values on the boundary is visible.}
    \label{fig:histogram_small_spin}
\end{figure}

For small spins, the proposal scheme for a larger step size requires more scrutiny. Then, the peak of the coherent intertwiner essentially covers the whole intertwiner range and each intertwiner value has a relevant probability. As a result, the potential to save on computational time by using Monte Carlo methods is reduced. Moreover, care is necessary at the boundary of the permitted intertwiner range: If we now choose a maximal step size that is e.g. half or a third of the total intertwiner range, we will propose values outside the permitted range. We remedy that by simply setting the value by hand to the value at the boundary. However, this choice leads to an overemphasis of those boundary values, which is shown in the histogram in fig.~\ref{fig:histogram_small_spin}. This is due to the fact that the proposal method violates detailed balance: we unintentionally propose intertwiner values at the boundary more frequently compared to the others without changing the acceptance conditions\footnote{We can fix this flaw by using ``periodic boundary conditions'' on the intertwiner values. Instead of ending at the minimal / maximal intertwiner range, we continue at the other end of the distribution.}. Here we cure this simply by setting the step size for new proposals to $\pm 1$.

%We propose a new intertwiner label by adding / subtracting a random natural number. We generate this number between $1$ and a fraction of the whole length of the intertwiner range. Let us consider an example from the simulations generated for the boundary data of an equilateral $4$-simplex: for all spins $j=50$, the possible intertwiner labels $\iota \in \{0,\dots,100\}$. We choose to generate the shift between $1$ and $\tfrac{101}{2}$ (rounded to an integer), which is a fairly large range. The acceptance rates of new configurations is a bit above $17\%$ and thus fairly low, yet enough of the configuration space is explored. For spins $j=50.5$ we choose instead a range up to $\tfrac{102}{3}$, and observe an acceptance rate of above $25\%$.

%Hence, the acceptance rate is directly tied to the proposal method of new configurations and crucially, large shifts of intertwiner labels must be proposed to explore a sufficiently large portion of all possible intertwiners. This can also be seen in a histogram of the sampling: with a low acceptance rate, we capture both the peak of the distribution as well as a fair part of the tails to either side. If the shift in label is small however, it is highly unlikely that the algorithm explores labels far beyond the peak, which then leads to a worse approximation of the actual result. Indeed, in \cite{Dona:2019dkf} the goal was to reduce computational costs by truncating the intertwiners labels with less than $30\%$ \textbf{CHECK!} value compared to the peak of the distribution. This truncation turned out to be too severe.

The important take away message is that the algorithm is capable of sampling values of the coherent state peak including the tails, where the sampling is only determined by the probability of the intertwiner labels; no cut-off is introduced by hand. Therefore, this is a different implementation of the idea introduced in \cite{Dona:2019dkf} to approximate the coherent states by truncating the intertwiner degrees of freedom. However, it is also clear that this sampling algorithm becomes efficient compared to the full calculation at large spins when the coherent states are strongly peaked.

\paragraph{Parameters for Monte Carlo algorithm} 
Beyond the discussion on how to deal with sampling close to the boundary of the intertwiner range, the algorithm is fairly standard. We start from a random configuration. To ensure that we are within the region of probable configurations and independent of the starting point, we let the system ``thermalize'' with respect to the probability distribution and run $10^4$ Monte Carlo steps. When plotting the intertwiner label as a function of thermalization steps, one will see that it fluctuates around the peak of the distribution as desired. Then, during the sampling process, we perform $\sim 2 \cdot 10^3$ Monte Carlo runs to ensure that the samples are independent of one another as in each step we only perform small changes. Finally, we take $N$ samples, where larger $N$ should lead to better approximations of the final result.

\subsubsection{Sampling multiple coherent intertwiners}

So far we have discussed the sampling process for a single intertwiner label governed by a probability distribution derived from a coherent intertwiner state. Since Monte Carlo methods are tailored towards many variables, let us briefly discuss how this idea can be generalized to many coherent intertwiners, e.g. for a spin foam calculation where the boundary consists of several tetrahedra described by coherent states. Since each of these intertwiners possesses their own boundary state, we can derive an \textit{independent} probability distribution for each of them. Thus, we can sample each intertwiner independently for the same (or similar) values for thermalization steps and steps between taking samples. This is significantly simpler than sampling all variables at the same time, in particular it takes longer for the full system to thermalize or to become uncorrelated. Hence, we expect numerical costs to scale only \textit{linearly} with the number of boundary intertwiners; additionally we can parallelize the sampling of these intertwiners. However, despite sampling a collection of one-dimensional systems, we will have to increase the number $N$ of samples for all intertwiners to capture sufficiently many configurations of the system. Indeed, under increasing the number of coherent intertwiners the total number of configurations grows with the product of the ranges. If the system converges quickly under increasing $N$, $N$ should still be much smaller than the number of all possible configurations. Thus, we expect this algorithm to be substantially more efficient in terms of computational and memory costs for a large number of boundary intertwinters and large intertwiner space dimensions and offer the possibility to explore larger spin foam 2-complexes more efficiently

In the next section we discuss how to apply this sampling algorithm to the coherent SU$(2)$ vertex amplitude as a first test and proof of principle and present results for different sets of boundary data. We compare these results to the full calculation, the asymptotic formula and results found by random sampling.

\section{Approximating coherent vertex amplitudes} \label{sec:approx-coherent}

So far, we have discussed how to define a probability distribution for a single coherent tetrahedron and how to sample from it. The coherent vertex amplitude is defined as the contraction of five such tetrahedra against the SU$(2)$ $\{15j\}$-symbol. We will approximate the coherent amplitude by sampling each intertwiner label individually. Similarly, we can straightforwardly generalize this method to arbitrarily large 2-complexes with boundary for coherent state boundary data.

The key point we must address is that the probability distribution we intend to use is not part of the coherent vertex amplitude. To explain how we overcome this, consider a simple example, a $1$d integral $\int_{0}^{1} dx \, f(x)$ over the finite interval $[0,1]$. We can approximate the integral by sampling a probability distribution $P$ with $\int_{0}^{1} dx \, P(x) = 1$ as follows:
\begin{equation}
    \int_{0}^{1} dx \, f(x) = \int_{0}^{1} \frac{f(x)}{P(x)} P(x) \approx \frac{1}{N} \sum_{i=1}^N \frac{f(x_i)}{P(x_i)} \quad .
\end{equation}
In the final step, we approximate the expression by summing over $N$ samples $\{x_i\}$ generated from $P$. Note that sampling from $P$ modifies the original expression: we sample points $x_i$ more frequently which are more probable according to $P$, while we sample points less frequently which are less probable. Since this is not present in the original expression, we compensate for this by evaluating $\frac{f}{P}$ for these samples, i.e. samples which are less probable in $P$ contribute more to the approximation. So far we have not specified the distribution $P$ and a priori it is not clear whether $P$ is suitable for efficiently approximating the desired expression. The ideal $P$ would be such that the fraction $\frac{f}{P}$ is constant $\forall x \in [0,1]$, resulting in an exact result for any number of samples. Yet this would imply that we have already solved the integral defeating the purpose of studying it with Monte Carlo methods in the first place. Numerical integration algorithms attempt to find such optimal distributions~\cite{Hahn:2005pf}. For rapidly oscillating functions, for which Monte Carlo methods suffer from the sign problem, probability distributions adapted to the functions are challenging to define and yet it is not clear whether this leads to a good convergence. Instead one can guess a simpler to define distribution, but convergence remains an open question.
A concrete example for a simple distribution is the constant one, which leads to random sampling.
\paragraph{Random sampling of intertwiners}

One of the simplest probability distributions we can propose is the constant probability for all intertwiner labels. For discrete variables such as the orthonormal intertwiner labels, the probability distribution is simply given by the inverse of the number of possible intertwiner labels. For a single intertwiner label, this reads:
\begin{equation}
    \sum_{\iota} f(\iota) = \sum_{\iota} \frac{f(\iota)}{P({\iota})} P({\iota}) = \sum_{\iota} n^{\iota} f(\iota) \frac{1}{n^{\iota}} \approx \frac{n^{\iota}}{N} \sum_{\i= 1}^N f(i) \quad ,
\end{equation}
where $n^\iota$ denotes the total number of possible intertwiner labels. $N$ is the number of samples; the larger $N$, the better the approximation will become, however convergence might be slow. From such a constant distribution, we can simply generate samples by randomly selecting one among all the possibilities, hence the name random sampling.

At first sight, random sampling appears paradoxical; it is geared towards approximating the sum / integral of a constant function, for which Monte Carlo methods are not necessary in the first place. From a practical point of view, a few advantages emerge common for Monte Carlo methods. It is straightforward to implement in most situations\footnote{Determining the total number of possibilities can be intricate, e.g. when implementing coupling rules of representations in spin foams, see also~\cite{Dona:2023myv}. Alternatively, we can simply allow all possible values of variables and set forbidden configurations to give vanishing contributions. Yet, this leads to slower convergence.} and the numerical costs scale with the number of samples rather than the number of variables in the system. In particular for systems with many variables, random sampling can provide a reasonable approximation at less costs than brute force summation. However, it is a priori not clear how many samples will be necessary for a convergent result. 
The expectation is that less samples are necessary to obtain a convergent result if one instead uses importance sampling of a distribution adapted to the problem at hand. In the next paragraph, we expand on the numerical challenge of computing the coherent vertex amplitude and how we choose a probability distribution to more efficiently approximate it. 

\paragraph{Importance sampling coherent intertwiners}

As we describe in detail in section \ref{sec:Monte-Carlo}, our aim is to sample intertwiner degrees of freedom with respect to the absolute value of coherent intertwiners $|c_\iota(j_i,\vec{n}_i)|$, precisely the coefficient of coherent intertwiners expressed in the orthonormal intertwiner basis. In general, we could choose any coherent state to define such a probability distribution\footnote{Using coherent states which describe tetrahedra with non-closing normals are likely not suitable, as their intertwiner norm is exponentially suppressed as one uniformly scales up its spins.}, but this would not be adapted to the calculation at hand. Instead we pick the states encoded in the boundary data and rewrite the vertex amplitude as follows:
\begin{align}
	& \includegraphics[width = 0.225 \textwidth, valign = c]{drawings/vertex} \; = \; \sum_{\{\iota_i\}}	\prod_{k=1}^5 c_{\iota_k}(j_i,\vec{n}_k)\includegraphics[width = 0.275 \textwidth, valign = c]{drawings/vertex_snw_15.pdf} \\ \nonumber
    \quad & = \sum_{\{\iota_i\}} \prod_{k=1}^5 \frac{n_k}{|c_{\iota_k}(j_i,\vec{n}_k)|} \frac{|c_{\iota_k}(j_i,\vec{n}_k)|}{n_k} \prod_{k=1}^5 c_{\iota_k}(j_i,\vec{n}_k)\includegraphics[width = 0.275 \textwidth, valign = c]{drawings/vertex_snw_15.pdf} \\ \nonumber
    \quad & \approx \frac{1}{N} \sum_{x=1}^N \prod_{k=1}^5 n_k \frac{c_{\iota_k(x)}(j_i,\vec{n}_k)}{|c_{\iota_k(x)}(j_i,\vec{n}_k)|} \includegraphics[width = 0.275 \textwidth, valign = c]{drawings/vertex_snw_15.pdf} \quad .
\end{align}
In the second line, we insert the probability distribution and its inverse. Then, in the third line, we approximate the full expression by sampling with respect to the distribution with $N$ samples in total. The final expression we evaluate for the samples of intertwiners depends then on the phase of the coefficients $c_\iota$ and the overall normalisations.

Before presenting results of this sampling method, let us briefly revisit the properties of the coherent vertex amplitude and its sign problem, which is not solved by this sampling method.

\subsection{Sign problem in the coherent vertex amplitude}

While the SU$(2)$ $\{15j\}$-symbol is defined to be real, the coherent vertex amplitude is generically complex. This is due to the introduction of the complex overcomplete basis of Perelomov coherent states, which are however essential to define coherent tetrahedra states that are peaked on the shape of classical polyhedra. Unfortunately, it is also the reason that we cannot simply use spin foam amplitudes itself to define a probability distribution\footnote{Technically this already applies to the $\{15j\}$-symbol, as it can be negative.}. On the other hand, coherent SU$(2)$ states are only defined up to a phase. Similarly the coherent vertex amplitude is defined up to a global phase~\cite{Dona:2017dvf}, and we choose it such that the coherent amplitude is purely real. In the numerical setting, this can be done as follows: we compute the full amplitude for a spin configuration and choice of coherent states and compute its phase. Under uniform scaling of spins (keeping the $\vec{n}_i$ fixed), this phase changes linearly, such that we use it to turn amplitude real for all spins. Thus, we numerically obtain a real amplitude for all spins with a tiny imaginary part limited by numerical precision. Moreover, for Regge-geometry boundary data, the real part oscillates with the Regge action~\cite{Regge:1961ct} of the described $4$-simplex (under uniform scaling of boundary areas).

The goal of our Monte Carlo algorithm is to approximate the full amplitude, i.e. reproduce the significant real part and a vanishing imaginary part.
As we will see below, this reveals the strengths and weaknesses of our algorithm. While the real part (for almost all cases) shows a good convergence and accuracy compared to the full calculation and the asymptotic approximation (for large spins), reproducing the oscillating behavior for Regge-geometric boundary data, the imaginary part is more subtle. 
While it is in most cases significantly smaller than the real part, typically up to three orders of magnitude, it is not as small as in the full calculation. This suggests that the convergence of the imaginary part is slower than the real part and more samples are necessary to approximate it better. 
We expect this behavior: the individual summands of the coherent vertex amplitude are generically complex. While the real parts sum up to a non-vanishing number, their imaginary parts exactly cancel due to the choice of global phase. This is the literal definition of the sign problem, and Monte Carlo methods suffer from slow convergence due to sampling of configurations. Still, for the coherent vertex amplitude the convergence of the imaginary part is acceptable. If we would only consider the absolute value of the amplitude, this convergence issue would remain unnoticed.

However, we will also see that for some boundary data that the real part suffers from the sing problem. These are precisely the boundary spins where the entire amplitude almost vanishes, i.e. close to the roots of the cosine of the Regge action. The same argument as for the imaginary part holds here as well, as (most of) the summands in the vertex amplitude cancel each other.

%The first test whether this idea holds any water is the coherent SU$(2)$ BF vertex amplitude for different boundary data, 
In the following we will introduce the boundary data to a few geometric Euclidean $4$-simplices, e.g. the equilateral $4$-simplex and an isosceles $4$-simplex, and compute the amplitude using importance sampling Monte Carlo. Since Monte Carlo methods are inherently random, we need to provide an error estimate. To do so, we take $10^5$ samples and repeat this process $30$ times. We compute the mean and variance of these estimates, thus using $3 \cdot 10^6$ samples in total for each set of boundary data. Additionally, for the equilateral $4$-simplex, we will also show results for random sampling to give an impression of convergence. Moreover, we will compare these results to the full calculation where it is computationally feasible and the asymptotic amplitude to leading order.

\subsection{Results}

%The results for various boundary data are presented as follows. In addition to the Monte Carlo results, which are split into the real part (the actual amplitude) and the imaginary part, which should vanish by choice of the phase, we present the full amplitude, calculated via by using the full numerical algorithm, and its asymptotic approximation valid for large spins such that we can compare them. 
The main results are shown in plots of the rescaled vertex amplitude, where we account for the asymptotic scaling behavior of the vertex amplitude ($\lambda^{-6}$ for all $j \rightarrow \lambda j$). These plots nicely show the oscillatory nature of the vertex amplitude, and, at a glance, we get a good impression of the accuracy of the Monte Carlo results in most cases including an error estimate. However, from these plots, it is difficult to see deviations if the real part of the amplitude is small, and how small the imaginary part of the amplitude is. Therefore, we add a logarithmic plot of the absolute value. Finally, we also add a plot of the relative error $\epsilon$ defined as:
\begin{equation}
    \epsilon = \left|\frac{\mathcal{A}^\text{MC}_v - \mathcal{A}_v}{\mathcal{A}_v}\right| \quad .
\end{equation}
The relative error is computed for the full amplitude as well as its asymptotic approximation.

For all the plots, we use the same colors. The full calculation is in orange, its asymptotic approximation in yellow; the same colors apply to the relative errors. Monte Carlo results are shown as crosses with error bars (exception the logarithmic plots), where blue crosses show the real part and purple crosses the imaginary part.

\subsubsection{Equilateral $4$-simplex}

The equilateral $4$-simplex is one of the simplest examples of the coherent vertex amplitude. In Regge calculus it is prescribed by having the same edge lengths for all its ten edges. All of its 4d dihedral angles are equal and also all of its sub-simplices are equilateral as well. In spin foams, it is described by choosing all ten spins, related to its areas, to be the same. We prescribe the normal vectors for the triangles in each tetrahedron as follows:
\begin{eqnarray} \label{eq:equilateral_n}
    \vec{n}_{12} = (0,0,1) &, \quad & \vec{n}_{13} = \left(0, 0.94280, -\frac{1}{3}\right) \quad , \nonumber \\
    \vec{n}_{14} = \left(0.816497, -0.47140, -\frac{1}{3}\right) &, \quad & \vec{n}_{15} = \left(-0.816497, -0.47140, -\frac{1}{3}\right) \quad .
\end{eqnarray}
Here we choose the same normal vectors for each tetrahedron. Instead one could also use the twisted spike configuration \cite{Dona:2017dvf}, where the normal vectors are chosen to be pairwise anti-parallel, i.e. $\vec{n}_{ab} = -\vec{n}_{ba}$. Both choices differ by a global phase, which we are free to choose.

Before presenting the results of our importance sampling algorithm, we briefly show results for random sampling to provide a comparison of accuracy and convergence of the results

\begin{figure}
    \centering
    \includegraphics[width = 0.45 \textwidth]{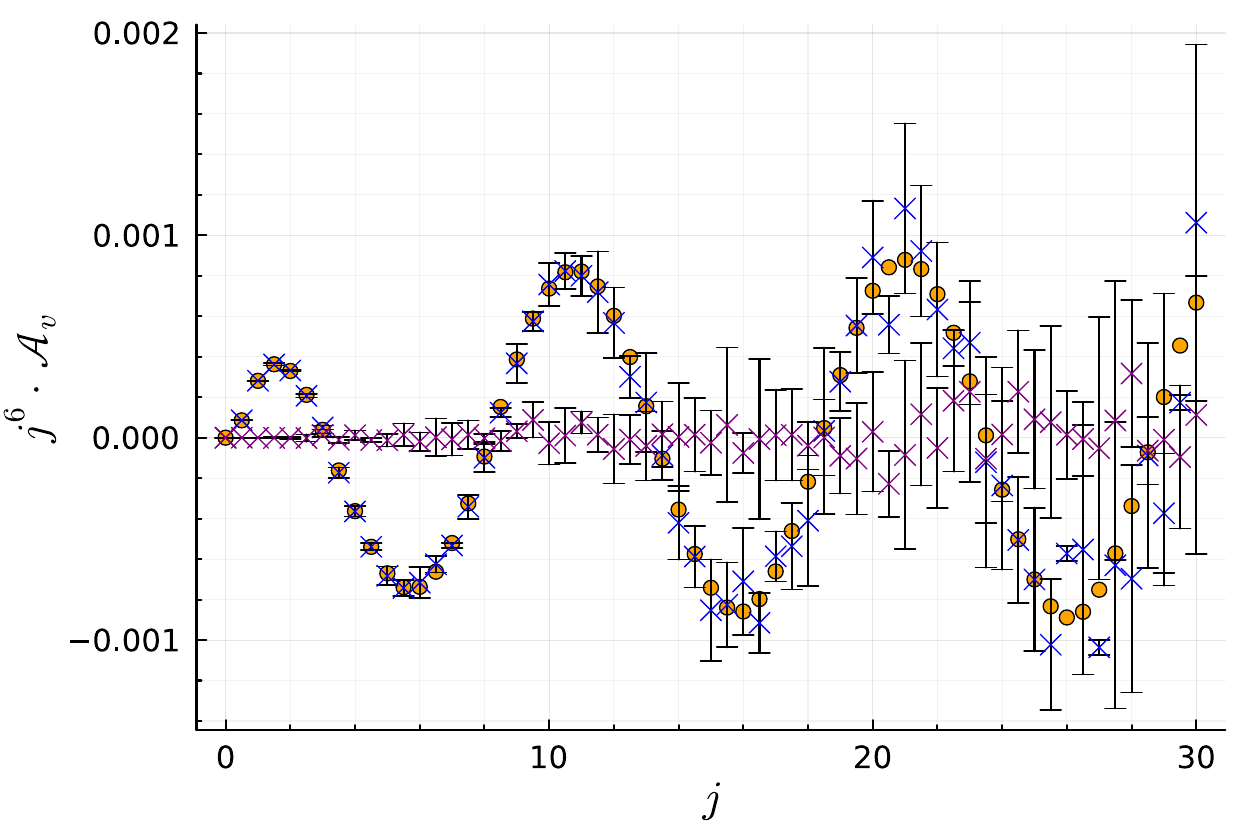}
    \includegraphics[width = 0.45 \textwidth]{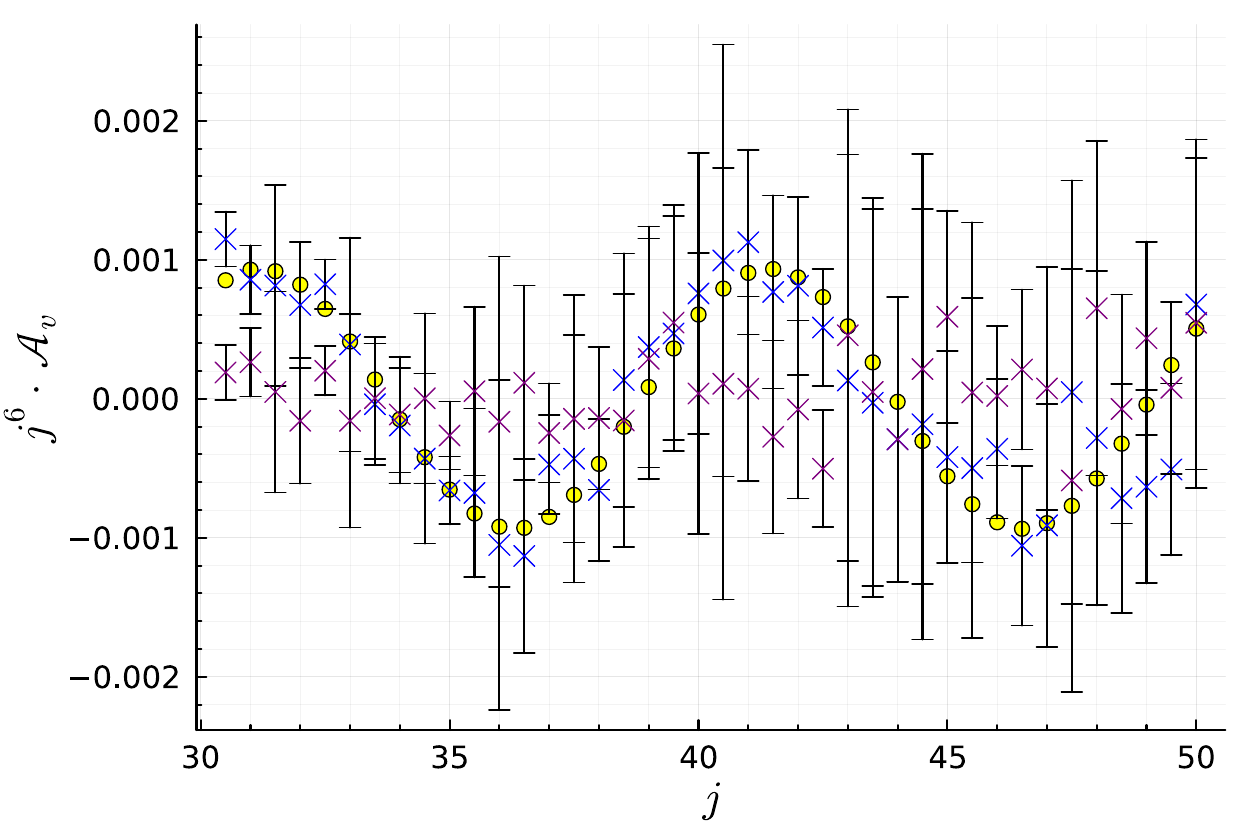} \\
    \includegraphics[width = 0.45 \textwidth]{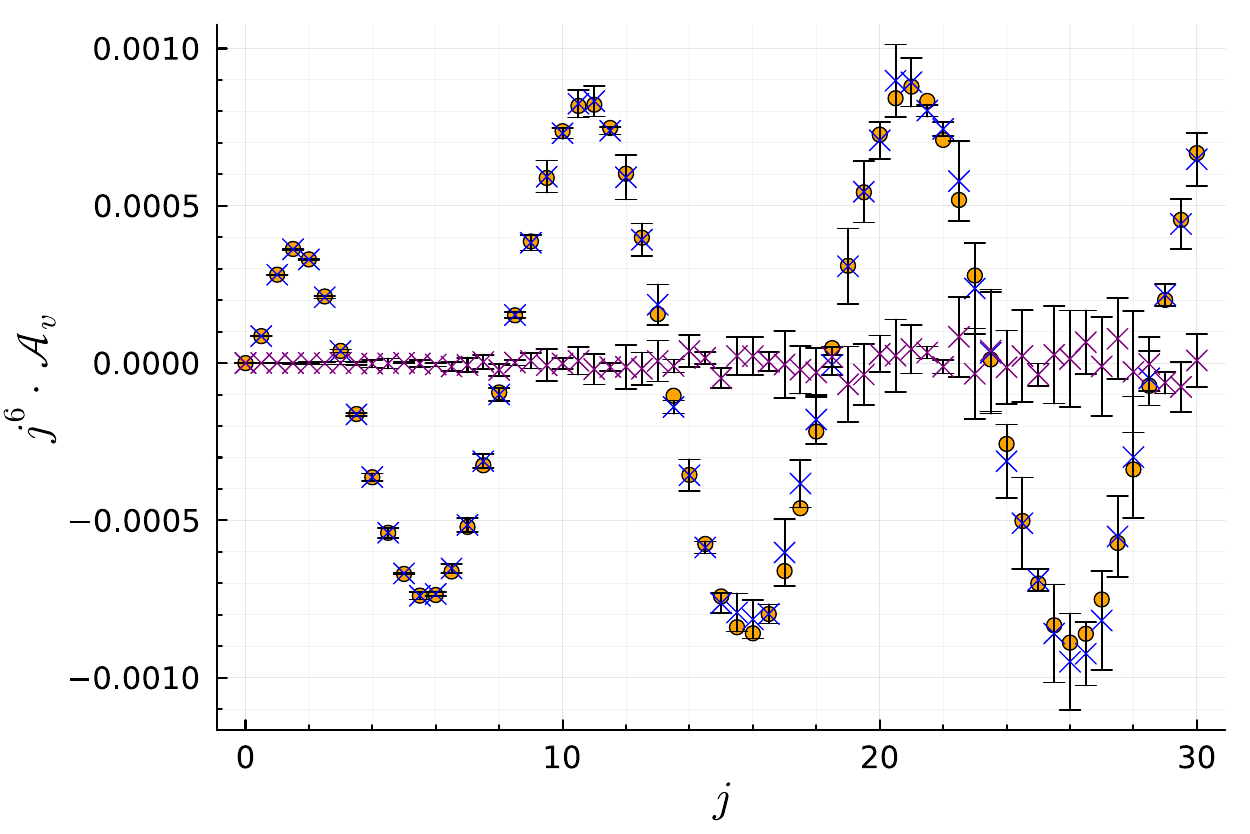}
    \includegraphics[width = 0.45 \textwidth]{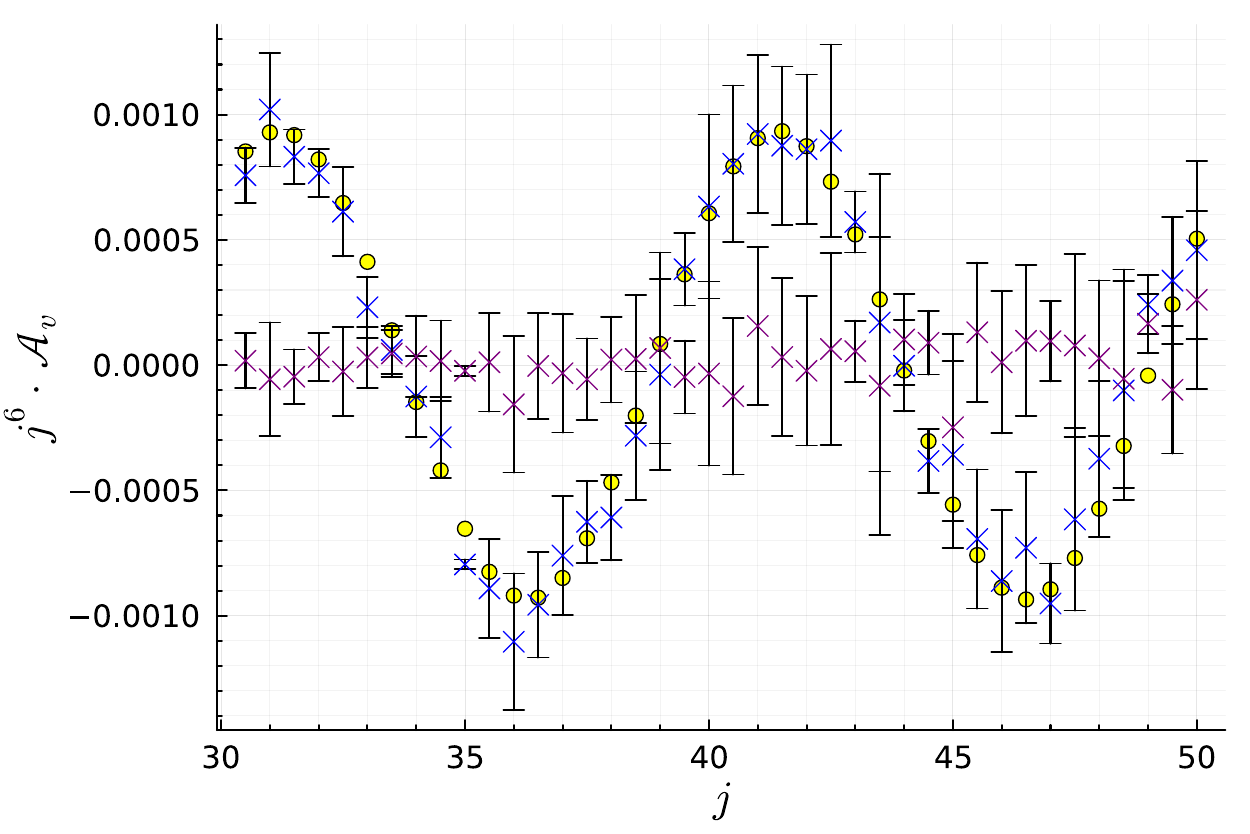}
    \caption{Random sampling results for equilateral coherent vertex amplitude for $30$ runs, top plots show $10^5$ samples per run, bottom plots $10^6$ samples per run. $j$ labels the spins. \textit{Left:} comparison for spins up to $j=30$ with full numerical calculation. \textit{Right:} Comparison from $j=30.5$ up to $j=50$ to asymptotic formula.}
    \label{fig:random_rescaled}
\end{figure}

\paragraph{Random sampling}
For random sampling we present the results of two runs: both have $30$ repetitions in total with $10^5$ and $10^6$ samples each respectively. From the results of these runs we compute the mean and the variance of the amplitude to estimate an error; the mean is thus computed from $3 \cdot 10^6$ and $3 \cdot 10^7$ samples in total respectively. In fig.~\ref{fig:random_rescaled} we plot the rescaled results from both runs and compare to the full numerical calculation and the semi-classical amplitude. Since the plots are busy in particular due to error bars, we split them into two intervals: from $j=0$ to $j=30$ to compare to the full amplitude and from $j=30.5$ to $j=50$ to compare to the asymptotic formula. For the simulations with $10^5$ samples per run, the real part agrees well with the full calculation up to spins $j \sim 15$, yet for larger spins deviations are visible and the variance is large. The same holds in comparison to the asymptotic formula. As expected, convergence for the imaginary part is worse and for spins $j > 20$ the imaginary part is frequently of the same order of magnitude as the real part. Thus, more samples are clearly necessary. Indeed, the runs with $10^6$ samples significantly improve the results and the agreement with full and semi-classical formula is decent. However, the variance of the real part becomes large for $j>40$. The imaginary part is improved as well, but shows non-vanishing values already early on. Nevertheless, while we observe inaccuracies, in most cases random sampling reproduces the correct order of magnitude (recall that the result are rescaled by $j^6$), and for $j>20$ the results are obtained at lower computational costs than the full calculation.

%\begin{figure}
%    \centering
%    \includegraphics[width = 0.45 \textwidth]{plots/random_1e6_small.pdf}
%    \includegraphics[width = 0.45 \textwidth]{plots/random_1e6_large.pdf}
%    \caption{Random sampling results equilateral coherent vertex amplitude for $30$ runs, $10^6$ samples per run. $j$ labels the spin. \textit{Left:} comparison for spins up to $j=30$ with full numerical calculation. \textit{Right:} Comparison up to $j=50$ to asymptotic formula.}
%    \label{fig:random_large}
%\end{figure}

\begin{figure}
    \centering
    \includegraphics[width = 0.45 \textwidth]{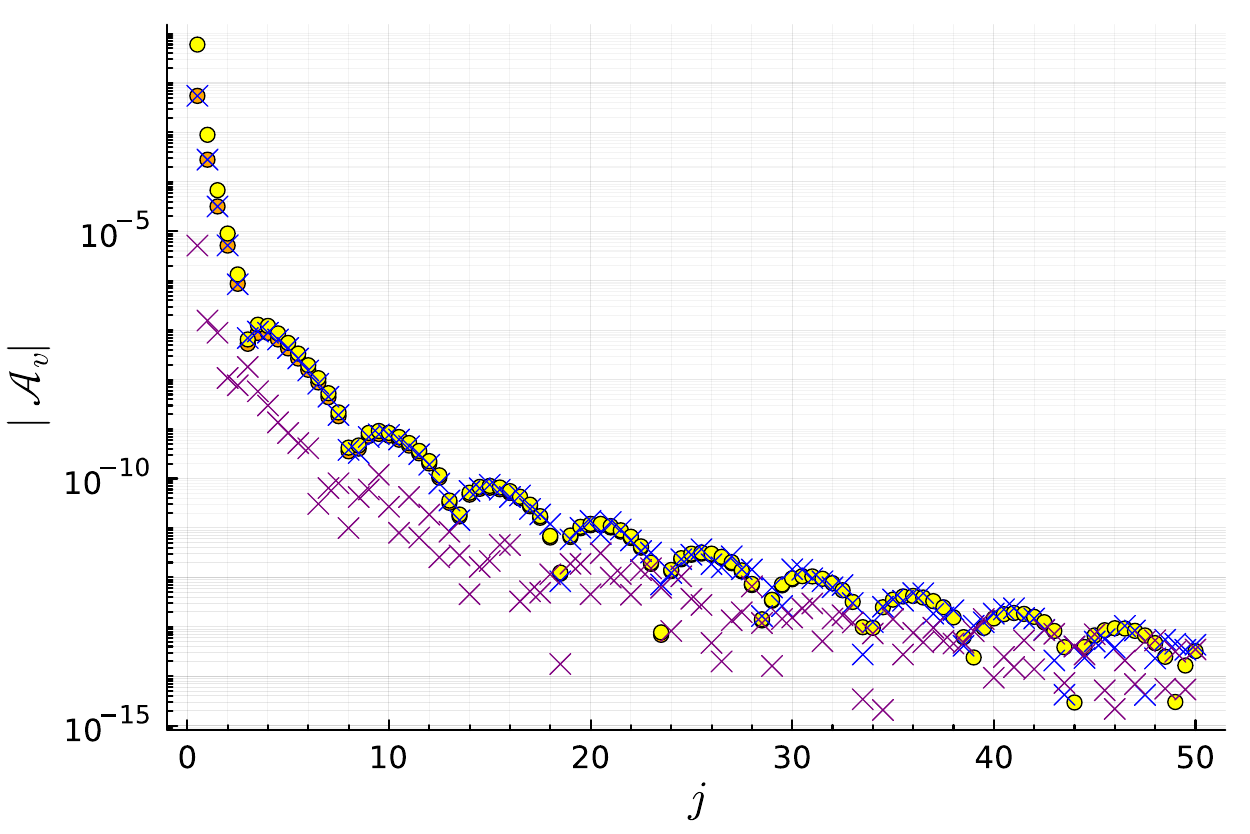}
    \includegraphics[width = 0.45 \textwidth]{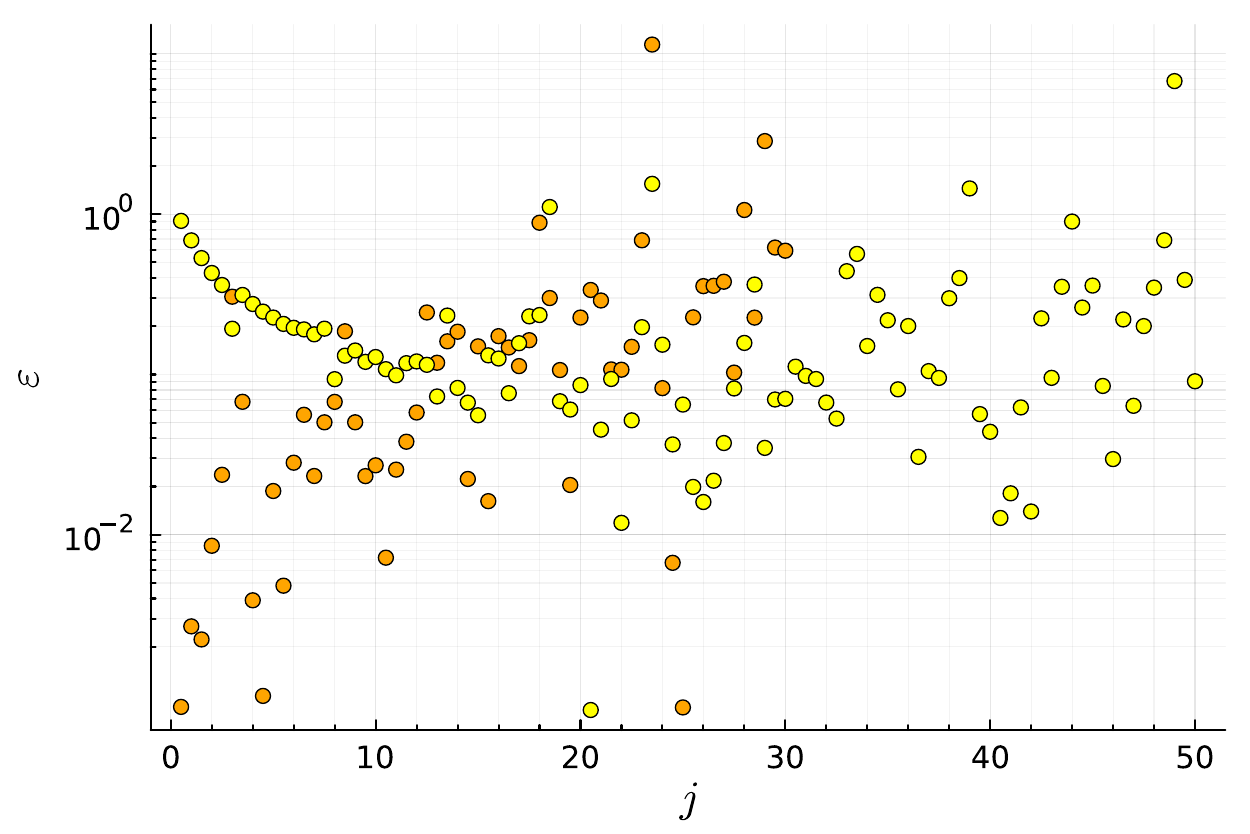} \\
    \includegraphics[width = 0.45 \textwidth]{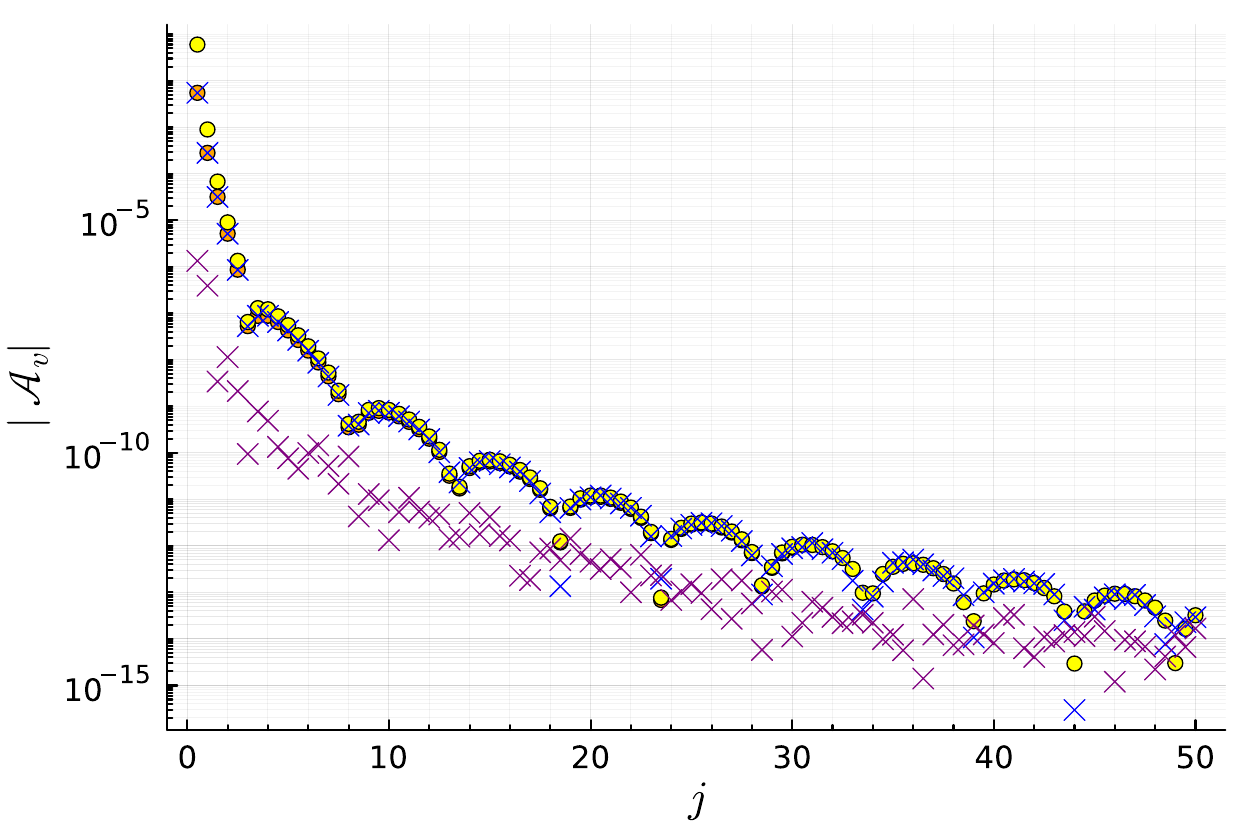}
    \includegraphics[width = 0.45 \textwidth]{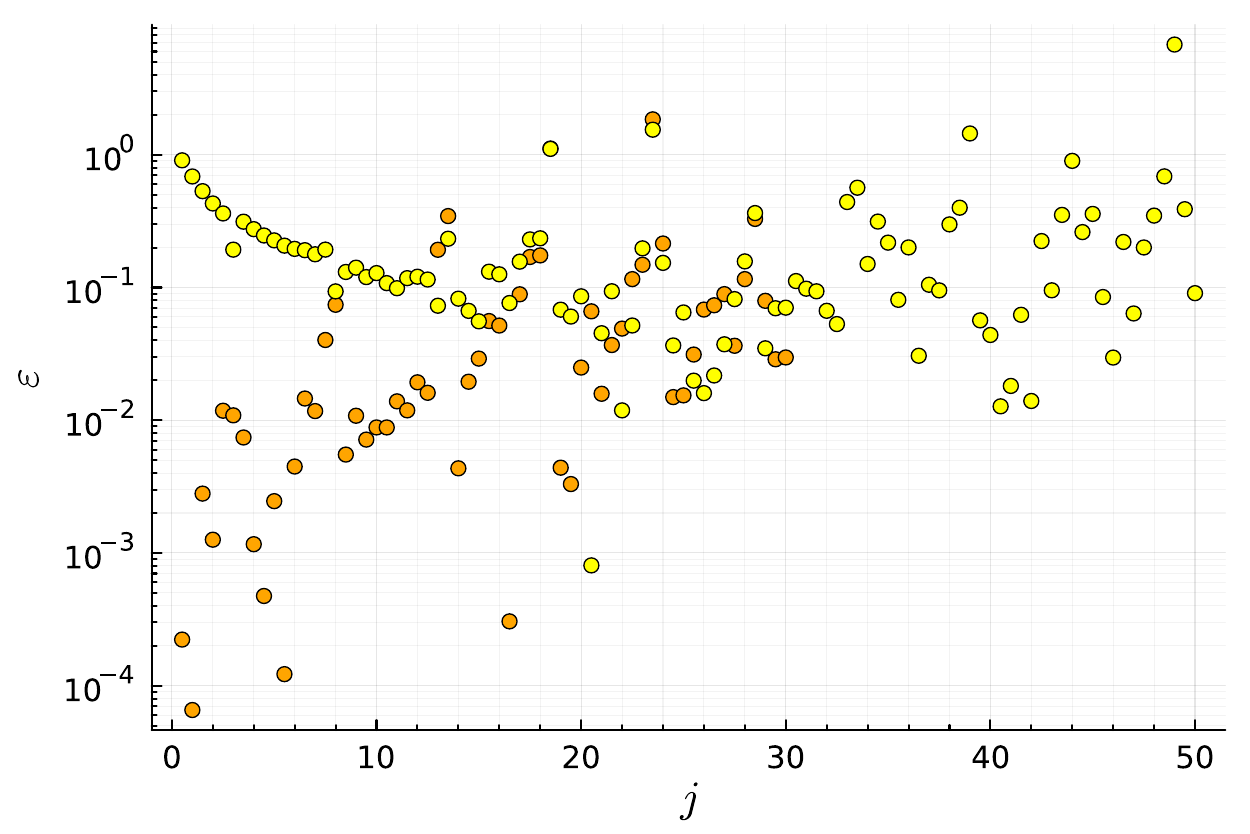}
    \caption{Random sampling of equilateral vertex amplitude, top shows results from $30$ runs with $10^5$ samples, top results for $30$ runs with $10^6$ samples. \textit{Left:} Logarithmic plot of absolute value of vertex amplitude $\mathcal{A}_v$. \textit{Right:} Relative error $\epsilon$ of real part of $\mathcal{A}_v$.}
    \label{fig:random_log_error}
\end{figure}

These results are also confirmed in the logarithmic plots and relative error, see fig. \ref{fig:random_log_error}. The logarithmic plots nicely show that random sampling adequately reproduces the real part, unless it almost vanishes compared to boundary spins of similar size; this indicates that the sign problem is more severe for these cases. The agreement gets worse as we increase the spins $j$, which must be compensated by increasing the number of samples. Convergence of the imaginary part is worse; in particular for large spins it is often of the same order of magnitude as the real part. We also see that the relative error for the real part grows for larger spins. For small spins, it is fairly low, but quickly increases for growing spins. Still in many cases it is at or below $10\%$, which could be improved by more samples. To improve on both the real and imaginary part, we must increase the number of samples further, in particular for large spins.

To summarize, we can draw two conclusions for random sampling. First, it is able to decently approximate the coherent vertex amplitude even for relatively small number of samples. While the variance is high and increases with larger spins, the relative error remains under control. This is already a achievement since these results are obtained at significantly lower costs compared to the full calculation. The second insight is that the sign problem does not appear to be too severe, at least for the real part of the amplitude.

In the following, we will improve on these results with our importance sampling algorithm using the boundary coherent states, in particular in terms of accuracy for the same number of samples. Furthermore, we also increase the boundary spins further.

%\paragraph{Single run}

%A single run consists of taking $10^5$ samples using importance sampling from the coherent state distribution. We include it mostly to show the impressive agreement to the full and asymptotic amplitude even at high spins. Since it is a single run only we cannot present 

\paragraph{Results of multiple runs and error estimate}

\begin{figure}
    \centering
    \includegraphics[width = 0.45 \textwidth]{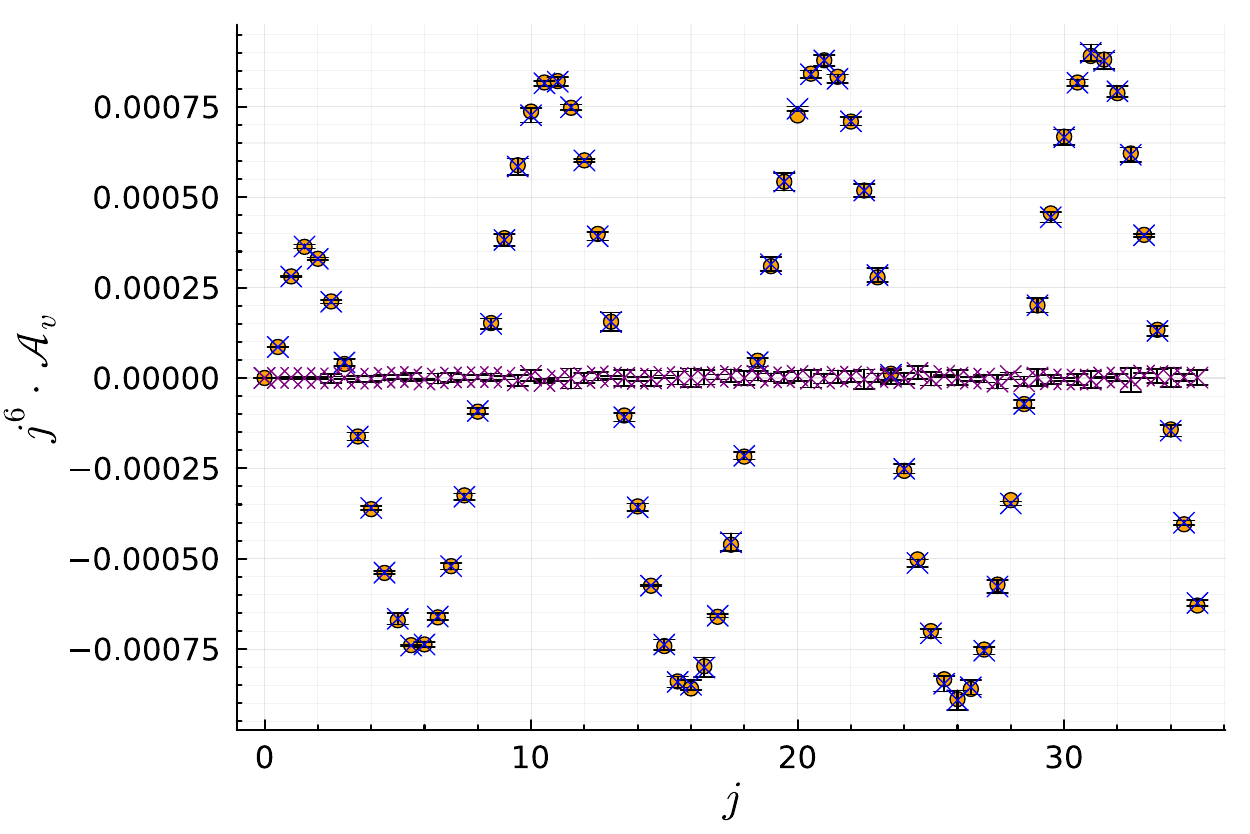}
    \includegraphics[width = 0.45 \textwidth]{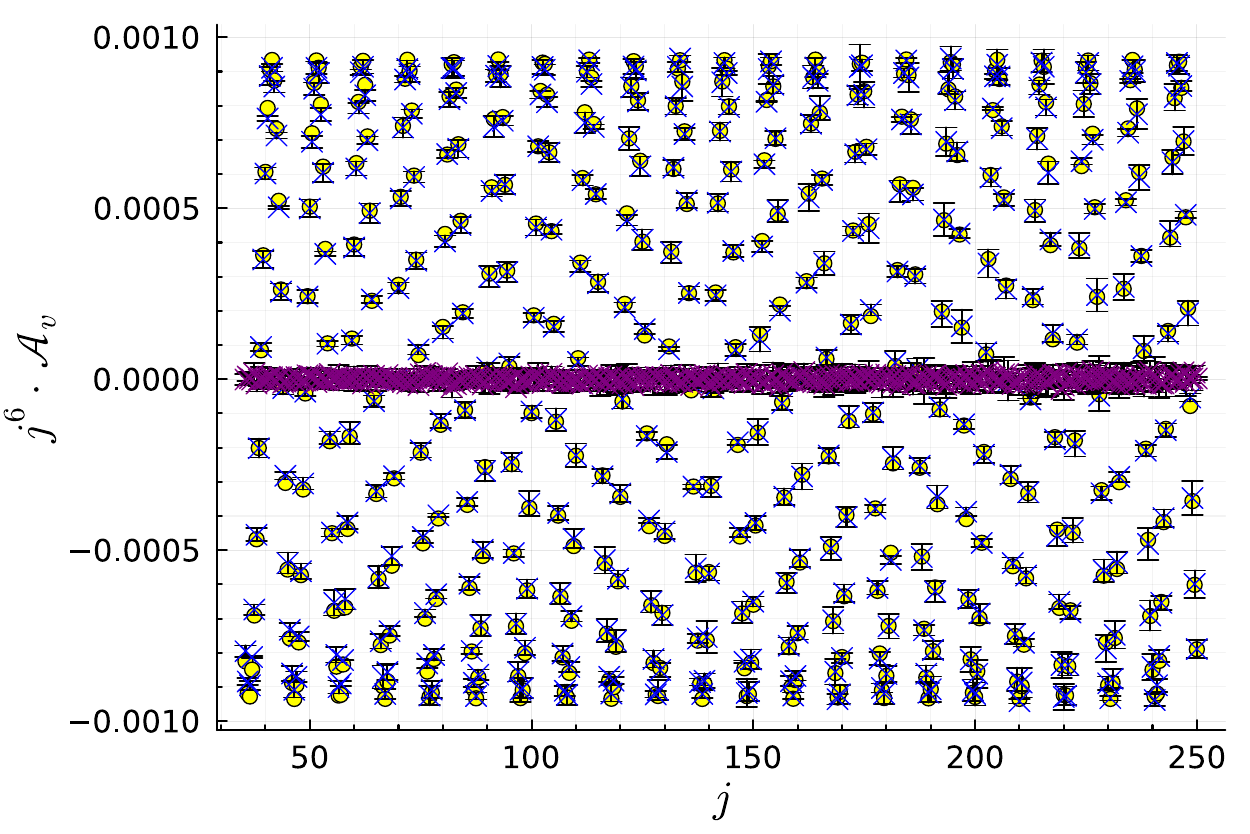} \\
    \includegraphics[width = 0.45 \textwidth]{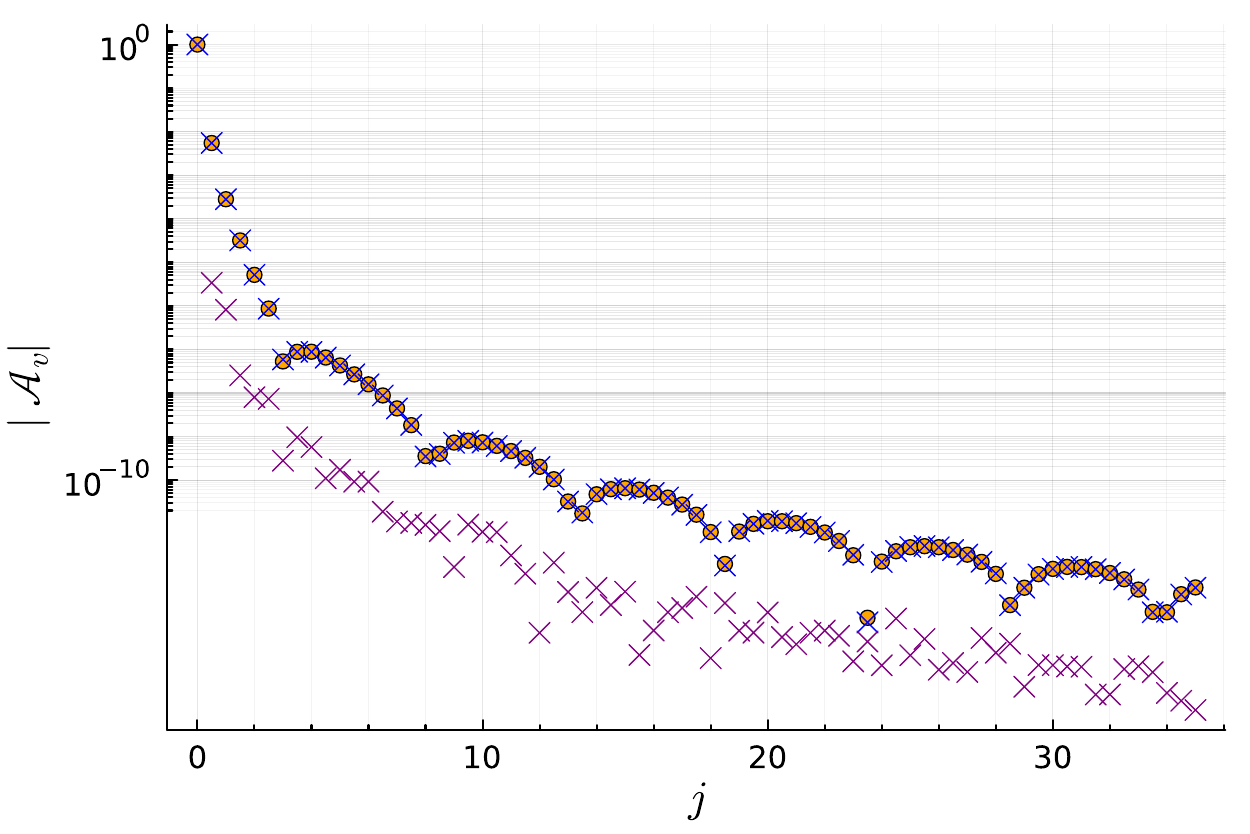}
    \includegraphics[width = 0.45 \textwidth]{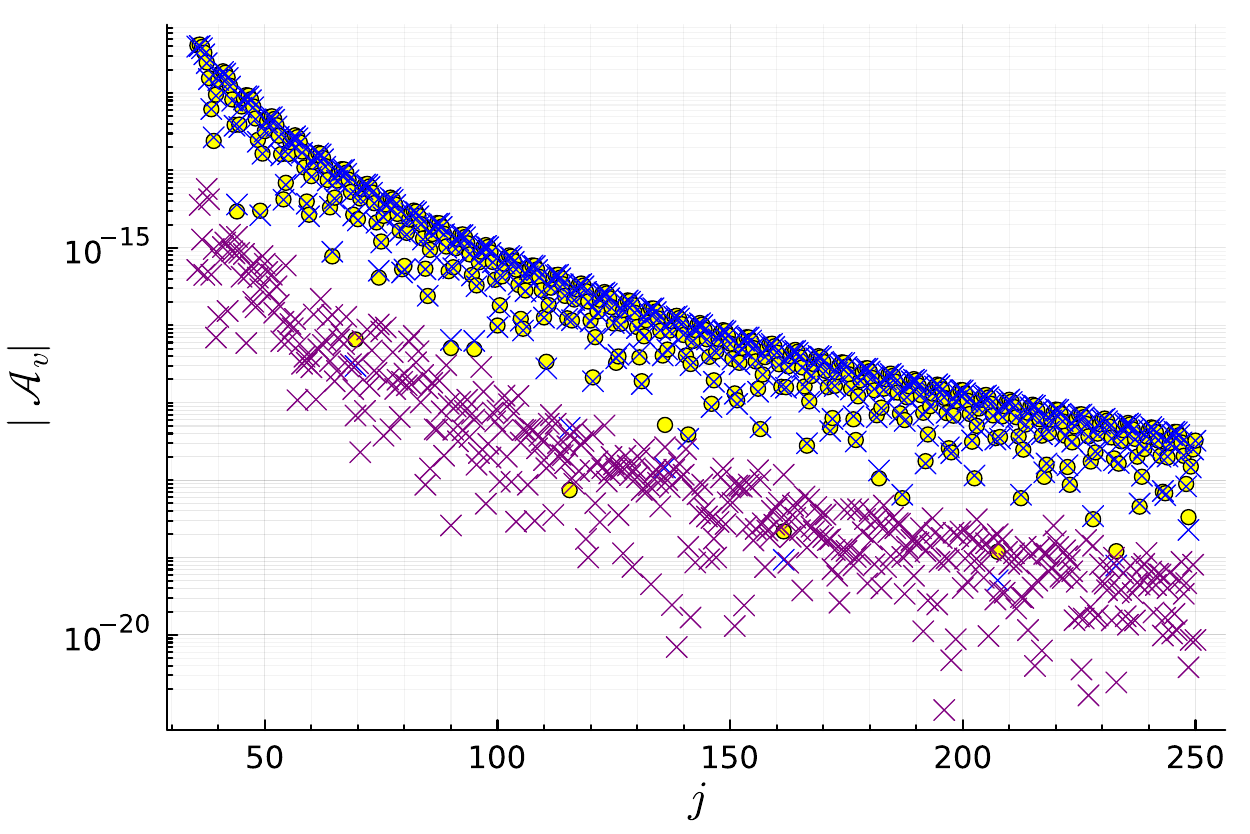}
    \caption{Equilateral vertex amplitude approximated via coherent state importance sampling. Top plots show the rescaled amplitude, bottom plots show logarithmic plots of absolute value. \textit{Left}: Spins $j$ from $0$ to $35$. \textit{Right}: Spins $j$ from $35.5$ to $250$. }
    \label{fig:equi_all}
\end{figure}

Similar to random sampling, we present the data for $30$ runs, each with $10^5$ samples; hence $3 \cdot 10^6$ samples in total. The results range from $j=0$ to $j=250$ and are presented in fig.~\ref{fig:equi_all}. We have split the plots around $j=35$ between the results for the full and the asymptotic amplitude. Let us discuss the top plots first, which show the rescaled amplitude. Let us start with real part.

Overall, the Monte Carlo results are barely distinguishable from the full calculation, and also when compared to the asymptotic formula the results agree well besides a few deviations. In particular for spin around $j\sim40$ we see a few deviations, probably because the asymptotic formula is not yet accurate enough. The excellent agreement can also be seen in the logarithmic plots, with a few exceptions. These are again the cases for which the amplitude itself almost vanishes and where the sign problem is more severe. Another remarkable feature are the relatively small variances, in particular for small spins. For larger spins, we irregularly see slightly larger variances; this is a first indication that more samples / more runs might be necessary to improve the results. We discuss the convergence of the Monte Carlo estimates for different sample sizes for the equilateral coherent vertex amplitude in appendix \ref{app:convergence}, where we show several positive and negative examples.

The behavior of the imaginary part is more intricate. In the rescaled plots it appears to be small, typically much smaller than the real part with rather small variances. This continues for large spins, yet we observe that it deviates further from zero and also that the variances grow, while typically being much smaller than the real part. To quantify this, we consider the logarithmic plots: here we observe that the (absolute value of the) imaginary part is typically a few orders of magnitude smaller than the real part. This continues also to large spins. The only exceptions are the amplitudes which almost vanish, where the Monte Carlo method suffers from the sign problem. Here real and imaginary part are of a similar magnitude and it is likely that neither has converged for the given samples. Thus, in most cases, we reproduce a significantly smaller imaginary part, but far away from the limits of numerical accuracy. It is clear that more samples are necessary to further improve the imaginary part, yet we suspect it might give diminishing returns due to the sign problem. At least for the coherent vertex amplitude, this level of accuracy appears acceptable.

\begin{figure}
    \centering
    \includegraphics[width = 0.45 \textwidth]{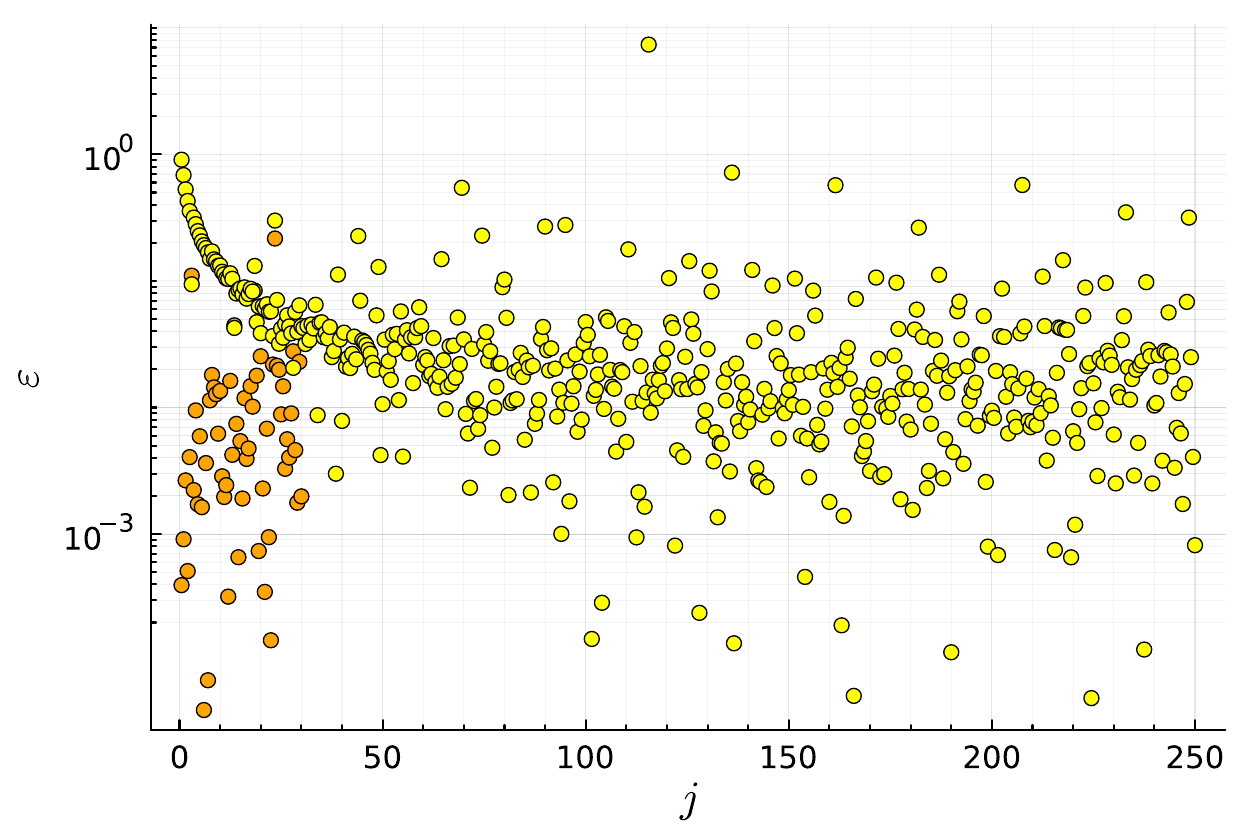}
    \caption{Relative error $\epsilon$ of real part of amplitude with respect to the full vertex amplitude calculation (orange) and its asymptotic approximation (yellow) for spins $j=0.5$ to $250$.}
    \label{fig:equi_error}
\end{figure}

As the final plot for the equilateral $4$-simplex, we show the relative error of the real part in fig. \ref{fig:equi_error}. For small spins, the relative error from importance sampling is comparable to the random sampling results for ten times the samples, and the relative error is below $1\%$ in most cases. For large spins compared to the asymptotic formula, we see again more randomness, but most errors are below $10\%$. Considering the size of the boundary spins this is an impressively accurate result, but it also shows that more samples are necessary to obtain a more precise result.

\subsubsection{Isosceles $4$-simplex}

An isosceles $4$-simplex possesses only two different edge lengths, four of its edges have one lengths, the remaining six the other. The latter six edges form an equilateral tetrahedron. All remaining tetrahedra are isosceles, i.e. they have an equilateral triangle as their base and the three remaining edges have equal lengths different from the base triangle. Here, we consider an isosceles tetrahedron, whose equilateral triangles have twice the area of the isosceles ones. To do so, we assign the following boundary data: first, we label the equilateral tetrahedron as $5$, whereas $i \in \{1,2,3,4\}$ denote isosceles tetrahedra. Thus, the spins are:
\begin{equation}
    j_{ab} = j \; \forall \; a,b < 5, \, a \neq b , \quad j_{a5} = 2j \; \forall \; a \neq 5 \quad.
\end{equation}
The smallest non-vanishing spin we can start from is $j=1$. Non-integer spins are violating the SU$(2)$ coupling rules in the isosceles tetrahedra. We assign the following normal vectors: For the equilateral $4$-simplex, we again assign the data for equilateral tetrahedra, see eq. \eqref{eq:equilateral_n}. The new data is for the isosceles tetrahedra:
\begin{eqnarray}
    \vec{n}_{12} = (0,0,1) &, \quad & \vec{n}_{13} = \left(0,0.9860,\frac{1}{6}\right) \quad, \nonumber \\
    \vec{n}_{14} = \left(0.9759, 0.14086, \frac{1}{6} \right) &, \quad & \vec{n}_{15} = \left(-0.48795, - 0.5635, - \frac{2}{3}\right) \quad .
\end{eqnarray}
The last vector is the one assigned to the equilateral triangle with twice the area compared to the isosceles one. Again, we are not using a twisted spike configuration.

\paragraph{Results}

\begin{figure}
    \centering
    \includegraphics[width = 0.45 \textwidth]{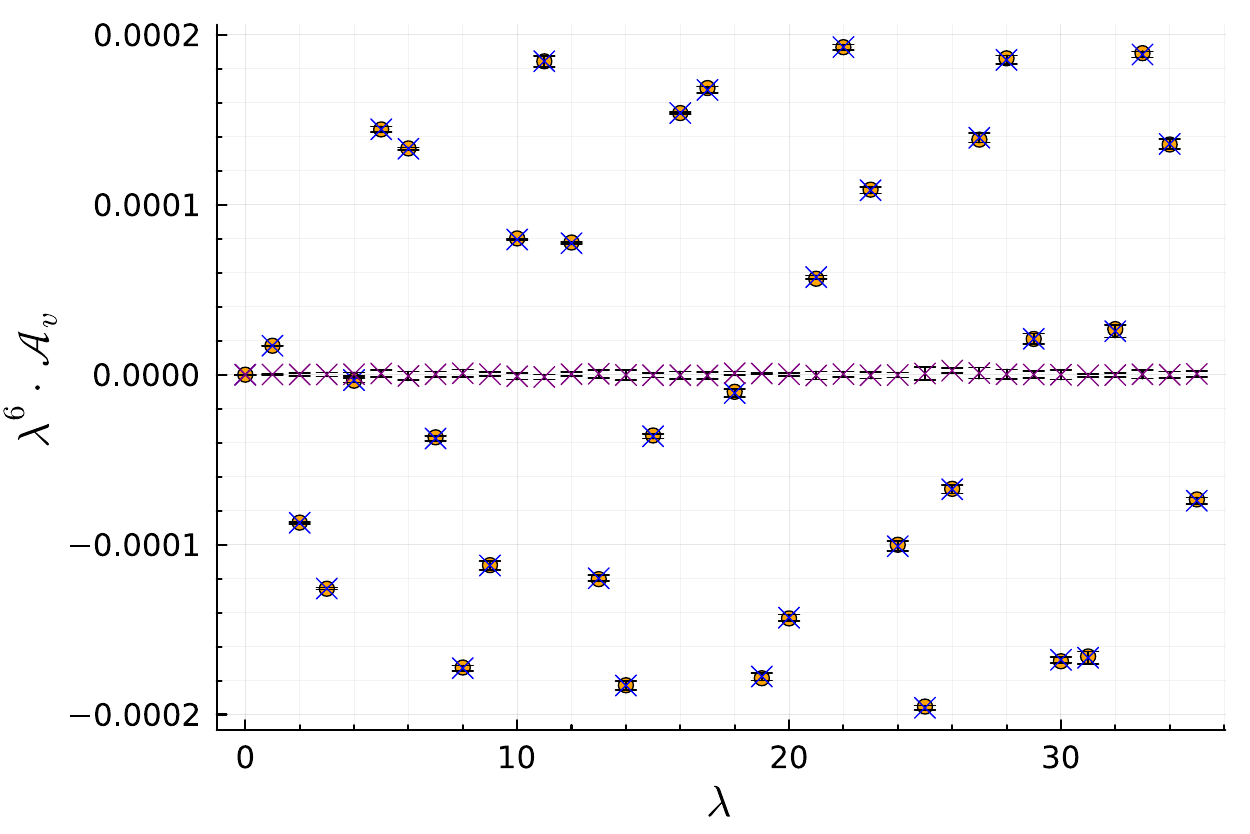}
    \includegraphics[width = 0.45 \textwidth]{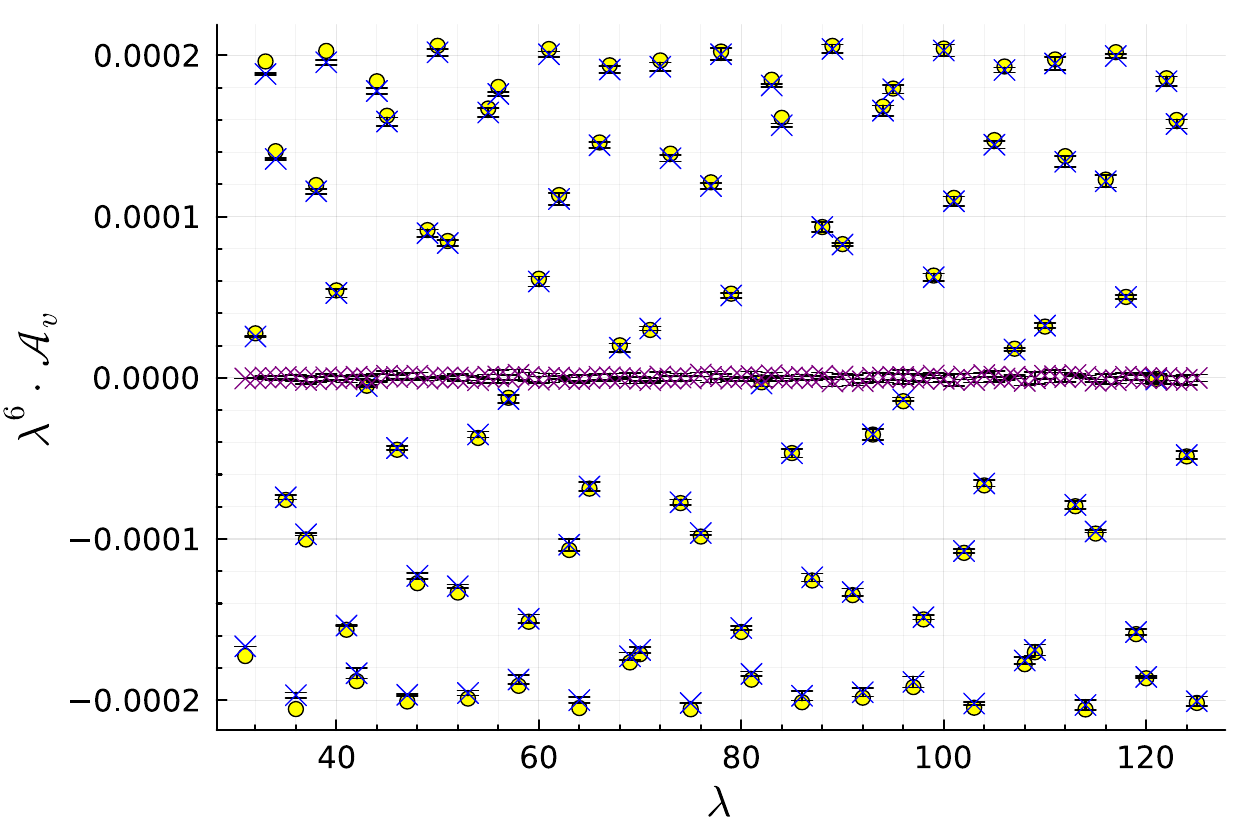} \\
    \includegraphics[width = 0.45 \textwidth]{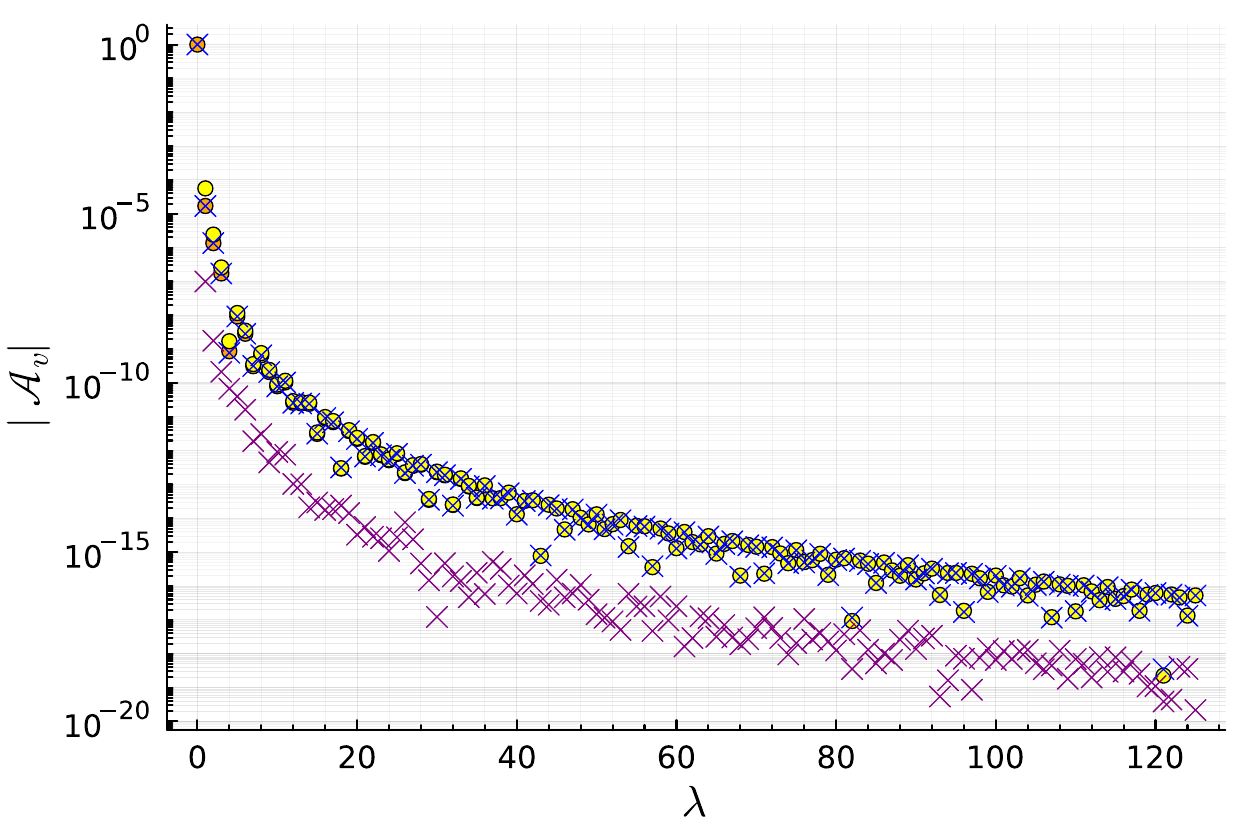}
    \includegraphics[width = 0.45 \textwidth]{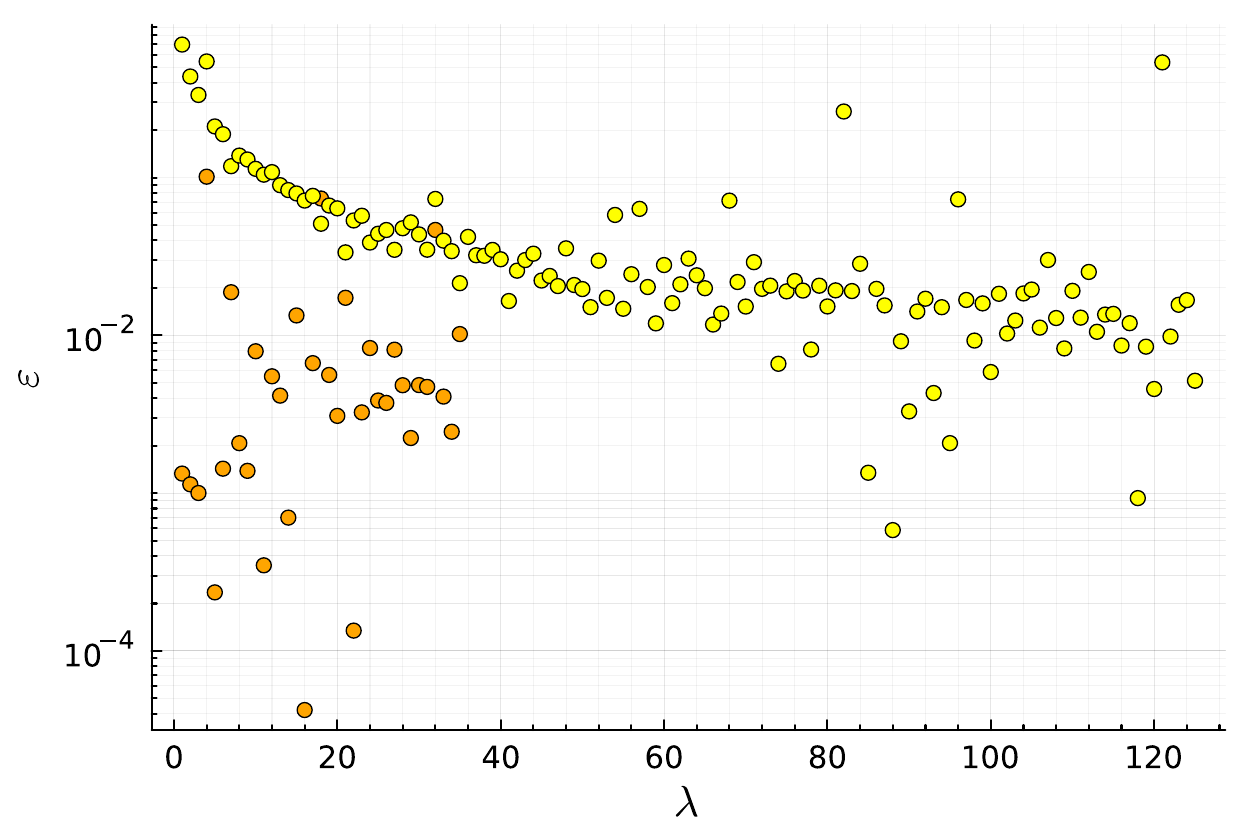}
    \caption{Importance sampling results for the isosceles case. \textit{Top left:} rescaled amplitude compared to full calculation up to $\lambda = 35$. \textit{Top left:} rescaled amplitude compared to asymptotic formula up to $\lambda = 125$. \textit{Bottom left:} logarithmic plot of absolute value of real and imaginary part and comparison to full and asymptotic vertex amplitude. \textit{Bottom right:} relative error with respect to full and asymptotic amplitude.}
    \label{fig:iso_all}
\end{figure}

As for the equilateral case, we consider $30$ runs with $10^5$ samples. The results including comparison to the full and asymptotic amplitude can be found in fig.~\ref{fig:iso_all}, where we plot the results over the scaling parameter $\lambda$, $(j,2j) \rightarrow (\lambda j,2 \lambda j)$, up to $\lambda = 125$. Since the results are qualitatively similar to the equilateral $4$-simplex case, we keep the discussion brief.

Let us begin with the plots of the rescaled amplitude, where we observe a good agreement to the full amplitude, small variances and small imaginary part close to zero. For $\lambda > 35$ we compare to the asymptotic approximation, which is closely but not fully matched for $\lambda < 60$. This might be because the asymptotic approximation is not good enough yet, and the matching improves for larger $\lambda$. In the logarithmic plot, we can barely see any deviations. For the isosceles $4$-simplex, we observe only few boundary data for which the coherent amplitude almost vanishes, thus we encounter less cases suffering from a severe sign problem leading to convergence issues. Additionally, the logarithmic plots show similar to the equilateral case that the imaginary part is a few orders of magnitude smaller than the real one, but not vanishing completely (or to numerical precision). The precision of the real part is confirmed by the relative error of the (real part of the) amplitude. Compared to the full calculation, the error is below $1\%$, except for one case where the amplitude is small. For the asymptotic formula, the relative error drops quickly and appears to asymptote to below $10\%$ for large $\lambda$, which shows the convergence of the asymptotic formula to the full one. Due to the randomness of Monte Carlo method, we see fluctuations of this error.

\subsubsection{Non-regular $4$-simplex}

The next example is an irregular $4$-simplex, which possesses two different triangle areas, namely $j$ and $2j$, but these are not assigned to form an isosceles $4$-simplex, but a non-regular configuration. For $j=1$, this $4$-simplex is prescribed by three different edge lengths: $l_1 \approx 1.58$, $l_2\approx 2.74$ and $l_3 \approx 1.52$. One of its five tetrahedra is equilateral with edge length $l_3$, one is isosceles with base length $l_3$ and $l_2$ as the other length. The three remaining tetrahedra have one equilateral, one isosceles and two non-regular triangles, where the remaining edge has length $l_1$. In the following we specify the five tetrahedra in terms of their spin foam boundary data (not in a twisted spike configuration) and add the edge lengths as further information. To keep the notation consistent with the numbering of tetrahedra, we label the edge lengths by the set of three tetrahedra sharing it in the $4$-simplex.
%\begin{equation}
%    l_{14} = l_1, \quad  l_{12} = l_{13} = l_{15} = l_2, \quad l_{23} = l_{24} = l_{25} = l_{34} = l_{35} = l_{45} = l_3 \quad .
%\end{equation}
\paragraph{Tetrahedron $1$ - non-regular:}
\begin{equation*}
    l_{134} = l_1, \quad  l_{123} = l_{124} = l_2, \quad l_{125} = l_{135} = l_{145} = l_3 \quad .
\end{equation*}
\begin{eqnarray}
    j_{12} = 2j \; &, \quad & \vec{n}_{12} = (0,0,1) \quad , \nonumber \\
    j_{13} = j \; &, \quad & \vec{n}_{13} = \left(0,0.83099, -0.55629\right) \quad, \nonumber \\
    j_{14} = j \; & \quad &  \vec{n}_{14} = \left(0.44287, -0.70315, -0.55629  \right) \quad , \nonumber \\
    j_{15} = j \; &, \quad & \vec{n}_{15} = \left(-0.44287, -0.12784, -0.88743 \right) \quad .
\end{eqnarray}

\paragraph{Tetrahedron $2$ - isosceles:}
\begin{equation*}
    l_{123} = l_{124} = l_{234} = l_2, \quad l_{125} = l_{235} = l_{245} = l_3 \quad .
\end{equation*}
\begin{eqnarray}
    j_{12} = 2j \; &, \quad & \vec{n}_{21} = (0,0,1) \quad , \nonumber \\
    j_{23} = 2j \; &, \quad & \vec{n}_{23} = \left(0,0.88878, -0.458333\right) \quad, \nonumber \\
    j_{24} = 2j \; & \quad &  \vec{n}_{24} = \left(0.473665, -0.75204, -0.458333  \right) \quad , \nonumber \\
    j_{25} = j \; &, \quad & \vec{n}_{25} = \left(-0.94733, - 0.27347, -0.16667 \right) \quad .
\end{eqnarray}

\paragraph{Tetrahedron $3$ - non-regular:}
\begin{equation*}
    l_{134} = l_1, \quad  l_{123} = l_{234} = l_2, \quad l_{135} = l_{235} = l_{345} = l_3 \quad .
\end{equation*}
\begin{eqnarray}
    j_{13} = j \; &, \quad & \vec{n}_{31} = (0,0,1) \quad , \nonumber \\
    j_{23} = 2j \; &, \quad & \vec{n}_{32} = \left(0,0.83099, -0.55629\right) \quad, \nonumber \\
    j_{34} = j \; & \quad &  \vec{n}_{34} = \left(0.44287, -0.85342, -0.27485  \right) \quad , \nonumber \\
    j_{35} = j \; &, \quad & \vec{n}_{35} = \left(-0.44287, - 0.80856, 0.38743 \right) \quad .
\end{eqnarray}

\paragraph{Tetrahedron $4$ - non-regular:}
\begin{equation*}
    l_{134} = l_1, \quad  l_{124} = l_{234} = l_2, \quad l_{145} = l_{245} = l_{345} = l_3 \quad .
\end{equation*}
\begin{eqnarray}
    j_{14} = j \; &, \quad & \vec{n}_{41} = (0,0,1) \quad , \nonumber \\
    j_{24} = 2j \; &, \quad & \vec{n}_{42} = \left(0,0.83099, -0.55629\right) \quad, \nonumber \\
    j_{34} = j \; & \quad &  \vec{n}_{43} = \left(0.44287, -0.85342, -0.27485  \right) \quad , \nonumber \\
    j_{45} = j \; &, \quad & \vec{n}_{45} = \left(-0.44287, - 0.80856, 0.38743 \right) \quad .
\end{eqnarray}

\paragraph{Tetrahedron $5$ - equilateral:}
\begin{equation*}
    l_{125} = l_{135} = l_{235} = l_{145} = l_{245} = l_{345} = l_3 \quad .
\end{equation*}
\begin{eqnarray}
    j_{15} = j \; &, \quad & \vec{n}_{51} = (0,0,1) \quad , \nonumber \\
    j_{25} = j \; &, \quad & \vec{n}_{52} = \left(0,0.94281, -\frac{1}{3}\right) \quad, \nonumber \\
    j_{35} = j \; & \quad &  \vec{n}_{53} = \left(0.8165, -0.47141, -\frac{1}{3}  \right) \quad , \nonumber \\
    j_{45} = j \; &, \quad & \vec{n}_{54} = \left(-0.8165, - 0.47141, -\frac{1}{3} \right) \quad .
\end{eqnarray}

\paragraph{Results}

\begin{figure}
    \centering
    \includegraphics[width = 0.45 \textwidth]{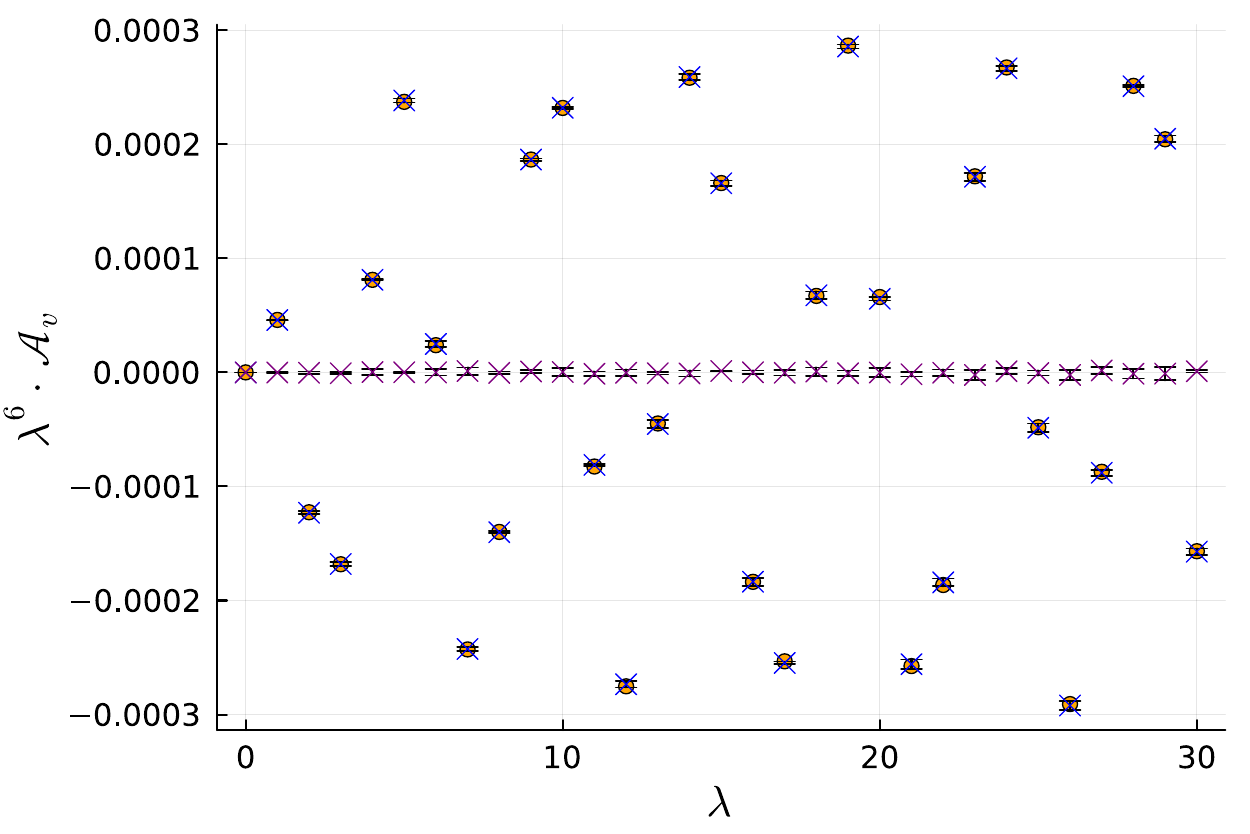}
    \includegraphics[width = 0.45 \textwidth]{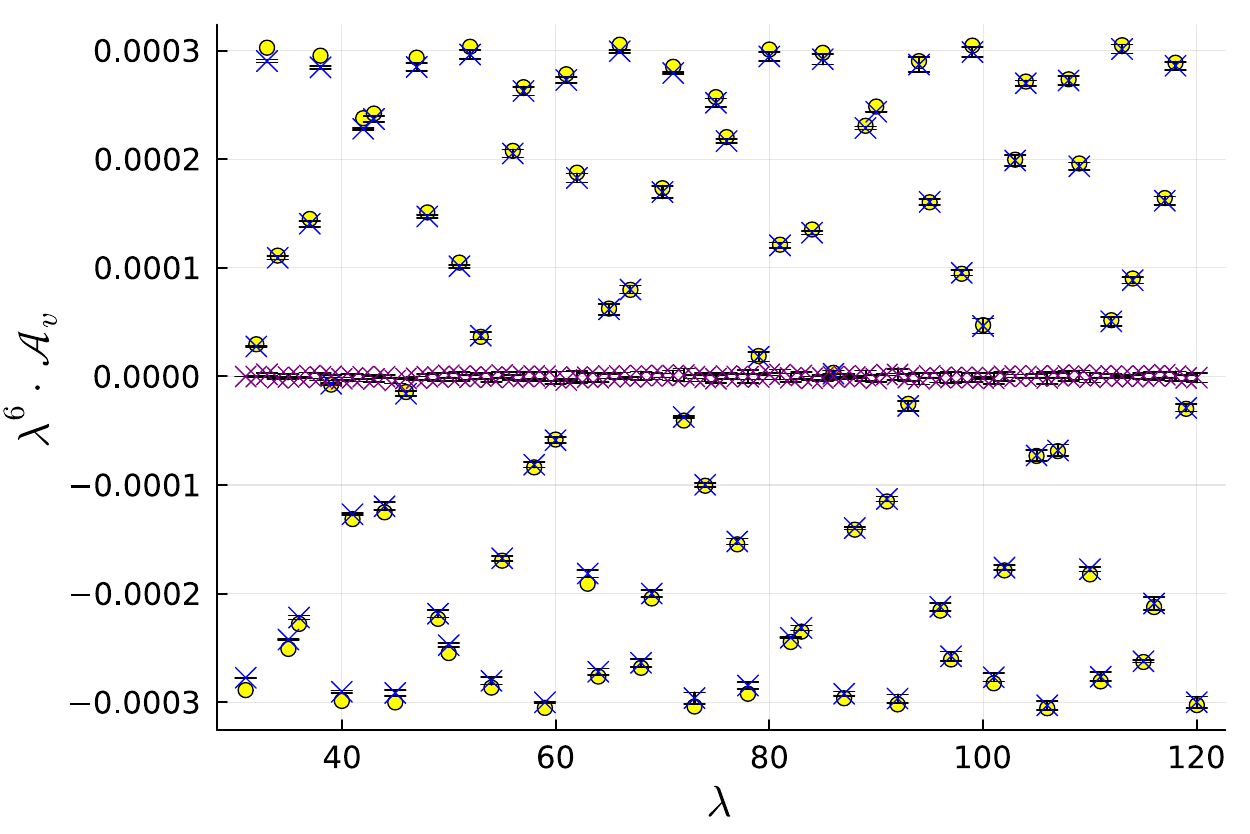} \\
    \includegraphics[width = 0.45 \textwidth]{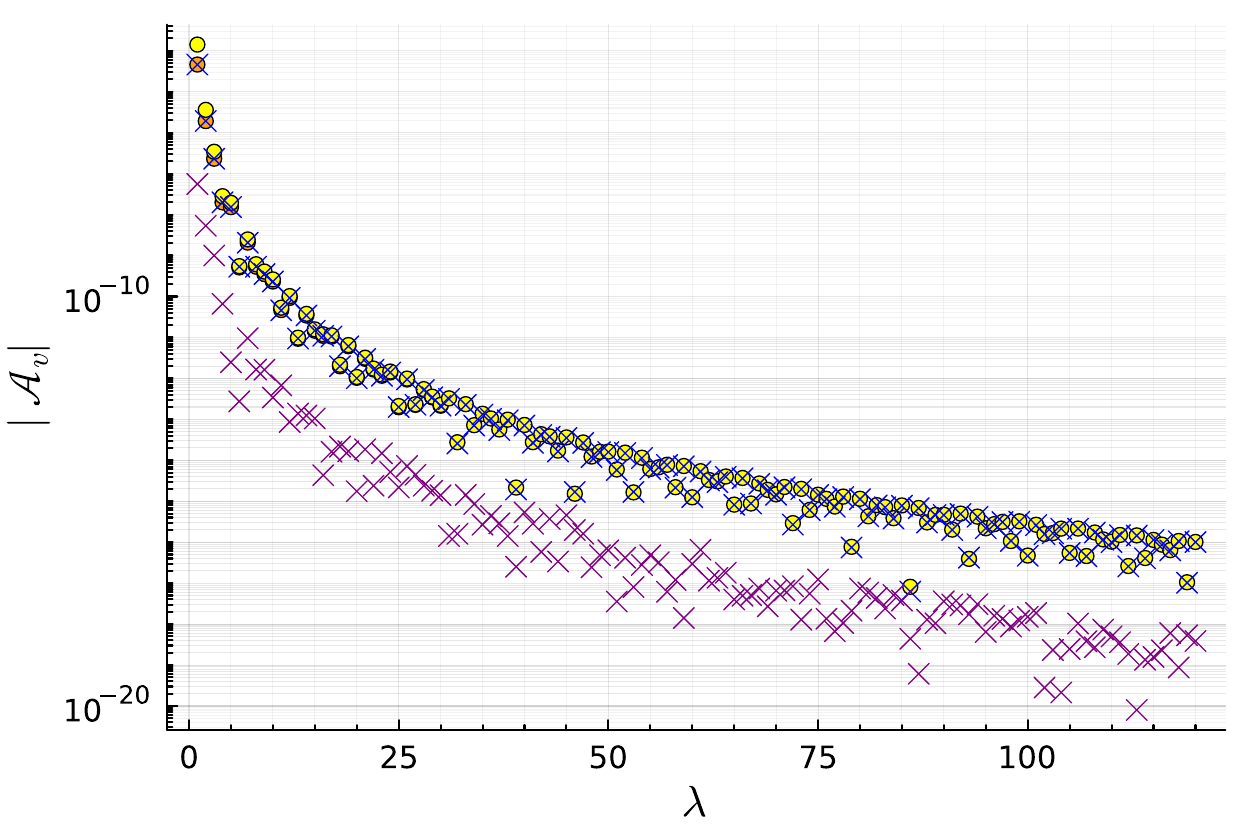}
    \includegraphics[width = 0.45 \textwidth]{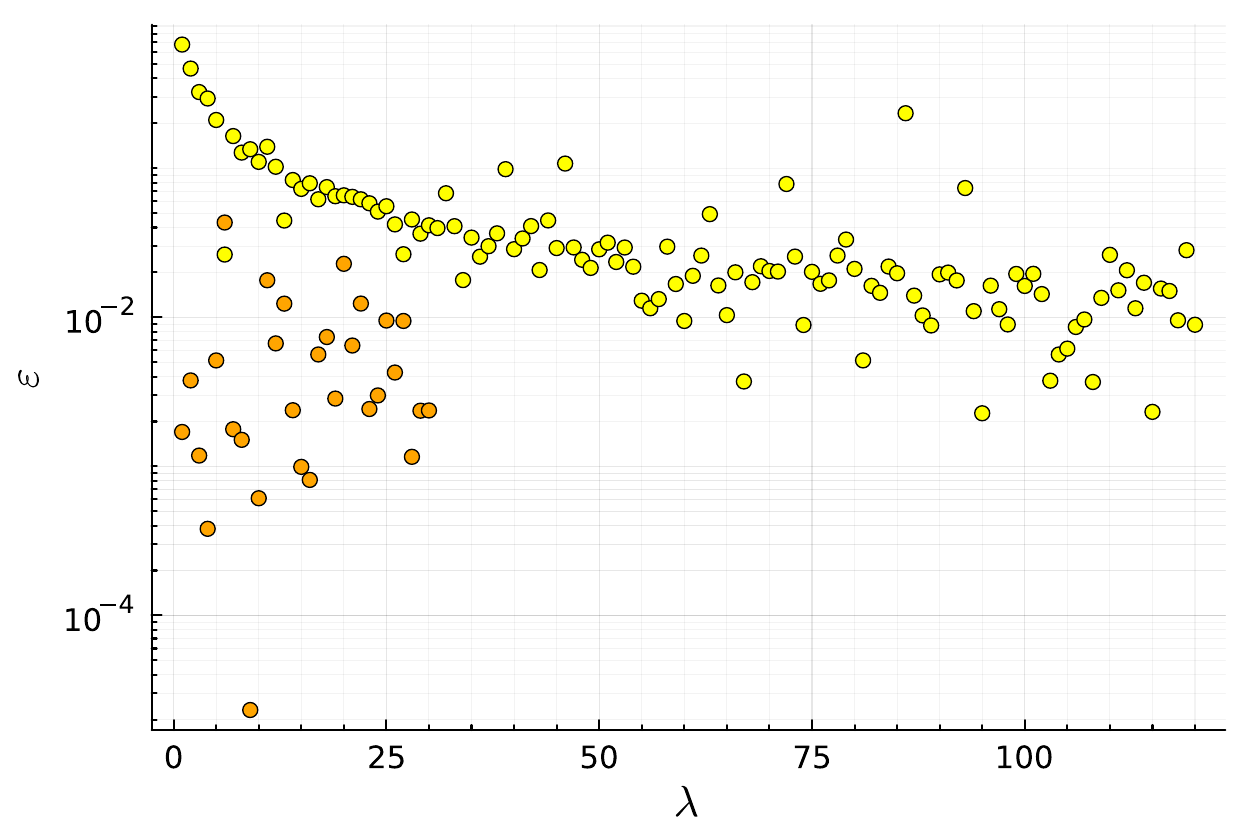}
    \caption{Importance sampling results for the non-regular case. \textit{Top left:} rescaled amplitude compared to full calculation up to $\lambda = 30$. \textit{Top left:} rescaled amplitude compared to asymptotic formula up to $\lambda = 120$. \textit{Bottom left:} logarithmic plot of absolute value of real and imaginary part and comparison to full and asymptotic vertex amplitude. \textit{Bottom right:} relative error with respect to full and asymptotic amplitude.}
    \label{fig:non-regular}
\end{figure}

The results for the non-regular $4$-simplex are summarised in fig. \ref{fig:non-regular}. All plots are over $\lambda$, which scales all spins according to $(j,2j) \rightarrow (\lambda j, \lambda 2j)$. Beginning with the plots of the rescaled amplitude, we observe again a very good agreement of the (real part of the) Monte Carlo simulations with the full coherent vertex amplitude up to $\lambda = 30$. For $\lambda > 30$ we compare to the asymptotic formula, where the results initially deviate slightly from the asymptotic formula and the agreement improves under increasing $\lambda$. Again, this signifies convergence of the asymptotic formula to the full result. For all values of $\lambda$ tested, namely up to $\lambda = 120$, the margins indicating the variance of the runs remain small. More importantly, the imaginary part, which is supposed to vanish, is small also with small error estimates. These findings are confirmed in the logarithmic plots, where we essentially see no deviations for the real part; it seems for the boundary data considered we do not encounter a severe sign problem and observe convergence for the used number of samples. Moreover, the imaginary part is again a few orders of magnitude smaller than the real part. Lastly, the relative error is also fairly low: for the full computation it is below $1\%$ in most cases ($\lambda<30$), while the relative error for the asymptotic formula asymptotes to a level fairly below $10\%$.

\subsection{Computational time}

Besides the accuracy of the algorithm, we must discuss the computational costs and compare them to existing algorithms, in particular the full calculation utilizing tensor network techniques. 
Already, we can infer that the Monte Carlo simulations are more efficient due to the fact that we could perform simulations for significantly larger values of boundary spins. Indeed, there are two limiting factors for the full calculation, memory and computational time. In an effort to reduce the computational time, tensor network methods, e.g. implemented in Julia in the package \verb|TensorOperations|\footnote{\url{https://jutho.github.io/TensorOperations.jl/stable/}}, utilize linear algebra techniques that are highly optimized, fast and can additionally utilize graphics processors (GPUs) for further acceleration. However, this comes with higher memory costs, e.g. for the coherent vertex amplitude we must store a five dimensional array storing all possible values of the $\{15j\}$-symbol, and can quickly go beyond the capabilities of consumer hardware\footnote{In particular memory costs can be optimized by rewriting a problem in terms of tensors with less indices. In general this also improves computational time and helps in adapting code to GPUs, which typically have more limited memory. Such an optimization for coherent vertex amplitudes is promising.}. Here instead, our main focus is on the computational time and we will try to compare the Monte Carlo simulations to a fast current algorithm.

\begin{figure}
    \centering
    \includegraphics[width = 0.45 \textwidth]{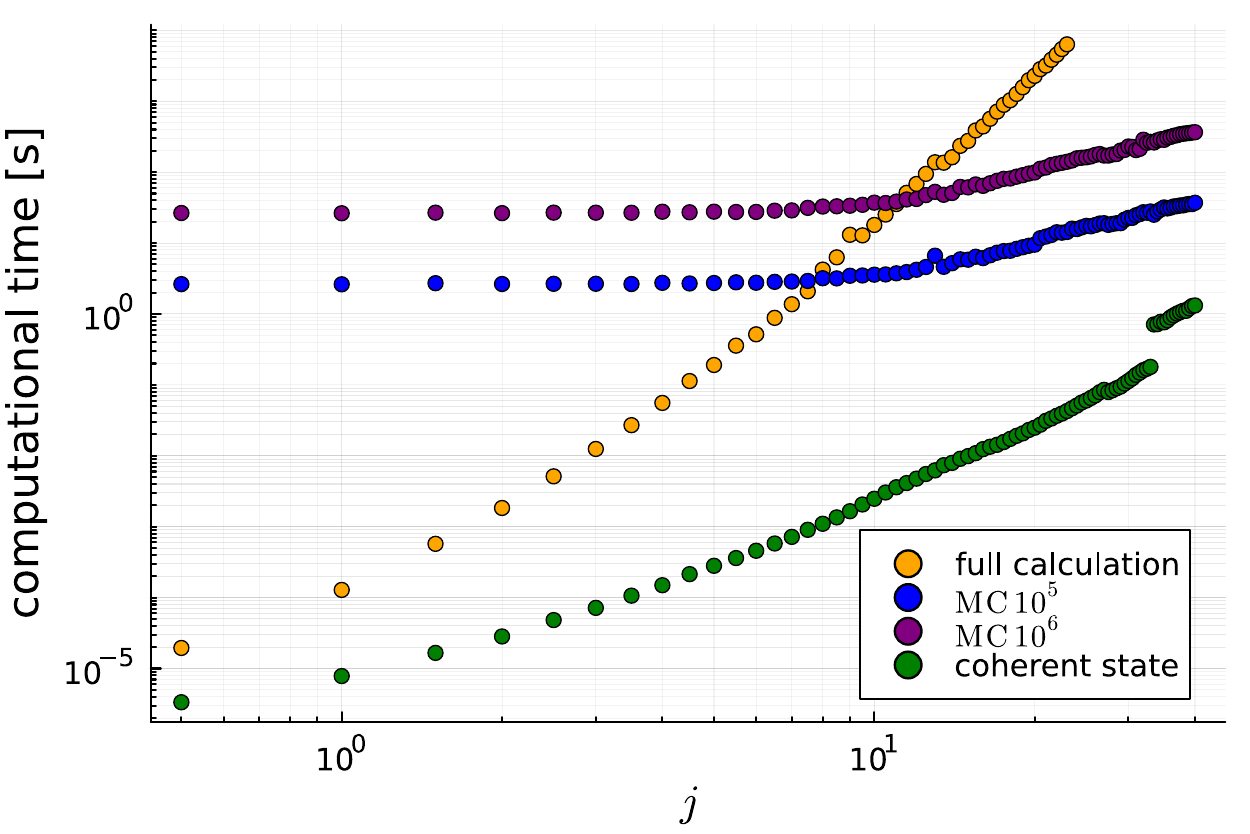}
    \caption{Computational time in seconds plotted over the spin $j$. Measurements for equilateral coherent amplitude / coherent state respectively.}
    \label{fig:loglog_comp}
\end{figure}

We measure all computational times on a desktop computer equipped with an Intel\textregistered~Core\texttrademark~i5-10400 and 32 GB of memory. To measure the computational time and avoid other influences like compilation time, we have used \verb|BenchmarkTools| in \verb|Julia|. To have a better comparison between the algorithms that specifically compute the vertex amplitude, we have precomputed (and benchmarked the computation of) the coherent states as well. For the Monte Carlo algorithm, we show the result for a single run with $10^5$ and $10^6$ samples (each with $10^4$ thermalization steps and $10^3$ steps between taking samples). The data shown in fig. \ref{fig:loglog_comp} are for an equilateral $4$-simplex.

As already suspected in sec.~\ref{sec:spin-foam-nutshell}, the full calculation clearly outperforms the Monte Carlo algorithms at small spins ($j \leq 10$). Recall that the number of intertwiners to sum over is $(2j+1)^5$ in this case, which is lower than or of a similar size as the number of samples used in the Monte Carlo algorithm. We could have lowered the number of samples to make the Monte Carlo algorithm more efficient, yet since the full calculation is fast and accurate in this regime, there is not reason to use Monte Carlo methods. The situation rapidly changes for $j > 10$, as the numerical costs for the full calculation grow exponentially and the Monte Carlo algorithm becomes more efficient. From the double-logarithmic plot we also see that the numerical costs of the Monte Carlo algorithm grow linearly with the number of samples, indicated by the linear shift. Note also that changing the number of thermalization steps and number of steps between samples will have an impact on the numerical costs. Moreover, the numerical costs of the Monte Carlo algorithms grow with growing spins. This is because the evaluation of $\{15j\}$-symbols becomes more costly as the range of the sum over internal auxiliary spins in eq.~\eqref{eq:15jsymbol} grows linearly under uniformly scaling the spins.

Despite this level of efficiency, we cannot use the Monte Carlo algorithm to go to arbitrarily high boundary spins. First, the computational costs grow with growing boundary spins. Second, with higher intertwiner ranges, we expect that we must increase the number of samples further, too. Lastly, increasing the boundary spins has revealed another source of computational complexity, the coherent states. As we can see from fig.~\ref{fig:loglog_comp}, the time to calculate a coherent states grows rapidly with growing boundary spins. It is defined as a sum over three magnetic indices, the fourth being fixed as a function of the remaining three, of four Wigner matrices and a $4jm$-symbol, see eq.~\eqref{eq:coefficient_c}. Since the range of magnetic indices grows with the size of the spin, the costs scale rapidly, but dwarf compared to the costs of the full calculation. Eventually, analytical methods such as asymptotic expansions should be accurate enough and more efficient to compute in such large spin regimes. As we have seen, Monte Carlo methods can bridge this gap between the full calculation and asymptotic formulae for a single vertex, and the gained efficiency can be used to explore larger 2-complexes.

\section{Discussion and conclusion} \label{sec:discussion}

Markov Chain Monte Carlo methods are a powerful tool to explore the dynamics of high dimensional statistical theories. To do so, one uses the partition function to sample probable configurations and approximate expectation values of observables. The same strategy cannot be applied to spin foams, as the amplitudes are typically complex. Thus, we cannot directly use the partition function to define a probability distrubition for sampling, and we must propose a new one. Yet, even if we do so, due to the oscillatory nature of the amplitudes, the results might not converge, which is known as the sign problem. However, we do not know yet how severe this sign problem is.

In this article we propose a probability distribution for sampling intertwiners derived from coherent states, concretely coherent (space-like) tetrahedra, which can be straightforwardly applied to coherent boundary data of spin foams. The idea is to sample with respect to the absolute value of the coefficient of the coherent states expressed in the orthonormal spin network basis. For large spins and coherent tetrahedra, this is a function of one intertwiner label and sharply peaked. The distribution is defined independently for each tetrahedron, and hence can be directly generalized to boundaries with arbitrarily many boundary tetrahedra. We then use this distribution to approximate the so-called coherent vertex amplitude for a $4$-simplex, where we sample the data from its five boundary tetrahedra.

We perform this study in SU$(2)$ BF theory for three different sets of Regge-type boundary data, an equilateral, an isosceles and a non-regular Euclidean $4$-simplex and compare the approximation to the full numerical calculation and the asymptotic formula. It is best to separately discuss the real and imaginary part of the amplitude. For the real part, we observe for almost all cases an excellent agreement to the full calculation and at sufficiently large spins to the asymptotic formula already at a moderate sample size. These results are very encouraging for two reasons: first, the results are obtained at substantially lower numerical costs compared to the full numerical calculation, which allows us to instead increase the size of the boundary spins by almost an order of magnitude. Second, the fact that the results have converged to the known results at a moderate number of samples suggests that the sign problem is not severe in these cases, at least for the real part in most cases. However, as the coherent amplitude oscillates with the Regge action under uniform scaling of the boundary spins, some amplitudes almost vanish. In these cases, the sign problem is more severe: the various terms summed over cancel almost completely, which is difficult to capture using Monte Carlo sampling methods.

In contrast to the real part, the sign problem is present in the imaginary part for all boundary data. This is the case for our choice of global phase, which renders the full amplitude purely real\footnote{This phase choice is convenient for demonstrating the sign problem. For other choices the sign problem manifests itself by the difficulty of determining the phase. Recall that we have used the phase from the full calculation and scaled it with the spins.}. Therefore, the imaginary part always vanishes by definition, i.e. the various imaginary parts of summands exactly cancel. Again, this sign problem is challenging for Monte Carlo algorithms, yet for the cases studied the imaginary part is usually a few orders of magnitude smaller than the real part. An exception are the cases where the entire amplitude is small (compared to boundary data of similar size). Moreover, we observe that convergence of the imaginary part is worse for larger boundary spins, but still under control. We expect that this can be improved by increasing the number of samples, yet it is not clear how efficient this is.

In terms of computational costs, the full calculation is only more efficient for small spins. Above spins of the order $10$ the costs grow so rapidly that the Monte Carlo method quickly becomes more efficient and provides an accurate approximation. Hence, it should be possible to use the freed resources for simulations of larger boundary spins and bridge the gap to the asymptotic approximations. The good accuracy is also an advantage of the importance sampling algorithm compared to randomly sampling coherent intertwiners. Random sampling provides decent and convergent results for real and imaginary parts, but requires more samples to do so. Additionally, the number of samples required to obtain a convergent results grow with growing boundary spins. Still random sampling is useful for sampling bulk spins and intertwiners~\cite{Dona:2023myv}.

To sum up, sampling coherent intertwiners using Monte Carlo techniques is efficient and accurate at simulating the coherent SU$(2)$ BF vertex amplitude. This is an important proof of principle and opens the door for more efficient studies of spin foams defined on larger 2-complexes and for larger boundary spins, such that we can better bridge the gap between the full calculation and asymptotic approximations, the latter e.g. via the complex critical points method~\cite{Han:2021kll,Han:2023cen}.

\subsection{Generalizing the algorithm}

 In the following, we discuss several directions in which the algorithm can be generalized, namely to larger 2-complexes, to the Lorentzian EPRL model and sampling bulk intertwiners.

\paragraph{Larger 2-complexes with boundary}

If the boundary states for spin foams defined on larger 2-complexes are given by coherent states, the algorithm presented in this article can be straightforwardly generalized to this setting. The probability distribution is defined individually for each intertwiner, i.e. they are explicitly independent from one another. Hence, this method can be extended to almost arbitrarily many coherent boundary tetrahedra. Of course, more boundary intertwiners readily imply increased numerical costs alone for computing coherent states (and the normalization to define the probability distribution). More importantly, more degrees of freedom generically require larger numbers of samples to obtain convergent results. Obviously, the same is true for the full calculation and, as is typical for Monte Carlo methods, we expect our algorithm to more beneficially scale with the size of the system and the size of boundary spins.

So far we have left bulk variables, i.e. representations and bulk intertwiners, unaddressed. The most direct option is to sum over these variables, yet this becomes inefficient quickly: if we have $N$ samples for all boundary data combined, we must sum over all bulk variables $N$ times. This is probably still more efficient than summing over all data exactly, but loses the beneficial scaling of Monte Carlo algorithms. Instead, one can use random sampling for the bulk variables as in~\cite{Dona:2023myv}.

\paragraph{Lorentzian (and Riemannian) EPRL coherent vertex amplitude}

BF theory written in the spin foam representation is computationally non-trivial, simpler than modern 4d spin foam models, yet similar enough to those models for methods to be transferable. Indeed, both the Riemannian and Lorentzian EPRL models use coherent states derived from the same SU$(2)$ boundary states. More precisely, the same coefficents $c_\iota(\{j_i\},\{\vec{n}_i\})$ appear in the coherent amplitudes, see e.g. the derivation for the Lorentzian model in the seminal paper by Speziale~\cite{Speziale:2016axj}.
Hence, the sampling algorithm should be straightforwardly applicable, where one has to replace the SU$(2)$ $\{15j\}$-symbol by the appropriate Spin$(4)$ / SL$(2,\mathbb{C})$ vertex amplitude with orthonormal boundary spin network data. Computing said amplitude is computationally non-trivial, in particular for the Lorentzian model~\cite{Dona:2019dkf,Gozzini:2021kbt}. Therefore, the here presented algorithm might be highly beneficial for the Lorentzian case, if it helps approximate coherent amplitudes with fewer samples.

\paragraph{Sampling bulk intertwiners}

Given the good convergence of the algorithm presented here, the question arises whether this method can also be applied to bulk variables. The following considerations are more speculative, but we believe this method to have potential there as well.

Without the boundary, we clearly cannot rely on external data to choose a probability distribution. Instead we have to guess one and at best it should be one adapted to the dynamics of the system we are considering. Already for the boundary, the coherent states work best as the spins are larger and one approaches the asymptotic regime of the vertex amplitude~\cite{Conrady:2008mk,Barrett:2009ci,Barrett:2009mw,Barrett:2009as}. For a single vertex amplitude, asymptotic analysis informs us for which boundary data the amplitude possesses critical points and thus which amplitudes dominate in the large spin limit. Hence, it might be possible to use the information of critical points in turn to guess a probability distribution for bulk intertwiners.

However this is easier said than done. Typically, there exist two types of solutions to the critical point equations for a single $4$-simplex, Regge geometries and so-called vector geometries~\cite{Conrady:2008mk,Barrett:2009ci,Barrett:2009mw,Barrett:2009as,Kaminski:2017eew,Liu:2018gfc}. The former have two inequivalent critical points, giving rise to the characteristic oscillations of the amplitude with the Regge action, while the latter only have one. For our considerations more relevant is the fact that Regge geometries span a $10$-dimensional space in the boundary Hilbert space of a $4$-simplex, while vector geometries span a $15$-dimensional space~\cite{Dona:2017dvf}. Indeed, for fixed spins a Regge geometry corresponds to an isolated point in that space, whereas vector geometries span an intricate $5$-dimensional space of configurations and are not isolated, i.e. one can continuously vary the parameters and obtain another vector geometry. Hence, it is unrealistic to guess a probability distribution for bulk intertwiners taking all critical configurations into account.

Instead, one can entertain the thought to devise a probability distribution defined only from Regge geometries of the vertices. For fixed spins, we determine the Regge critical points by looking for length configurations of flat simplices compatible with the assigned areas~\cite{Asante:2018wqy,Asante:2024rrd}. There can be multiple configurations fulfilling these requirements. For these Regge geometries, we can compute the spin foam boundary data for each of the tetrahedra and derive a probability distribution from their coherent states. In case there are multiple Regge critical points, we can choose one, e.g. as it best fits prescribed boundary data; most general would be to consider a superposition of the critical coherent states. However, so far we have only considered the information from one vertex, yet each bulk edge is shared by two vertices. Therefore we propose to use a (suitably normalized) product of distributions, which is informed by both vertices.
This could lead to situations, where the critical points associated to the two vertices do not agree and the coherent states describe non-matching tetrahedra. Hence, the algorithm would mostly sample intertwiners where the distributions overlap, which however do not fit well with neither critical point(s) and would lead to an exponential suppression. Such scenarios strongly resonate with the ideas of gluing constraints introduced in effective spin foams~\cite{Asante:2020qpa,Asante:2021zzh} and computed for spin foams models in the context of a hybrid algorithm~\cite{Asante:2022lnp}.

This idea has a few drawbacks. For the numerical implementation, there are two challenges. The first one is to compute the Regge geometries for various spin assignments, the spin foam boundary data and the coherent states. Indeed, we have seen that the latter can become costly for large spins as well. Second, even though we would then sample over intertwiners, the sum over bulk spins is unaddressed. Thus, one has to repeat this procedure and then sample for each spin assignment in the bulk. Besides these practical considerations, this choice represents also a substantial truncation of the spin foam partition function as we are excluding many bulk intertwiner degrees of freedom. We expect that some of these will play a role in the asymptotic regime as vector geometries. Conversely, this setup allows us to investigate the relevance of vector geometries in the partition function, e.g. whether they are suppressed for geometric boundary data in larger complexes. Moreover, in the Lorentzian theory we can choose to only sample Regge boundary data for critical points corresponding to Lorentzian $4$-simplices, i.e. exclude vector geometries including those Euclidean Regge geometric boundary data~\cite{Barrett:2009mw,Dona:2019dkf}. In a way, its underlying idea is similar to effective spin foams~\cite{Asante:2020qpa,Asante:2021zzh}, but implemented in the full theory. Thus, a comparison to this method as well as complex critical points~\cite{Han:2021kll,Han:2023cen} would be interesting.

\paragraph{Closing remarks}
To conclude, we have presented an algorithm that samples boundary intertwiners according to their associated coherent boundary data. For the example of the SU$(2)$ coherent vertex amplitde considered here it offers a good approximation at much lower numerical costs compared to the full numerical calculation and is able to bridge the gap to the regime where the asymptotic formula is valid. It is straightforward to generalize this to larger 2-complexes / triangulations with boundary. While the sign problem is present, it is under control in the vast majority of cases and most results show a good convergence. Therefore, we argue that our algorithm is ideal in identifying the relevant boundary intertwiners, allowing us to well approximate the full amplitude at lower costs in computational time and memory. We expect this advantage to grow even more for calculations with more boundary intertwiners and larger boundary spins, allowing us to tackle previously unfeasible computations. Furthermore, we are optimistic that the algorithm can be generalized to and will perform well for the Lorentzian EPRL model. We also suggest an application to bulk intertwiners, which however corresponds to a big truncation excluding vector geometries and must be justified. What is still missing is a dynamics informed method to sample bulk representations. We leave these specific questions and developments to further optimize spin foam numerics for future research.

\section*{Acknowledgements}
The author would like to thank Seth Asante, Jos\'e Sim\~ao, Alexander Jercher, Pietro Don\`a, L Glaser and Kevin Siebert for fruitful discussions and clarifying questions. Additionally, the author would like to thank the anonymous referee for constructive suggestions on how to improve the manuscript, in particular a discussion of convergence properties.
The author also gratefully acknowledges access to the Ara-Cluster at FSU Jena, which was used to produce most of the numerical results presented in this article. 

The author gratefully acknowledges support by the Deutsche Forschungsgemeinschaft (DFG, German Research Foundation) project number 422809950. 

% Appendices ==================================================================

% ==================================================================

\appendix

\section{Convergence of algorithm for coherent vertex amplitude} \label{app:convergence}
In this appendix, we want to briefly show and highlight the convergence properties of the algorithm for the SU$(2)$ BF coherent vertex amplitude, more precisely for equilateral boundary data and different boundary spins $j$. In fact, as discussed in the main text, the convergence behavior for real and imaginary part of the amplitude is starkly different, where the imaginary part always suffers from the sign problem. Moreover, for some boundary spins this also applies to the real part. We will demonstrate these different behaviors by showing $30$ Monte Carlo estimates for different numbers of samples for different boundary spins $j$. The number of samples range from $10^4$ to $10^6$. To keep the results comparable for different boundary spins, we rescale the amplitude by $j^6$.

\subsection{Examples of good convergence}
\begin{figure}
    \centering
    \includegraphics[width = 0.45\textwidth]{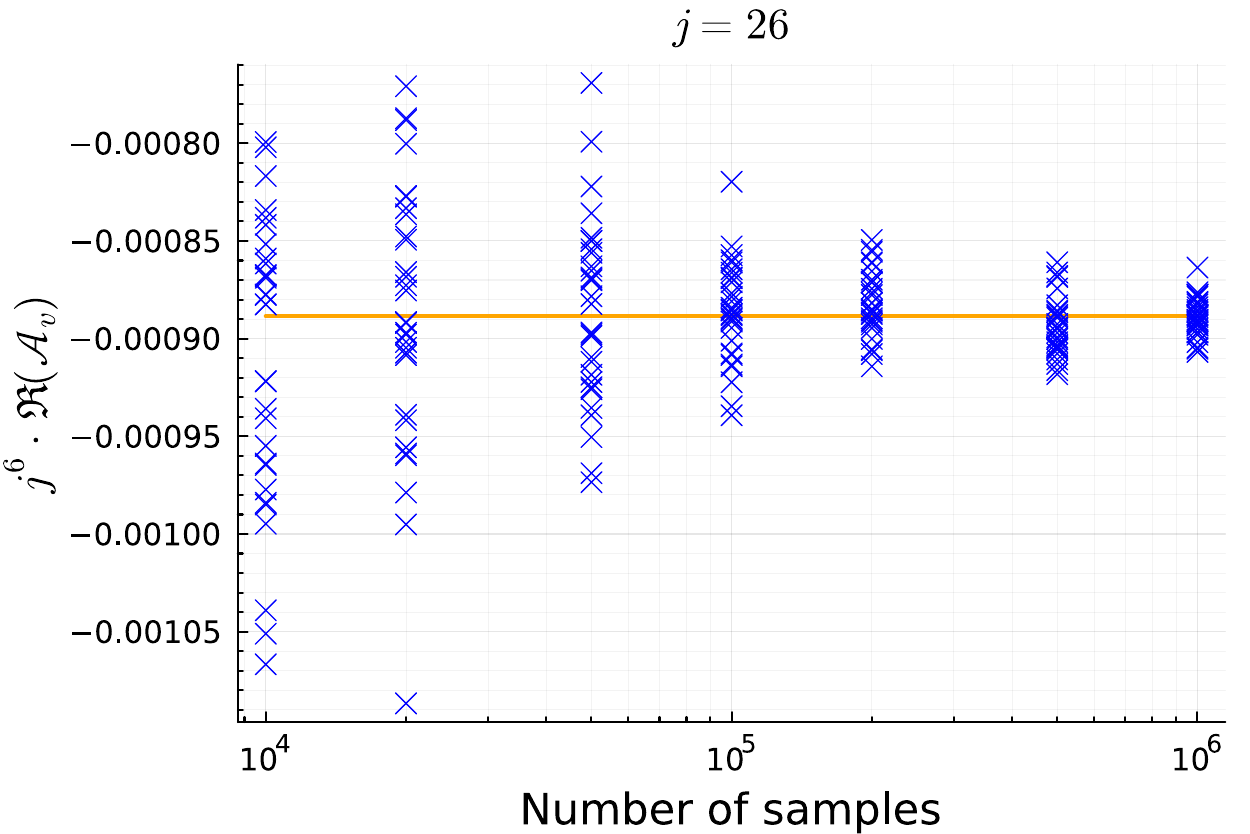} \quad 
    \includegraphics[width = 0.45\textwidth]{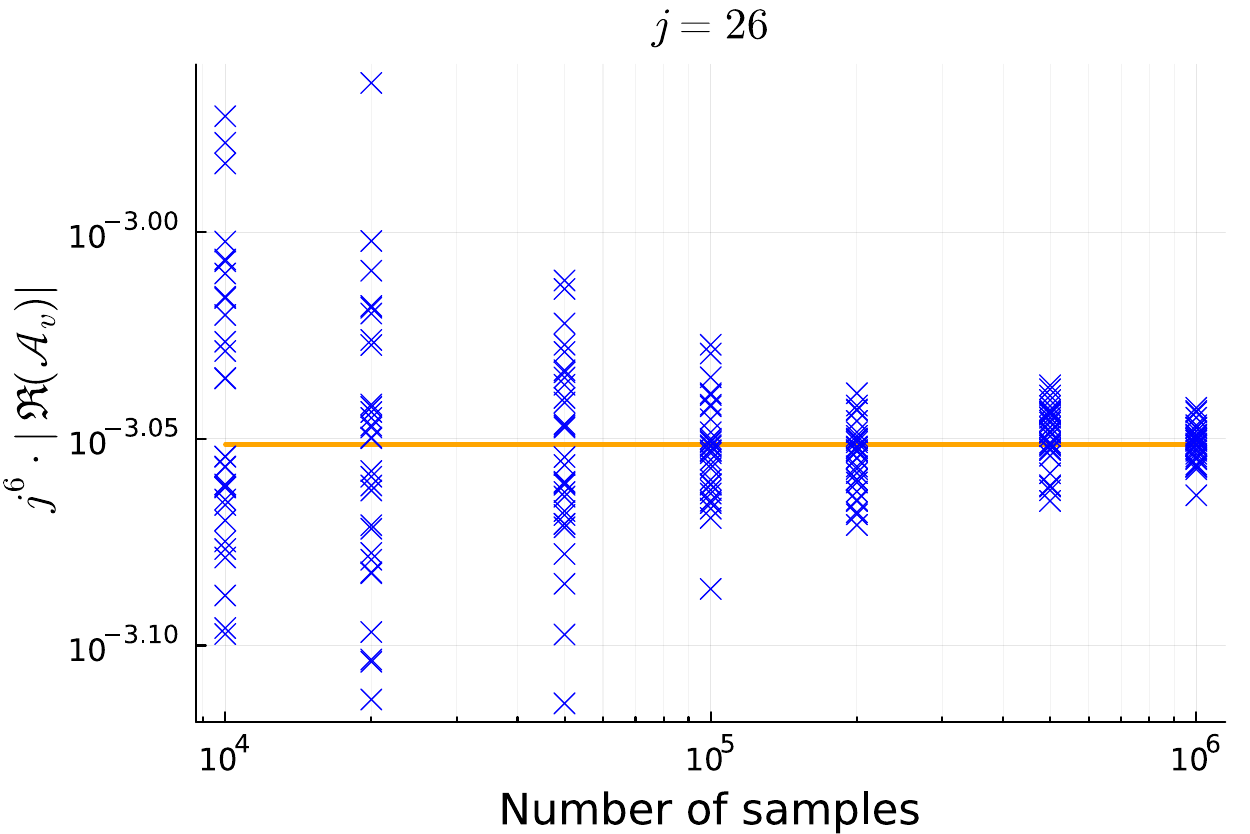}
    \caption{Plots of 30 Monte Carlo estimates of the real part of the rescaled equilateral vertex amplitude for $j=26$. \textit{Left:} Real part. \textit{Right:} Absolute value of the real part in logarithmic scale.}
    \label{fig:plot_26}
\end{figure}

\begin{figure}
    \centering
    \includegraphics[width = 0.45\textwidth]{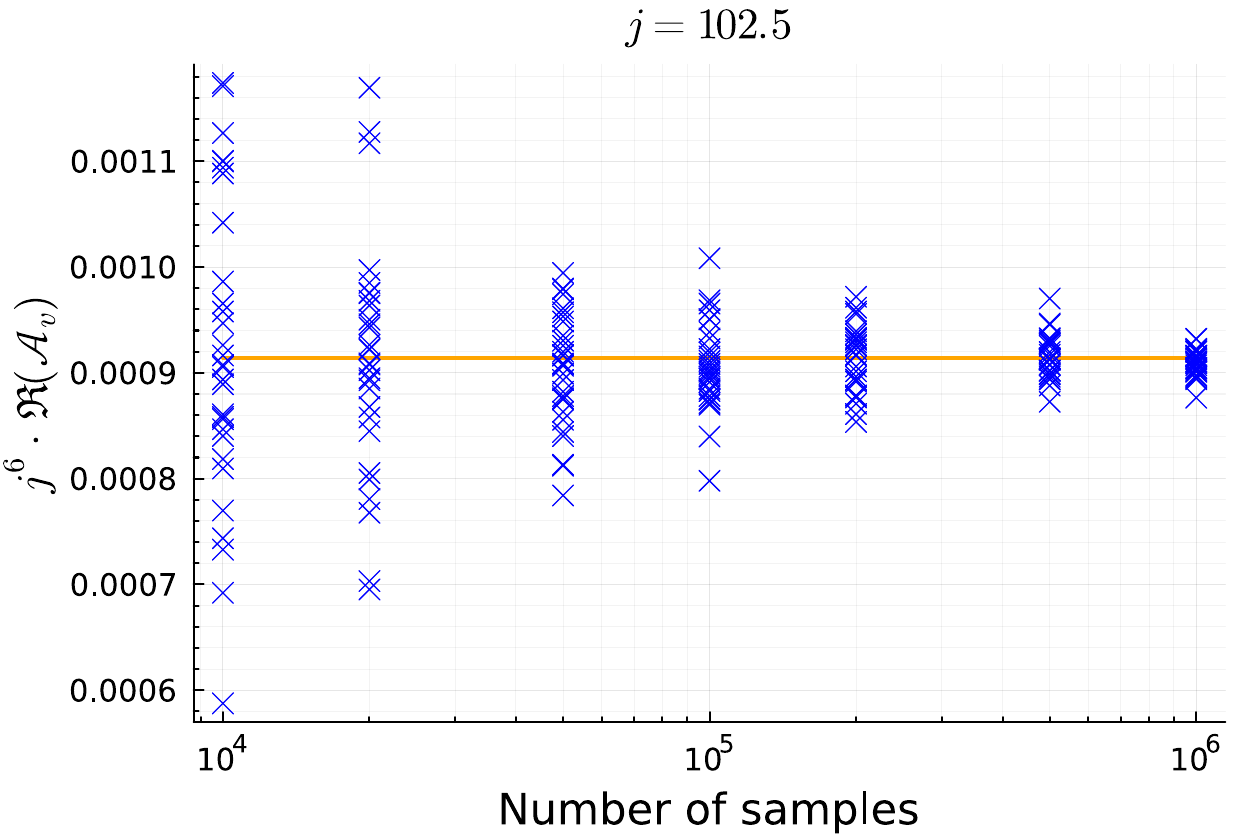} \quad
    \includegraphics[width = 0.45\textwidth]{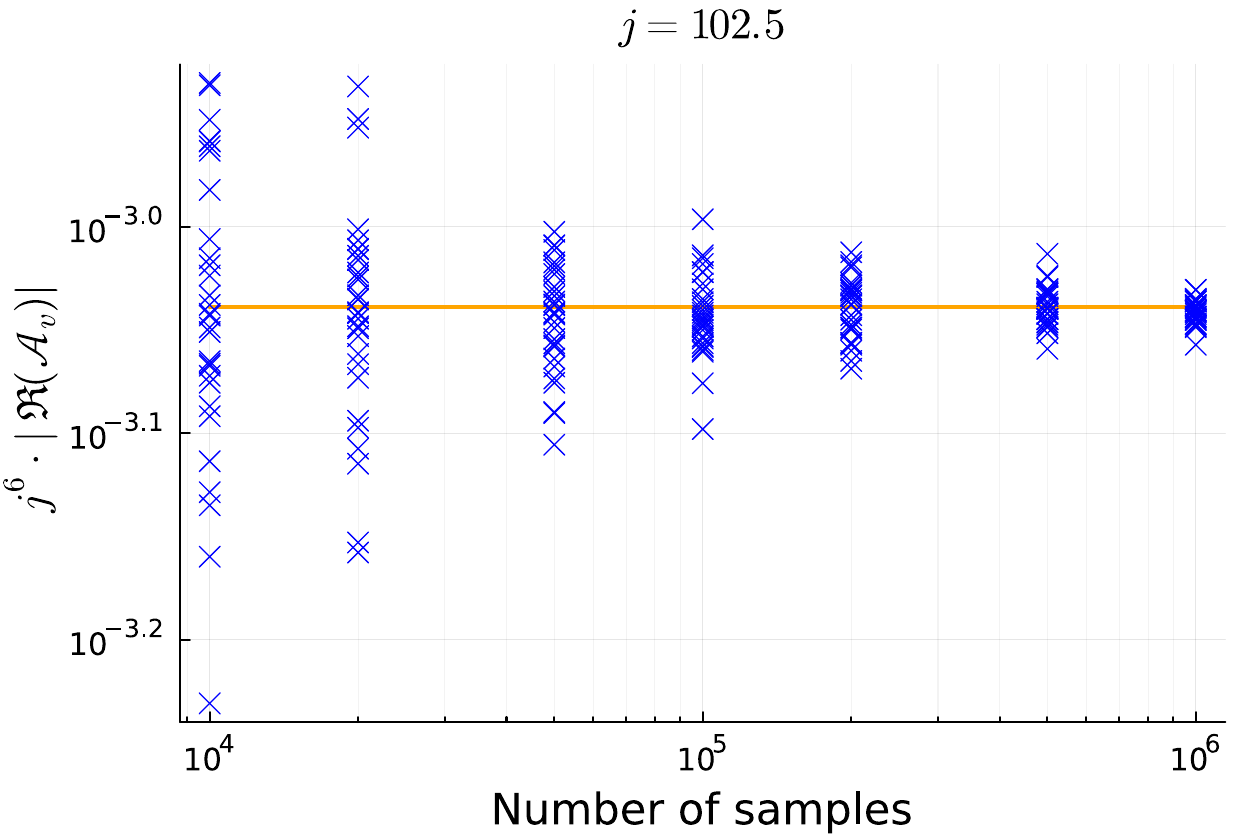}
    \caption{Plots of 30 Monte Carlo estimates of the real part of the rescaled equilateral vertex amplitude for $j=102.5$. \textit{Left:} Real part. \textit{Right:} Absolute value of the real part in logarithmic scale.}
    \label{fig:plot_1025}
\end{figure}

\begin{figure}
    \centering
    \includegraphics[width = 0.45\textwidth]{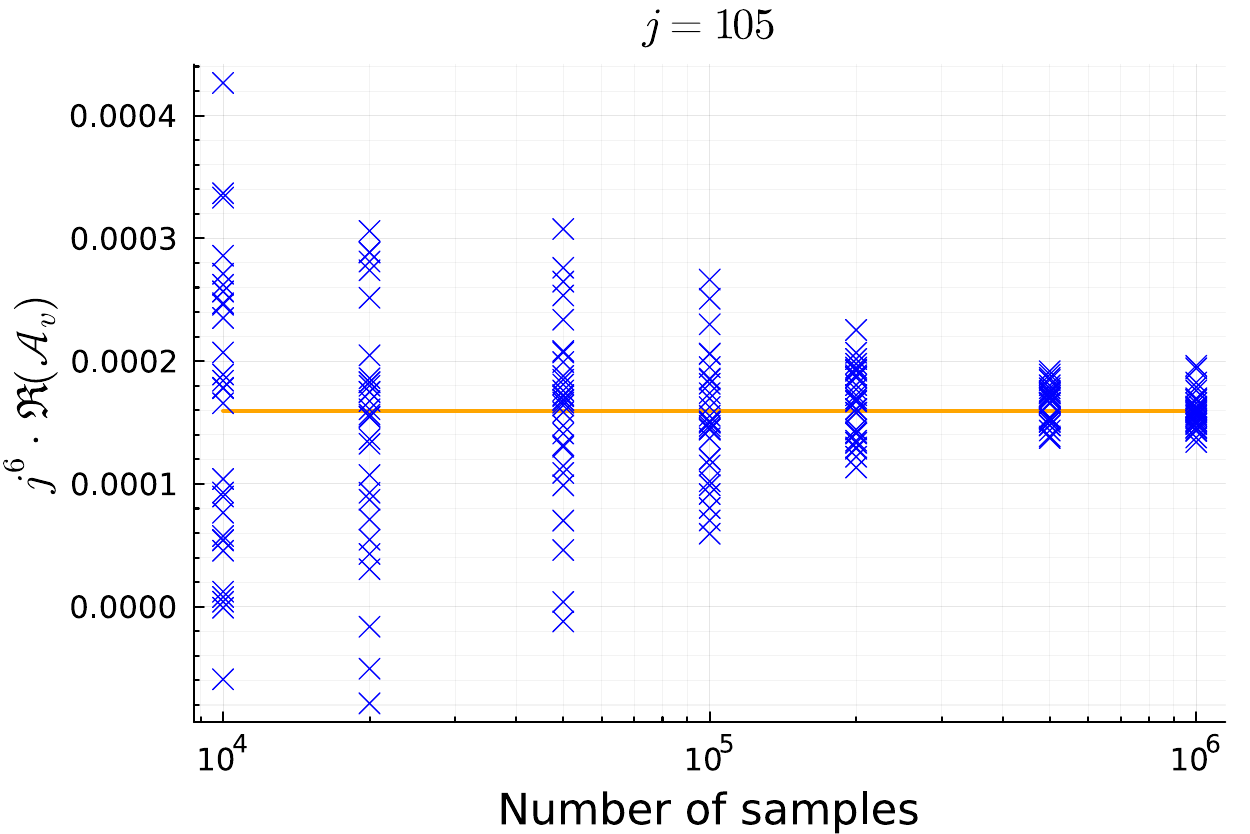} \quad
    \includegraphics[width = 0.45\textwidth]{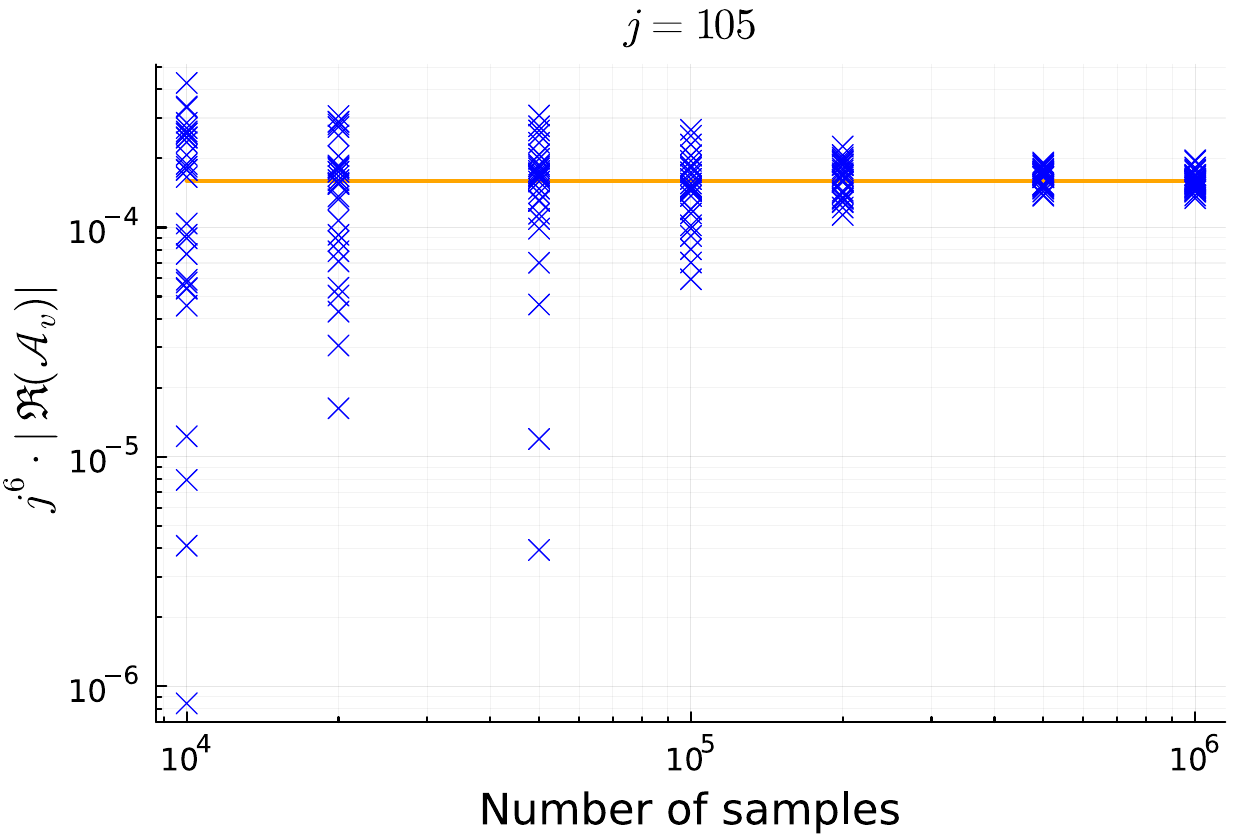}
    \caption{Plots of 30 Monte Carlo estimates of the real part of the rescaled equilateral vertex amplitude for $j=102.5$. \textit{Left:} Real part. \textit{Right:} Absolute value of the real part in logarithmic scale.}
    \label{fig:plot_105}
\end{figure}

As examples for good convergence, we present the real part of the amplitude for three choices of boundary spins, $j=26$, $j=102.5$ and $j=105$. The first two examples, see fig. \ref{fig:plot_26} and \ref{fig:plot_1025} respectively, are close to the minimum / maximum of the oscillations of the vertex amplitude respectively, and show differences in convergence due to the size of spins. The latter example, $j=105$ in fig. \ref{fig:plot_105}, is closer to a root of the oscillation, and roughly one order of magnitude smaller than the value for $j=102.5$ (after rescaling).

Because their behavior is very similar, we discuss the cases $j=26$ and $j=102.5$ at the same time. Clearly, we can see a good convergence of results to the exact value as we increase the number of samples, i.e. the estimates land closer to the exact value. Already for small numbers of samples the agreement is fairly good, as the estimates have the same order of magnitude and the correct sign. Actually, the rescaled amplitude has a similar value in both cases, such that we can compare the convergence in both cases. We note that the case for $j=102.5$ in fig. \ref{fig:plot_1025} shows larger absolute deviations. This is to be expected, as the configuration space grows exponentially in dimension as the boundary spins are increased. To compute this amplitude by brute force, we would have to sum over $(2 \cdot 102.5 + 1)^5 \sim 3.7 \cdot 10^{11}$ configurations. We achieve an excellent approximation for a much smaller number of samples.

The third example for good convergence is for $j=105$, which compared to $j=102.5$, is closer to the root of oscicalltions; the real part of rescaled amplitude is roughly one order of magnitude smaller than for $j=102.5$. In fig. \ref{fig:plot_105} we still observe a good convergence for a sufficienlty large sample size. Indeed, for $10^4$ samples, the results can be quite off and may rarely even produce the wrong sign. However, when increasing the sample size the convergence drastically improves albeit later than for optimal example.

\subsection{Examples of bad convergence}
Bad convergence can be observed for the imaginary part of the amplitude for all choices of boundary spins and for the real part of the amplitude if it almost vanishes, i.e. it is close to a root of the (rescaled) amplitude. The example for the real part is $j=23.5$ and for the imaginary part we take $j=102.5$ (to contrast it with the good convergence of its real part). Indeed, as the imaginary part is supposed to vanish for all boundary spins, the convergence plots for the imaginary part essentially look the same.

\begin{figure}
    \centering
    \includegraphics[width = 0.45\textwidth]{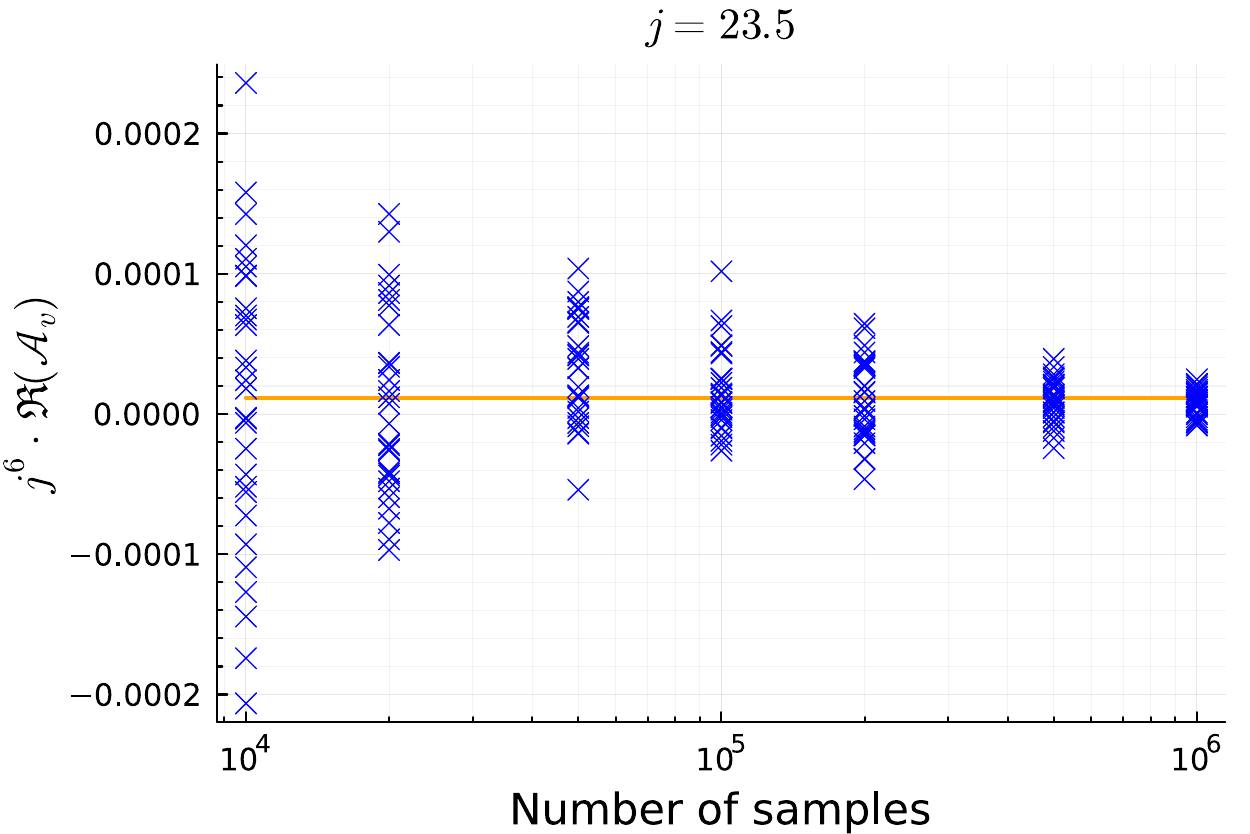} \quad 
    \includegraphics[width = 0.45\textwidth]{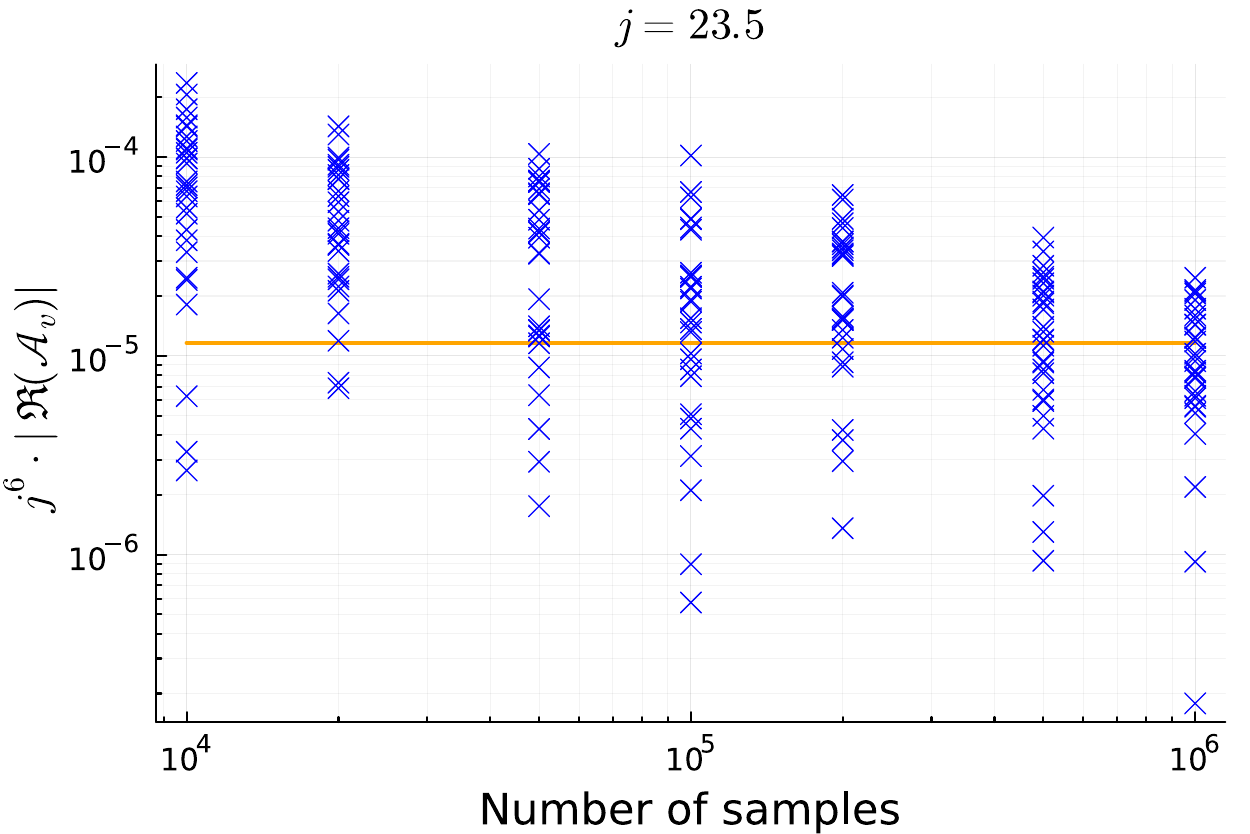}
    \caption{Plots of 30 Monte Carlo estimates of the real part of the rescaled equilateral vertex amplitude for $j=23.5$. \textit{Left:} Real part, \textit{Right:} Absolute value of the real part in logarithmic scale.}
    \label{fig:plot_23_5}
\end{figure}

$j=23.5$ in fig. \ref{fig:plot_23_5} is an example for slow convergence of the real part of the amplitude. Again, the rescaled amplitude allows us to compare results for different spins. Compared to the case $j=105$, which already showed slightly worse convergence, the result is one order of magnitude smaller and we suspect destructive interference between different summands. This is reflected in the convergence under increasing number of samples. For small number of samples the estimates fluctuate around zero, and while we see a convergence of the estimates to the exact value, the order of magnitude and sign can still be wrong at the maximal number of samples considered here.

\begin{figure}
    \centering
    \includegraphics[width = 0.45\textwidth]{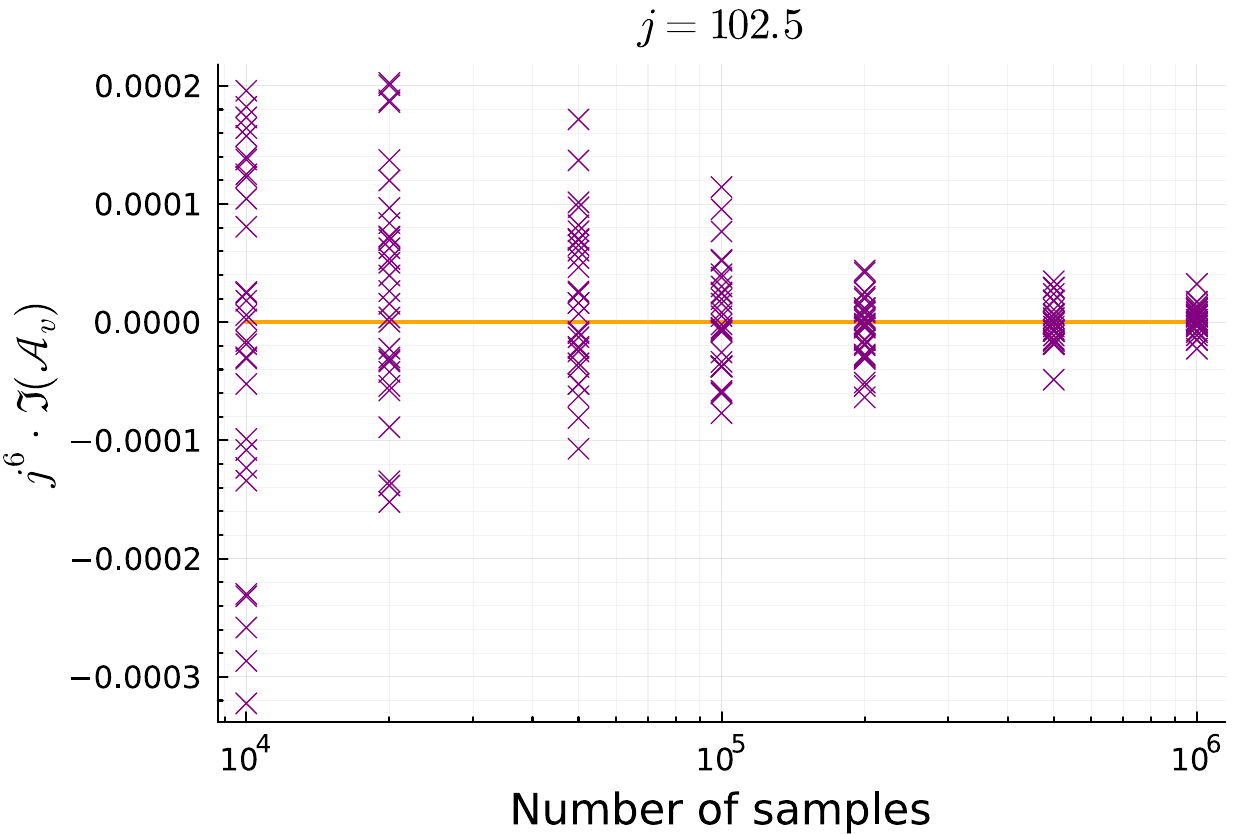} \quad 
    \includegraphics[width = 0.45\textwidth]{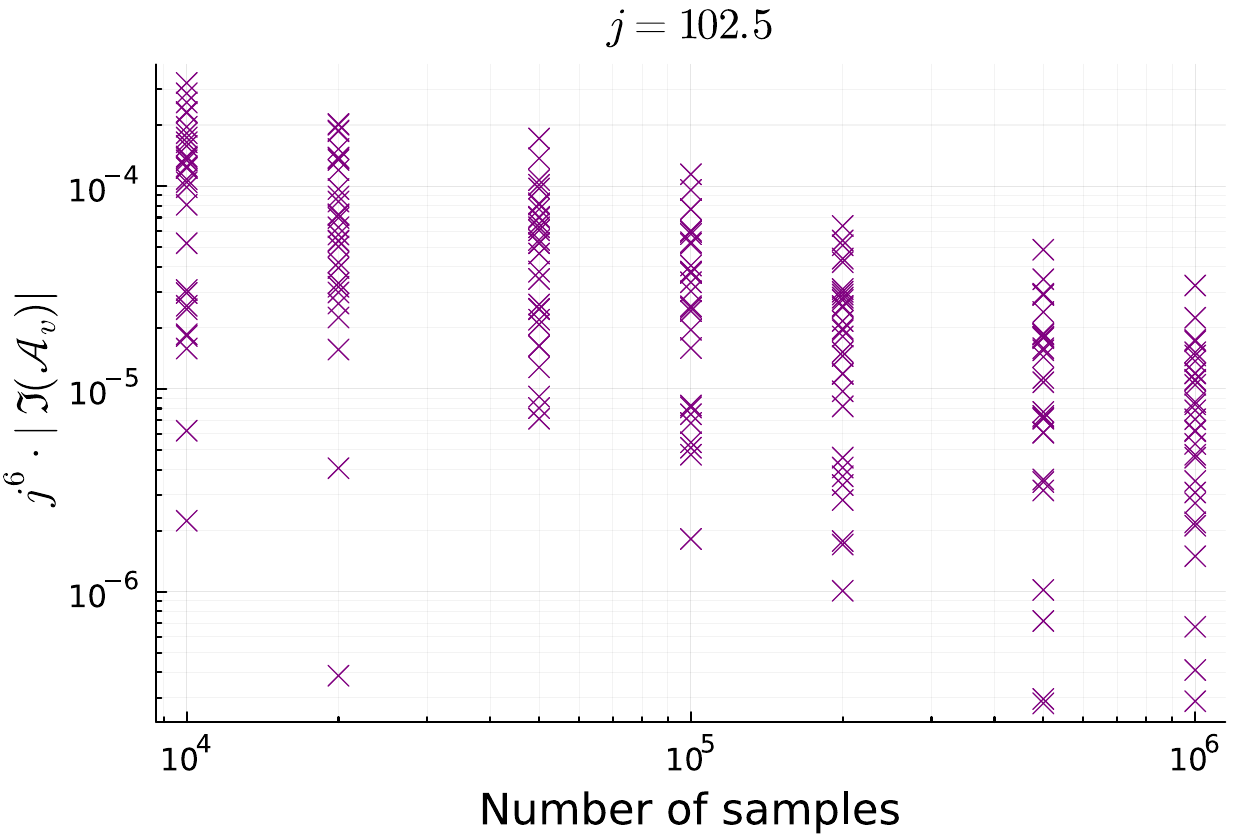}
    \caption{Plots of 30 Monte Carlo estimates of the imaginary part of the rescaled equilateral vertex amplitude for $j=23.5$. \textit{Left:} Real part, \textit{Right:} Absolute value of the real part in logarithmic scale.}
    \label{fig:plot_1025_imag}
\end{figure}

Lastly, we consider the imaginary amplitude for $j=102.5$, an example for which the real parts converges well. For all possible values of boundary spins, the imaginary part of the amplitude always vanishes by our choice of phase of coherent states. Thus, all summands contributing to this result must exactly cancel (at least to numerical accuracy), which is the very definition of the sign problem. So, while we see the estimates decreasing in size when increasing the number of samples, we need roughly $100\times$ more samples to lower the estimates by an order of magnitude. This is costly indeed and we suspect to be inefficient to reach a vanishing imaginary part (up to numerical accuracy).

\bibliographystyle{JHEP}
\bibliography{main}

\end{document}